\documentclass[structabstract]{aa}  
\usepackage{graphicx}
\usepackage{txfonts}
\begin{document}
    \title{New insights on Stephan's Quintet: exploring the shock in three dimensions\thanks{Based on observations taken at the 3.5m telescope at Calar Alto Observatory}}

   \author{
          J. Iglesias-P\'{a}ramo
          \inst{1,2},
          L. L\'{o}pez-Mart\'{\i}n\inst{3,4},
          J.M. V\'{\i}lchez\inst{1},
          V. Petropoulou\inst{1}
          \and
          J.W. Sulentic\inst{1}
          }

   \institute{
             Instituto de Astrof\'{\i}sica de Andaluc\'{\i}a (CSIC), Glorieta de la Astronom\'{\i}a s.n., 18008 Granada, SPAIN\\
              \email{jiglesia@iaa.es, jvm@iaa.es, vasiliki@iaa.es, sulentic@iaa.es}
         \and
Centro Astron\'{o}mico Hispano Alem\'{a}n, C/ Jes\'{u}s Durb\'{a}n Rem\'{o}n, 2-2, 04004 Almer\'{\i}a, SPAIN\\
         \and
             Instituto de Astrof\'{\i}sica de Canarias, C/ V\'{\i}a L\'{a}ctea s.n., 38200 La Laguna, SPAIN\\
             \email{luislm@iac.es}
         \and
             Departamento de Astrof\'{\i}sica, Universidad de La Laguna, E-38205 La Laguna, Tenerife, SPAIN\\
             }


 
  \abstract
   {}
   {In this paper we study the ionized gas emission from the large scale shock region of
Stephan's Quintet (SQ).}
   {
We carried out IFU optical spectroscopy on three pointings in and near
the SQ shock. We used PMAS on the 3.5m Calar Alto telescope to obtain
measures of emission lines that provide insight into physical properties
of the gas. Severe blending of H$\alpha$ and [N{\sc
ii}]$\lambda$6548,6583\AA\ emission lines in many spaxels required the
assumption of at least two kinematical components in order to extract
fluxes for the individual lines.
}
  {
Main results from our study include:
(a) detection of discrete emission features in the new intruder velocity
range 5400-6000~km~s$^{-1}$ showing properties consistent with H{\sc ii}
regions,
(b) detection of a low velocity component spanning the range
$5800-6300$~km~s$^{-1}$ with properties resembling a solar metallicity
shocked gas and
(c) detection of a high velocity component at $\approx 6600$~km~s$^{-1}$
with properties consistent with those of a low metallicity shocked gas.
}
   {
The two shocked components are interpreted as products of a collision
between NGC~7318b new intruder and a debris field in its path. This has
given rise to a complex structure of ionized gas where several components
with
different kinematical and physical properties coexist although part of
the original ISM associated with NGC~7318b is still present and remains
unaltered. Our observations suggest that the low velocity ionized component
might have existed before the new intruder collision and  could be
associated with the NW-LV H{\sc i} component of Williams et al. (2002).
The high velocity ionized component might fill the gap between
the H{\sc i} complexes observed in SQ-A and NGC~7319's tidal filament
(NW-HV, Arc-N and Arc-S in Williams et al. 2002).
}


   \maketitle
%

\section{Introduction}

Stephan's Quintet (SQ) is one of the most spectacular and intriguing
galaxy systems in the local Universe. Discovered by Stephan in 1877 it has
become a ``Crab Nebula"
in extragalactic astronomy as the subject of a multitude of studies across
the electromagnetic spectrum. It was know since the 60s that one of the
galaxies - NGC~7320 - shows a highly discordant redshift (Burbidge \&
Burbidge 1961) that
converted it from a quintet into a quartet. It has long since regained 
quintet status with the discovery of two tidal tails extending toward
accordant redshift galaxy NGC~7320c (Arp 1973). A dynamical analysis
revealed that SQ consists of a core of three galaxies that have
experienced several episodes of dynamical harassment from an ``old
intruder" NGC~7320c (Moles et al. 1997) and now from NGC~7318b.
The morphology of SQ galaxies reveals many signs of interaction including:
(1) an apparently unrelaxed stellar halo comprising 30\% of the optical
light (Moles et al. 1998), (2) twin tidal tails of apparently different age
(Sulentic et al. 2001) pointing toward the old intruder, (3) two spiral
galaxies (NGC~7319 and old intruder) stripped of the bulk of their
non-stellar material (Sulentic et al. 2001).
The IGM of SQ also reveals a  complex debris field including hot
($\approx 10^{6}~K$, Trinchieri et al. 2005), warm ($\approx 10^{4}~K$, Xu et al. 1999; Sulentic et al. 2001) and
cold -- atomic ($\approx 10^{2}~K$, Williams et al. 2002) and molecular ($\approx 10-1000~K$, Lisenfeld et al.
2004; Appleton et al. 2006) -- gas.
Also, recent star formation activity throughout a very extended disk shaped 
region around
galaxies NGC~7318b and NGC~7318a has been reported from
UV GALEX (Xu et al. 2005) and H$\alpha$ (Moles et al. 1997; V\'{\i}lchez \& Iglesias-P\'{a}ramo 1998; 
Iglesias-P\'{a}ramo \& V\'{\i}lchez 1999) images.

But perhaps the most conspicuous and interesting feature of SQ involves
the close pair of galaxies NGC~7318a and NGC~7318b. Their recession
velocities differ by
almost 1000~km~s$^{-1}$ (6630 and 5774~km~s$^{-1}$ respectively). Four galaxies in SQ
show radial velocities very close to the former value\footnote{Hickson et al. (1992) 
proposed a value of $v = 6450$~km~s$^{-1}$ as the recession velocity of the group.}
implying that
NGC~7318b is a new intruder entering the group from the far side and
colliding with NGC~7318a, NGC~7319 and a complex debris field between them. The
reality of this collision was confirmed by detection of an extended shock
front in X-ray, optical and radio continuum light. 
The observational evidences suggest that
NGC~7318b has passed through
SQ after entering from the far side with a high line-of-sight velocity. 
This large velocity difference argues that
the bulk of its motion should be along the line of sight.
All likely smaller transverse component of motion is indicated by the
diffuse X-ray morphology: the West edge of the shock is somewhat sharper
than the East (see e.g. E-W profile cuts in Trinchieri et al. 2005).
We therefore assume that NGC~7318b has already passed through the group and
is now in front of SQ.

The short crossing time of NGC~7318b\footnote{The crossing time of NGC~7318b was estimated as $t_{c} = D/\Delta v_{rad}$, where $D$ is the diameter of the smallest circle containing the nuclei of all the galaxies (from Hickson et al. (1982), and $\Delta v_{rad}$ is the difference in the radial velocities of SQ (6446~km~s$^{-1}$, from NED) and NGC~7318b (7445~km~s$^{-1}$, also from NED). The assumed distance for SQ is 88.6~Mpc (from NED).} 
($t_{c} \approx 1.2 \times 10^{8}$~yr) implies that we are seeing the collision
{\it in flagrante delicto} or {\it post flagrantem delictum}. 
The latter possibility would mean that the intruder is seen projected on SQ.

Optical (long slit) spectroscopy of the shock-front has been reported in
several papers (Ohyama et al. 1998; Xu et al. 2003). The latter authors
identify at least three different emission components in this region but
detailed study was restricted to the brightest regions. A complete mapping
of physical and dynamical properties along the shock region is still
lacking but has become feasible  with  Integral Field Unit (IFU)
spectrographs. We present IFU observations at three positions in or near to
the shock front. Our goal is to unravel the complex kinematics and
excitation properties of the ionized gas along the path of the collision.
The paper is organized as follows: 
Section~2 presents the instrumental setup and details about the observations
and data reduction. A description of the data analysis and main results of
the line fitting procedure are contained in Section~3.
Section~4 presents discussion of the implications of our results and a
summary of the main conclusions is presented in Section~5.
Finally, Appendix~A contains the details on the fitting procedure.


\section{Data acquisition and reduction}

\begin{figure}
   \centering
   \includegraphics[width=12cm,angle=-90]{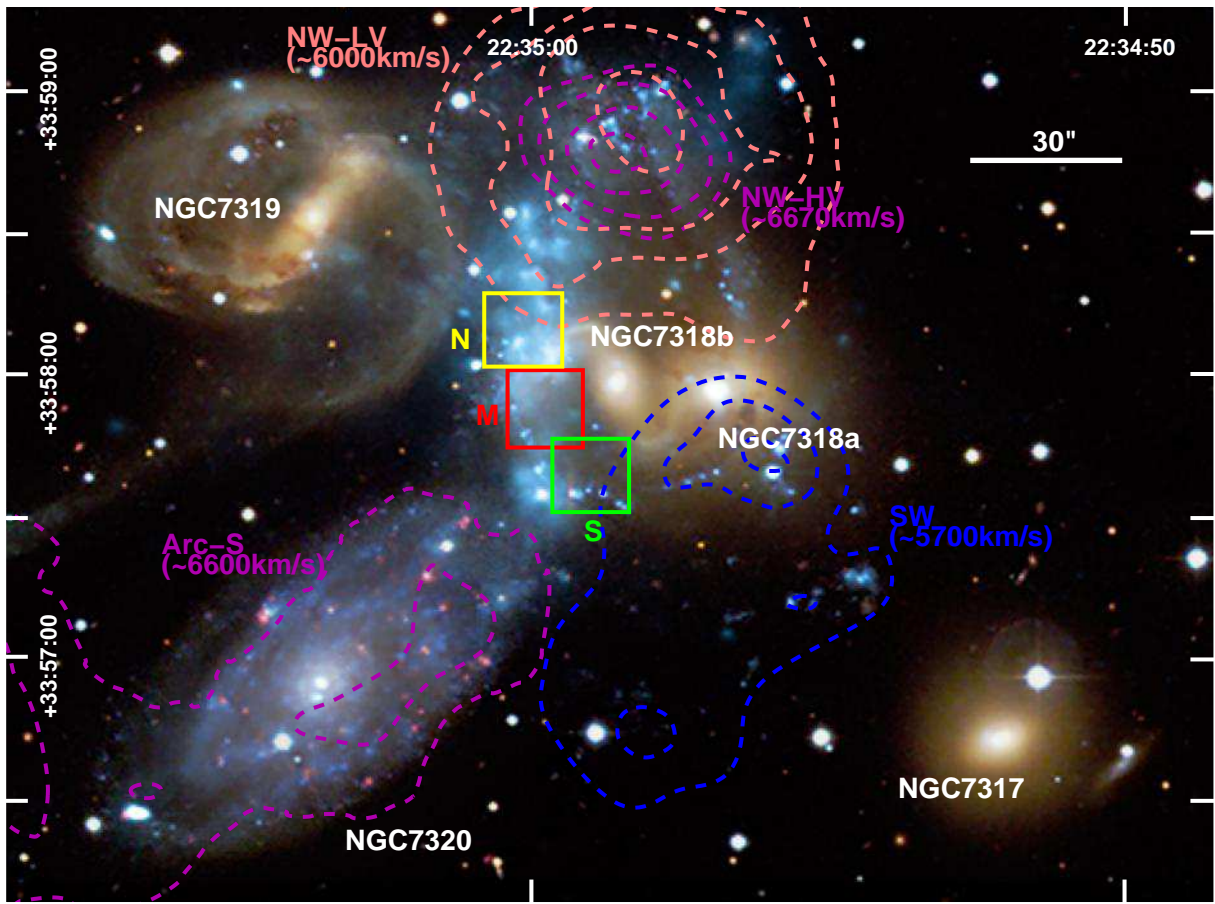}
      \caption{False color image of SQ. The color code is as follows: X-ray from CHANDRA (cyan), optical light (red, yellow, blue and white) from the CFHT. 
The image width is 6.3~arcmin. North is up. Overimposed we show our three PMAS pointings (yellow, red and green for the N, M and S pointings respectively).
We also overlaid the H{\sc i} contours from Williams et al. (2002) and labeled the four main structures reported in that paper as well as their approximate 
recession velocities.
The bar representing an angular distance of 30'' corresponds to a linear scale of 12.89~kpc.
}
         \label{sq_cfht}
   \end{figure}

SQ was observed on 2009 August 21-25 at Calar Alto Observatory (Almer\'{\i}a, Spain), using the
3.5m Telescope with the Potsdam Multi-Aperture Spectrometer (PMAS, Roth et al. 2005). The
standard lens array integral field unit (IFU) of 16$\arcsec\times$16$\arcsec$ field of view (FOV)
was used with a sampling of 1$\arcsec$. 
The position of the IFU covering different regions of SQ -- hereafter we will refer to them as pointings N, M and S -- is shown in Figure~\ref{sq_cfht}.
The figure also shows the position of the most remarkable H{\sc i} structures reported in Williams et al. (2002).

Most of the optical range was covered with the R600 grating using two grating rotator angles: 
143.3, covering from 3810 to 5394\AA; and 146.05, covering from 5305 to 6809\AA. 
The effective spectral resolution was 3.6\AA~FWHM ($\approx 166$~km~s$^{-1}$ at H$\alpha$).
The blue and red spectra have a total integration time of 3600 seconds (split into three 1200~s individual dithered exposures), 
respectively, in fields N and S and 4200 seconds (split into three 1200~s and one 600~s dithered exposures) in the red one for the M field.
Table~\ref{LOG} gives the coordinates, observation date, integration time, seeing conditions, 
as well as the median airmass during the observations and the overall photometric conditions during each observing night for the three pointings. 
Observations were taken under photometric conditions during the night of August 22, and under non photometric conditions during the rest of the nights.

The data were reduced following the standard steps for fiber-based integral field spectroscopy 
using the {\sc iraf}\footnote{IRAF is distributed by the National Optical Astronomy Observatory, 
which is operated by the Association of Universities for Research in Astronomy (AURA) under cooperative agreement with the National Science Foundation.} reduction package {\sc specred}. 
Bias was removed using a master-bias made out of combination of individual bias. 
Continuum lamps were  taken before science exposure  needed to identify the location of the 256 individual spectra on the CCD and extract them. 
We then performed a wavelength calibration using the HgNe lamp exposures taken before each science exposure.  
The continuum lamp and sky flats were
used to determine the response of the instrument for each fiber and wavelength. 
Observations of the spectrophotometric standard stars BD~$+$33$\rm ^o$2642 and Cyg~Ob2-9 were used 
for flux calibration co-adding  the spectra of the central fibers and compared them with the tabulated one-dimensional spectra. 
The error of this calibration is of the order of 5\%. 
The typical seeing during the observations was between 1.2$\arcsec$ and 1.5$\arcsec$.\\ 

The final products of our data reduction process are three $16 \times 16$ arrays each one 
composed of 256 spaxels containing the (blue and red) spectra of each pointing.
However, given the low signal-to-noise ratio of the individual spectra we preferred to perform 
a $2 \times 2$ binning that improves the quality of the resulting spectra and possibilities a more detailed analysis.
Figures~\ref{hcg92s_only_todos} to \ref{hcg92n_only_todos} show the flux of all spaxels of each pointing 
along the spectral region $6650\AA < \lambda< 6750\AA$, which corresponds to the expected position 
of the [N{\sc ii}]$\lambda$6548,6583\AA\ and H$\alpha$ lines according to the typical recession velocities of the galaxies in SQ.
At the end, each pointing corresponds to a $8 \times 8$ array, with spaxels of size $2'' \times 2''$.
In what follows we will always refer to these binned spectra and 
we will work with the three $8 \times 8$ arrays corresponding to our three pointings.
The spaxels will be named as $A$[$x$,$y$], where $A$ is one of the three pointings 
(S, M or N), $x$ and $y$ corresponds to the cartesian coordinates in the $8 \times 8$ array
(as shown in figures~\ref{hcg92s_only_todos} to \ref{hcg92n_only_todos}), 
where [1,1] and [8,8] correspond to the Southeast and the Northwest corners respectively.

\section{Results}

A visual inspection of figures~\ref{hcg92s_only_todos} to \ref{hcg92n_only_todos}
reveals that in pointing S the flux is concentrated in few spaxels where the H$\alpha$ and [N{\sc ii}] lines are clearly visible.
This situation is not the same in pointings M and N where the flux from emission lines is distributed is a more homogeneous way.
Another interesting feature is that while the brightest spectra of pointing S show 
three narrow emission lines clearly resolved, unequivocally identified as the H$\alpha$ and [N{\sc ii}]6548,6583\AA\ lines, 
the spectra of pointings M and N show broader features severely blended. 
We notice that Xu et al. (2003) also found variable linewidths spanning the range 200 to 1000~km~s$^{-1}$
in their long slit spectroscopic data of the shocked region. 
However, the non coincidence in the spatial position of the spectra and the different instrumental resolution
prevents a detailed and quantitative comparison between both sets of data.

The physical properties of the ionized gas were estimated after individual fits to the most conspicuous emission lines of each spaxel.
As the blue spectra are dominated by the background noise\footnote{Specially towards wavelengths bluewards 4500\AA. 
In particular, the [O{\sc ii}]3727,3729\AA\ lines were not detected in any of the spectra.}, 
and the [S{\sc ii}]$\lambda$6717,6731\AA\ doublet did not fit in the our red spectral range, 
we use only the most intense lines 
of the red spectra for the fit: [O{\sc i}]$\lambda$6300\AA, [N{\sc ii}]$\lambda$6548,6583\AA\ and H$\alpha$. 
Details about the fitting procedure are explained in Appendix~A.

The main results of the fit are shown in Tables~\ref{tabla_fit_hcg92s} to \ref{tabla_fit_hcg92n}. 
The recession velocities and integrated fluxes of each emission line are those resulting from the fit.
The velocity dispersions have been estimated by assuming that the observed width of the line is the quadratic sum of different components (Richer et al. 2010):
\begin{equation}
\sigma^{2} = \sigma_{obs}^{2} - \sigma_{fs}^{2} - \sigma_{ins}^{2} - \sigma_{ther}^{2}
\end{equation}
where, $\sigma_{obs}$ is the value obtained from the fit, 
$\sigma_{fs}$ corresponds to the fine structure broadening and is taken equal to 3.199~km~s$^{-1}$ for H$\alpha$ (Garc\'{\i}a D\'{\i}az et al. 2008), 
$\sigma_{ins}$ corresponds to the instrumental broadening and is equal to 61.7~km~s$^{-1}$, 
and $\sigma_{ther}$ corresponds to the thermal broadening and is taken to be equal to 9.1~km~s$^{-1}$ assuming an electronic temperature of 10000~K (Osterbrock 1989).
The spectra corresponding to the components S[7,5]b, S[7,7]a and N[6,1]a
give negative values of $\sigma^{2}$, probably due to small uncertainties in the determination of $\sigma_{ins}$.
The uncertainty in the measured value of $\sigma_{ins}$ is $\approx$5\%, 
as determined from the widths of the arc lines.
Thus, for these spectra giving negative values of $\sigma^{2}$, we assumed a value of $\sigma_{ins}$
equal to 0.95 the average value and computed an upper limit of $\sigma = 16.7$~km~s$^{-1}$.

As it is explained in the Appendix, we have performed a 2-component fit to all the spaxels in the three pointings.
In order to avoid spurious results, only those components for which 
the intensity peak of the H$\alpha$ line is above $5 \times \Sigma_{bkg}$ are considered in this work. 
This means that after the fit, each spaxel can have associated 0, 1 or 2 components, depending on the 
corresponding signal to noise.
For the rest of the lines, we measure the fluxes of those for which the intensity peak is above
$\Sigma_{bkg}$, and give the corresponding upper limits otherwise.
In what follows, we will refer to the low and high velocity components as A and B.

\subsection{Properties of the emission lines}

Figures~\ref{hcg92s_compo} to \ref{hcg92n_compo} show some properties of the main emission lines of the three pointings
based on the results issued of our fitting procedure.
As it can be seen, the emission coming from pointing S is dominated by
component A (where component B is almost absent), 
contrarily to what occurs in pointings M and N.
The velocity of component A is constrained between 5500 and 6000~km~s$^{-1}$
in pointing S, it spans a range between 5800 and 6300~km~s$^{-1}$ in pointing M
and it concentrates between 6000 and 6300~km~s$^{-1}$ in pointing N.
On the contrary, component B ranges between 6500 and 6800~km~s$^{-1}$
in most spaxels of the three pointings.
The velocity dispersions of component A are low and typical of
those of H{\sc ii} regions ($20 \leq \sigma \leq 40$~km~s$^{-1}$)
for the brightest spaxels of pointings S and M (S[2,2], S[2,3], S[5,2] and M[2,2])
but show larger values for fainter spaxels of pointings S and M,
suggesting that the ionization source of these spectra is not dominated by recent star forming regions.
In pointings M and N the velocity dispersion of most component A spaxels 
ranges between 100 and 200~km~s$^{-1}$.
Concerning component B, the velocity dispersions show values
larger than 100~km~s$^{-1}$ for most spaxels.
Finally, the [N{\sc ii}]$\lambda$6583\AA/H$\alpha$ ratio show values between 0.3 and 0.4
for the brightest spaxels in pointings S and M, which correspond to component A.
Again, these values are consistent with the ones found for H{\sc ii} regions.
The rest of the low velocity component spaxels, including those of pointing N,
show values $\geq 0.4$ in most cases.
Conversely, spectra corresponding to component B tend to show values of the order
or lower than 0.3 in the three pointings.

Figure~\ref{hcg92s_maps} shows the H$\alpha$ intensity maps and radial velocity
of components A and B for the three pointings\footnote{The maps of component B corresponding to pointing S are not included since this component is almost undetected at this pointing.} and illustrate in a graphical way what we have
previously stated.
We remark the fact that component A increases its recession velocity as we move from pointing S to N,
that is, from the southern edge to the core of the shock region.

Spaxels showing H{\sc ii}-like spectra (S[2,2], S[2,3], S[5,2] and M[2,2]) 
correspond with 3 H{\sc ii} regions clearly seen in the HST image.
Optical spectra of these H{\sc ii} regions were previously obtained with the HET (Gallagher et al. 2001, their figure~14)
and show recession velocities similar to the ones we find.
These H{\sc ii} regions are associated with the Southern arm of NGC~7318b and either they have survived
the shock or they passed through SQ without any interaction.
Their oxygen abundances are shown in Table~\ref{metalicidades}. They have been estimated following of Pettini \& Pagel (2004)
using the O3N2 indicator.
Instead of a single value, the abundances show a gradient from the tip of the spiral arm (S[5,2], O/H$\approx 58$\% the solar value)
to the inner region (M[2,2], O/H$\approx 83$\% the solar value).
We assume the solar value is $12 + \log$[O/H]$ = 8.69 \pm 0.05$ (Allende-Prieto et al. 2001).

\subsection{Diagnostic diagrams}

As we have seen in the previous subsection, many of our optical spectra show velocity dispersions
and [N{\sc ii}]/H$\alpha$ ratios too large to be consistent with those of star forming regions.
In fact, very few of them show clear signatures of being H{\sc ii} regions.
There is also a non negligible fraction of spectra whose properties are not easily assignable to either
of the two classes, because of a combination of broad emission lines and low [N{\sc ii}]/H$\alpha$ ratios.

For this reason
we compare our data to two sets of theoretical models describing the properties of emission lines in star formation regions and shocks.
Figures~\ref{hcg92s_o1n2} to \ref{hcg92n_o1n2} show the [N{\sc ii}]$\lambda$6583\AA/H$\alpha$ vs. [O{\sc i}]$\lambda$6300\AA/H$\alpha$ for the spectra of the three pointings. 
This diagnostic diagram has revealed to be suitable to disentangle the nature of the 
process responsible for the properties of the ionized gas at each spaxel.
We have overplotted in the diagram two sets of models of ionized gas:
\begin{itemize}
\item Star formation models from Dopita et al. (2006): 
These models are parametrized by the metallicity of the ionizing stars and gas, the age of the stellar population and the parameter $R$, 
that accounts for the weak coupling between the ionization parameter and both 
the pressure of the ISM and the mass of the stellar cluster (see Dopita et al. 2006 for details).
We have included in the plot models with solar and 0.4 solar metallicity, values of $R = 2, -2$ and $-6$, along a time range between 0.1 and 6~Myr.
\item Shock$+$precursor models from Allen et al. (2008):
The complete library of models includes the radiative shock component as well as the photoionized precursor. 
In this work we selected only the models corresponding to solar and SMC metallicity, pre-shock density of 0.1, 1 and 10~cm$^{-3}$
(only 1~cm$^{-3}$ for the SMC models), 
and shock velocity values of $v_{s} = 100$, 300 and 1000~km~s$^{-1}$, for a range of values of the magnetic field between 10$^{-4}$ to 1000~$\mu$G.
\end{itemize}

The figures show that for solar metallicity the shock and star formation models are well separated from each other, 
overlapping only for the shock models with the lowest shock velocities (i.e. $v_{s} = 100$~km~s$^{-1}$ in this work).
However, this does not hold when including lower metallicities.
In particular, the SMC shock models of low and intermediate shock velocity ($v_{s} \leq 300$~km~s$^{-1}$) overlap with the solar metallicity star formation models.
This fact underlines the point that information on the metallicity (which is not available with the present data) 
is required in order to analyze the nature of emitting gas when shocks are likely to be present.

The first thing we notice is that, with the exception of the spaxels corresponding
to the H{\sc ii} regions and some fainter spaxels of pointing S,
the position of components A and B lay well above the H{\sc ii} regions from spiral 
and irregular galaxies from van Zee et al. (1998) and van Zee \& Haynes (2006), supporting that
the main source of ionization is shock rather than star formation.
This result is also consistent with the conclusions of Cluver et al. (2010) from H$_{2}$
observations in the shocked region.

The brightest components of pointing S occupy the locus of the solar metallicity star formation models, 
whereas the fainter ones are located towards larger values of [O{\sc i}]/H$\alpha$, 
fairly consistent with the solar metallicity shock models with $v_{s}$ increasing up to $\approx 300$~km~s$^{-1}$, or
with the low metallicity (SMC) shock models with $v_{s}$ between 300 and 1000~km~s$^{-1}$.
The velocity dispersions tend to be larger for those components closer to 
the shock models than to the star formation ones, although a strong relation is not observed.
The position of pointing S with respect to the X-ray emission delineating the shock (e.g. Trinchieri et al. 2005),
the radial velocities of most components and their velocity dispersions not very high as compared to those measured in
pointings M and N, suggest that only a moderate fraction of the spaxels in pointing S could be affected by the shock
induced by the collision of NGC~7318b with SQ.
The rest of the spaxels of pointing S must correspond to the diffuse ionized medium 
present between the H{\sc ii} regions of normal spiral galaxies.

Concerning pointing M, the low velocity component of the bright spaxel M[2,2] is located between the solar metallicity 
star formation and shock models and it shows a low velocity dispersion, as it would be expected for an H{\sc ii} region.
The rest of the components delineate a strip overlapping with the solar metallicity shock models 
($100 \leq v_{s} \leq 300$~km~s$^{-1}$), and the SMC shock models ($300 \leq v_{s} \leq 1000$~km~s$^{-1}$).
The velocity dispersions of most components are large and of the same order
as expected for shock ionized spectra.

Finally, pointing N shows a very similar picture as pointing M, but without any spaxel showing an H{\sc ii}-like spectrum,
most of the components showing properties of shock ionized spectra.

A remarkable point emerging from diagrams corresponding to pointings M and N is that 
component A spectra are preferentially located in the locus occupied by the solar
metallicity shock models, whereas component B spectra are shifted towards
the low metallicity (SMC) shock models.
This suggests that the shock induced by the passage of NGC~7318b through SQ 
has involved at least two filaments of different recession
velocities and metallicities.

\label{o1n2}

\section{Discussion}

\subsection{H{\sc i} filaments and ionized gas components}

The overall picture emerging from our observations can be
summarized as follows:
in most spaxels of our three pointings, two components are detected.
A low velocity one (component A) -- which shows H{\sc ii}-like spectra in a reduced
number of spaxels of pointings S and M, and shock-like spectra in most spaxels
of pointings M and N -- and a high velocity one (component B) -- which shows basically shock-like
spectra in most spaxels of pointings M and N --.
That is, the shock mostly affects pointings M and N but not pointing S.
The H{\sc ii}-like spectra correspond to star forming regions clearly seen in the
HST optical images, what suggest that they passed through SQ with minimal
disruption.
This is consistent with what we see on the full resolution HST image of SQ and with a slit
spectrum obtained with HET\footnote{The HET spectrum includes three relatively
normal H{\sc ii} regions in the new intruder followed by a velocity smeared
region at $\sim$6050~km~s$^{-1}$ and an uncertain detection at $\sim$6500~km~s$^{-1}$.} 
(Gallagher et al. 2001; figure 14) which partially
overlaps with our pointing S. 

Observations at other wavelengths show a link between different gaseous phases in the
shock region:
whereas the H{\sc i} observations of Williams et al. (2002) show almost no gas along
the ridge of the shock, H$_{2}$ was detected coincident with the ionized
emission observed in our spectra (Appleton et al. 2006; Cluver et al. 2010).
Despite the presence of dust signatures in this region (Guillard et al. 2010),
no active star formation has been detected in the shock ridge, in agreement with
the properties of the emission lines of the ionized gas.

An interesting point is that our observations show that the shock
involves two filaments of (ionized) gas with recession velocities
similar to the H{\sc i} filaments reported by Williams et al. (2002). 
Together with the detection of H$_{2}$ gas in the ridge of the 
shock region, this suggests that the shock region was occupied
by H{\sc i} filaments that were converted into molecular and ionized gas
respectively after the collision between NGC~7318b and SQ (Sulentic et al. 2001, Guillard et al. 2009).

One of these shocked ionized components shows a recession velocity between 
6000 and 6300~km~s$^{-1}$ (component A in pointings M and N), which is in good agreement with
the NW-LV H{\sc i} cloud of Williams et al. (2002) to the North of
the shocked region.
The high velocity shocked ionized component shows a recession velocity
of $\approx 6500$~km~s$^{-1}$ (component B in pointings M and N), 
which is consistent with those of the NW-HV, Arc-S and Arc-N
H{\sc i} filaments of Williams et al. (2002), and its emission line ratios are
consistent with shock models of metallicities lower than solar 
(close to the value of the SMC).

The recent hydrodynamical simulations of Hwang et al. (2011) suggest that the gas
dynamically associated to our two shocked ionized components comes mostly from
NGC~7318b (although a small fraction also from NGC~7319) in the case of component A, 
and from NGC~7319 in the case of component B.
Thus, we must address the point of whether these simulations are consistent with the
information we have from the observations.

\subsection{Metal abundances of the different components}

Concerning component A, we know that the oxygen abundances of the H{\sc ii} regions
detected in pointings S and M present a gradient where the oxygen abundance increases
from pointing S towards pointing M.
In particular, we estimated a value of 83\% the solar value for the H{\sc ii} region
detected in pointing M.
As this region is located to the South of this pointing, we argue that the
oxygen abundance of the shocked gas must be close to the solar value.
Thus, if the gas at this velocity comes from the inner regions of NGC~7318b, as the 
hydrodynamical simulations suggest, the metallicity should be close to the
solar value, according to our findings from the emission line properties
of this component.

The case of component B is more complicated since we do not have direct information
on the metallicity of the galaxy NGC~7319.
For this reason, we look for indirect estimators of the metal content of
the H{\sc i} filaments dynamically associated to this component:
Xu et al. (2003) performed long slit optical spectroscopy of a region (their spectrum M1)
very close to NW-HV and to the infrared starburst SQ-A (Xu et al. 1999). 
Using the calibration of Kewley \& Dopita (2002) they obtained a value of $12 + \log$O/H$= 8.76$ 
for values of [N{\sc ii}]$\lambda$6583\AA/H$\alpha = 0.16$ and
[O{\sc iii}]$\lambda$5007\AA/H$\beta = 2.48$.
The estimations of the gas metallicity strongly depend on the calibration used, 
so a direct comparison between abundances estimated with different methods
is not fair.
When applying the calibration of Pettini \& Pagel (2004) with the O3N2 indicator, 
we obtain $12 + \log$O/H$= 8.35$, which corresponds to 46\% the solar value.
Also, Lisenfeld et al. (2004) reported optical spectroscopic measurements of 
intergalactic star forming regions located in the tidal tails stemming from
NGC~7319 and spatially coincident with the Arc-N filament, whose recession
velocity is similar to that of Arc-S and NW-HV of Williams et al. (2002).
These authors obtained $12 + \log$O/H$= 8.7$ using the calibration of
van Zee et al. (1998), which implies a value of
[N{\sc ii}]$\lambda$6583\AA/H$\alpha \approx 0.24$ for the brightest region (SQ-B)
at the tip of the tidal tail (P.-A. Duc, private communication).
According to the calibration of Pettini \& Pagel (2004), a lower value of [N{\sc ii}]$\lambda$6583\AA/H$\alpha$
implies a lower value of the oxygen abundance.
Comparing with the value measured for the H{\sc ii} region in S[5,2]
([N{\sc ii}]$\lambda$6583\AA/H$\alpha \approx 0.30$) this means that 
the metallicity of SQ-B must be lower than 60\% the solar value
(which is the value estimated for S[5,2]).
These two observational evidences together suggest that the high velocity
gas filaments (NW-HV, Arc-S and Arc-N) must have metallicity significantly
lower than solar, as the properties of the ionized component indicate.

But, is this result consistent with the suggestions of the simulations that
the gas at this recession velocity comes from the galaxy NGC~7319?
This question is relevant since other authors have suggested that the high
velocity H{\sc i} filaments could be the relics of the group formation process
(Williams et al. 2002) instead of being tidal debris resulting from interactions among
galaxies in the group (Moles et al. 1998).
NGC~7319 is a high luminosity\footnote{$M_{B} = -21.68$ according to LEDA.} spiral galaxy that hosts an AGN in its nucleus.
The fact that no H{\sc i} was detected in the disk of this galaxy (Williams et al. 2002)
supports the idea that this gas was removed from the disk in one of the interactions
experienced by this galaxy and can now be found in the form of the neutral filaments
seen around the main bodies of the galaxies.
The metallicity of galaxies as luminous as NGC~7319 is difficult to estimate since it is well known
the existence of gradients of abundances in spiral galaxies, so usually the external parts of the disk
present lower abundances than the inner ones.
Pilyugin et al. (2004) presented a study of oxygen gradients along a sample of spiral galaxies where
they established a relation between the absolute magnitude of spirals and their characteristic metallicity,
which is the metallicity measured at a distance from the center of 0.4$R_{25}$.
According to these authors, this average characteristic metallicity for a typical spiral as luminous as NGC~7319 is 
$12 + \log$O/H$\approx 8.6$.
These authors also found that the oxygen abundance
can reach levels as low as $12 + \log$O/H$= 8.2$ at distances from the galaxy nucleus
between 0.5 and 0.9$R_{25}$, even for galaxies as luminous as NGC~7319.

Putting all these arguments together, we have shown that the two ionized components
detected in our bidimensional spectroscopic study of the shock region are kinematically
consistent with the H{\sc i} filaments and that their estimated metallicities are not
inconsistent with the origin of these filaments, according to the results of 
the recent hydrodynamical simulations. 
The final picture of SQ emerging from our observations suggests the existence of a low velocity component
showing normal H{\sc ii} emission at pointing S (where the recession velocity is consistent with that of NGC~7318b) 
and shocked emission in pointings M and N where the recession velocity is larger.
In between the recession velocity this component shows a smooth gradient suggesting that this component
was initially associated to the intruder galaxy NGC~7318b before the shock and that 
the northern part is currently shocked whereas the southern one has survived the shock
probably because it has not encountered material in its path through SQ.

However, an overestimation of the role of overlap in
our emission line decompositions cannot be completely discarded. 
Almost certainly there is SQ gas in the
range 6600-6700~km~s$^{-1}$ and new intruder gas from 5400-5800~km~s$^{-1}$. 
Gas in the range 5800-6600~km~s$^{-1}$ is the most difficult to interpret. The deep
HST image suggest impressive destruction of structure (from spiral arms to
H{\sc ii} regions) in the region of our pointings. As one example we could
consider spaxel M[3,4] (not an extreme example, see Figure~\ref{hcg92m_03_04}) where we think to
see the signature of a new intruder H{\sc ii} region at $\approx 5900$~km~s$^{-1}$ with high
[N{\sc ii}]/H$\alpha$. We model a second component with $v \approx 6400$~km~s$^{-1}$ with more
typical [N{\sc ii}]/H$\alpha \approx 0.2$. This is not a unique solution and rather than a
second component the strong red wing on the H$\alpha$ profile may be the
signature of an emission tail due to ablation of the new intruder H{\sc ii}
region. Data with higher signal-to-noise would make possible a more detailed consideration of
shock related effects.

\section{Summary and Conclusions}

This work has presented a 3-D spectroscopic study of the ionized gas properties of three regions along 
the shocked bar-like structure in front of the pair of galaxies NGC~7318b and NGC~7318a.
The analysis of the red part of the spectra around the H$\alpha$ line has revealed the existence of the following kinematical regimes:
\begin{itemize}
\item Spectra corresponding to 3 discrete emission features in pointings S and M: their recession velocities are consistent with that of NGC~7318b.
The estimated oxygen abundances, $8.45 < 12 + \log$O/H $< 8.61$, are lower than the solar value and show a gradient along the spiral arm of NGC~7318b.
We claim that these H{\sc ii} regions survived the collision of NGC~7318b with SQ.
\item A low velocity shocked component, showing recession velocities in the range $5800-6300$~km~s$^{-1}$, and partially
covering pointings S and M:
its recession velocity increases in a smooth way as we move Northwards (from S to M) and the properties of this shocked ionized gas
resemble those of solar metallicity shocked gas with velocity around 300~km~s$^{-1}$.
We propose that prior to the collision this ionized component was
associated to the NW-LV H{\sc i} filament of Williams et al. (2002).
\item A high velocity shocked ionized component, present in the three pointings studied in this work although at a very low surface brightness in pointing S:
this component shows a fairly constant recession velocity of $\approx 6600$~km~s$^{-1}$.
Unlike the low velocity component, the properties of the high velocity component are consistent with 
those of a low metallicity shocked gas with velocity in the range $300-1000$~km~s$^{-1}$.
This shocked ionized component is proposed to be the link between the 
NW-HV and Arc-S and Arc-N H{\sc i} structures reported by Williams et al. (2002),
and its oxygen abundance is consistent with being formed out of gas
formerly in the disk of NGC~7319.
\end{itemize}

The analysis presented in this work has shown the intriguing kinematical structure of this region, as it has been stated by 
the presence of blended components with different properties.
Although some light has been shed in the relations between the ionized and neutral gaseous components detected around this
region, their nature and origin is still far from being totally understood.
For this reason, further observations with higher spatial and spectral resolution are needed to clarify in detail the
state of the complex physical processes that are taking place in SQ.

\begin{acknowledgements}
We thank the anonymous referee for useful comments and suggestions that contributed to improve
this paper.
This work has been funded by the Spanish MICINN under the
AYA2007-67965-C03-02 and the Consolider-Ingenio 2010
Program CSD2006-0070 ``First Science with the GTC'' grants.
This research has made use of the NASA/IPAC Extragalactic Database (NED) 
which is operated by the Jet Propulsion Laboratory, California Institute of Technology, 
under contract with the National Aeronautics and Space Administration. 
We acknowledge the usage of the HyperLeda database (http://leda.univ-lyon1.fr)
\end{acknowledgements}

\newpage

\begin{appendix}

\section{Details on the fitting procedure}

The fitting procedure was performed as follows:
for each pointing, we performed a $2 \times 2$ binning of all the spaxels in order to improve the signal to noise ratio of the resulting spectra.
Then, for each spaxel, we fit the spectral range [6620,6780]\AA\ to a set of gaussians corresponding to 
the emission lines [N{\sc ii}]$\lambda$6548,6583\AA\ and H$\alpha$, under the following conditions:
\begin{itemize}
\item The intensities of the four emission lines must always be positive or zero (no underlying absorption in H$\alpha$ is assumed). 
In the case of H$\alpha$, we only allow intensity larger than zero.
\item The recession velocity of the lines is restricted to the interval [5400,6900]~km~s$^{-1}$ 
(this seems reasonable since the recession velocities of all the galaxies of SQ are contained within this range).
\item The theoretical relation [N{\sc ii}]$\lambda 6548/6583=0.333$ is preserved.
\item The velocity dispersion ($\sigma$) of the three emission lines is forced to be the same, $>1.35$\AA\ 
(which is the nominal dispersion of the spectra), and $<4.5$\AA.
This last restriction is imposed to avoid non physical solutions that lead to arbitrarily broad lines for some spectra.
\end{itemize}
The continuum was estimated before the fit, it was assumed to be the same for the three emission lines 
and equal to the median flux in the spectral interval [6600,6660]\AA~$\cup$~[6750,6800]\AA. 
The background uncertainty ($\Sigma_{bkg}$) was taken to be equal to the standard deviation of the flux along this interval.
After several tries, this was found to be a proper estimation given the almost null slope of the continuum along this spectral range.
The fit was performed with the IDL based routine MPFITEXPR (Markwardt 2009).
This code requires a set of initial input parameters ($I_{\rm{H}\alpha}$, $I_{6548}$, $v_{rec}$ and $\sigma$) and iteratively finds the solution that best matches the spectrum.

\begin{figure}
   \centering
   \includegraphics[width=20cm]{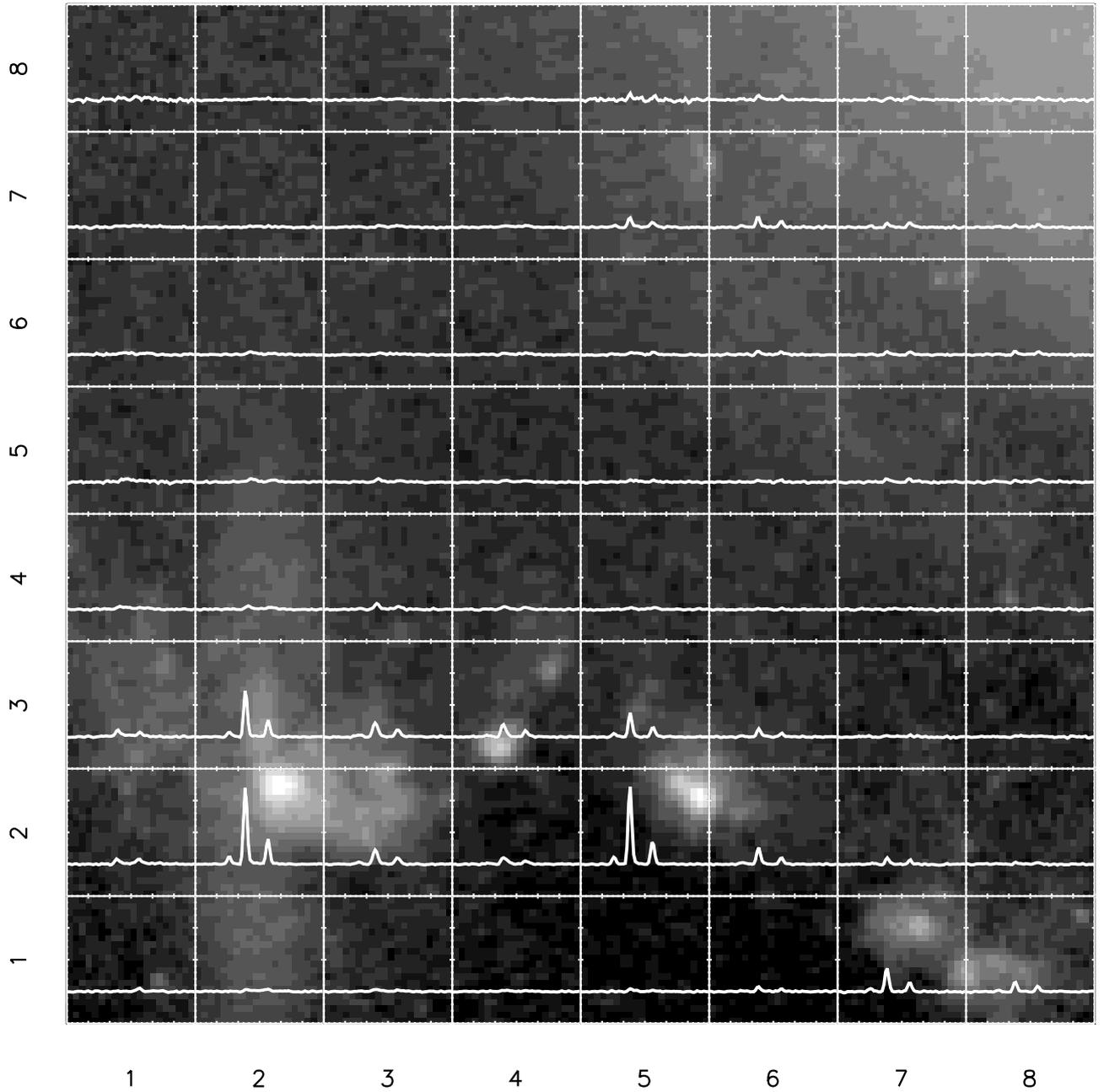}      
\caption{Spatial arrangement of the spectra of pointing S after a $2 \times 2$ binning of the spaxels
overlaid on the HST $V$-band image.
The X-axis of all spectra ranges from 6650\AA\ to 6750\AA.
The Y-axis scale is the same for all the spectra and ranges from $-1.53 \times 10^{-17}$ to $1.53 \times 10^{-16}$~erg~s$^{-1}$~cm$^{-2}$~\AA$^{-1}$.
Each spectrum is univocally identified by two numbers indicating the row and column occupied in the two dimensional array.
The orientation of the array is such that North is up and East is left. 
}
         \label{hcg92s_only_todos}
   \end{figure}

\begin{figure}
   \centering
   \includegraphics[width=20cm]{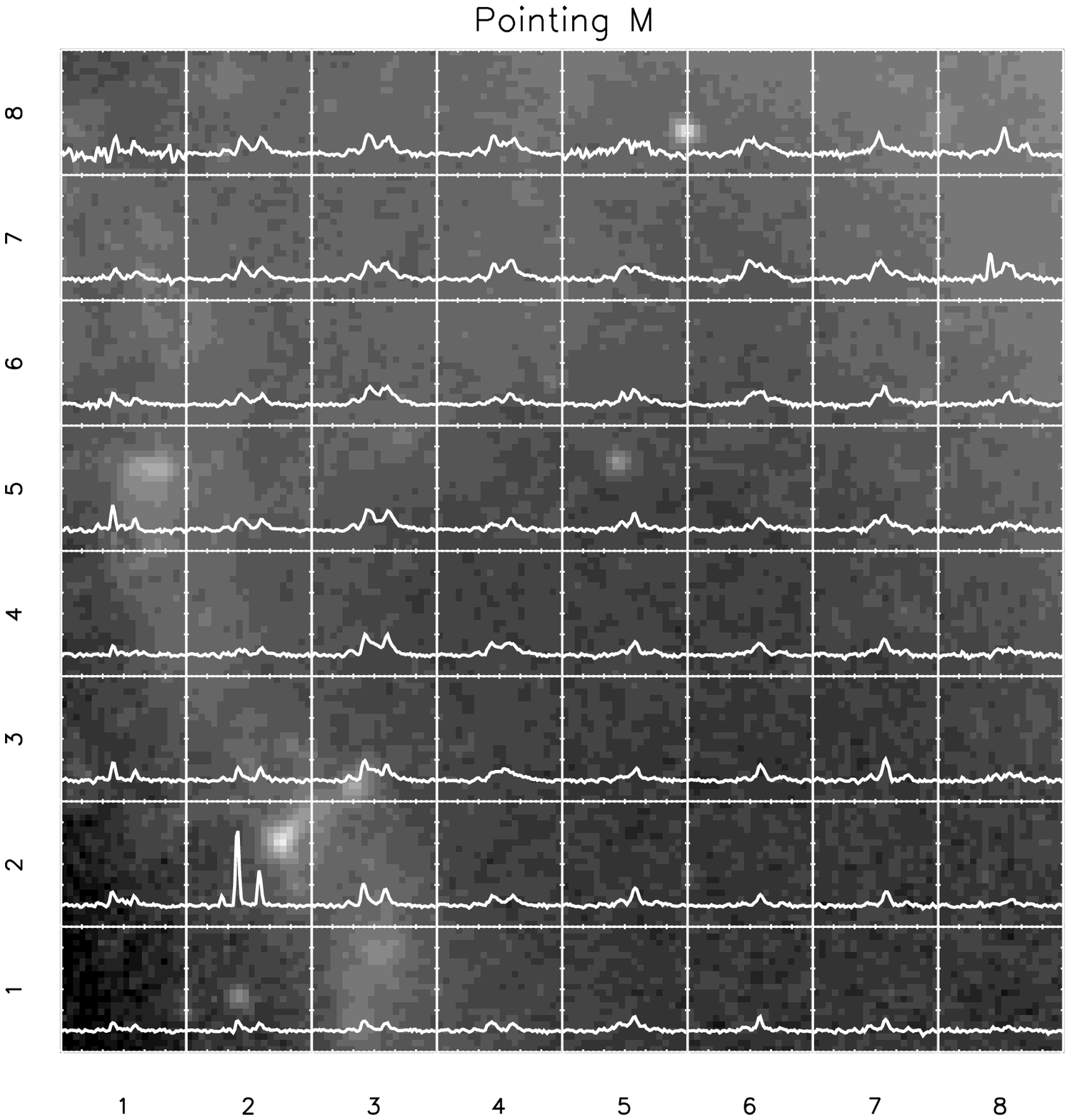}
      \caption{Same as Figure~\ref{hcg92s_only_todos} for pointing M.
The Y-axis scale is the same for all the spectra and ranges from $-4.39 \times 10^{-17}$ to $4.39 \times 10^{-16}$~erg~s$^{-1}$~cm$^{-2}$~\AA$^{-1}$.
}
         \label{hcg92m_only_todos}
   \end{figure}

\begin{figure}
   \centering
   \includegraphics[width=20cm]{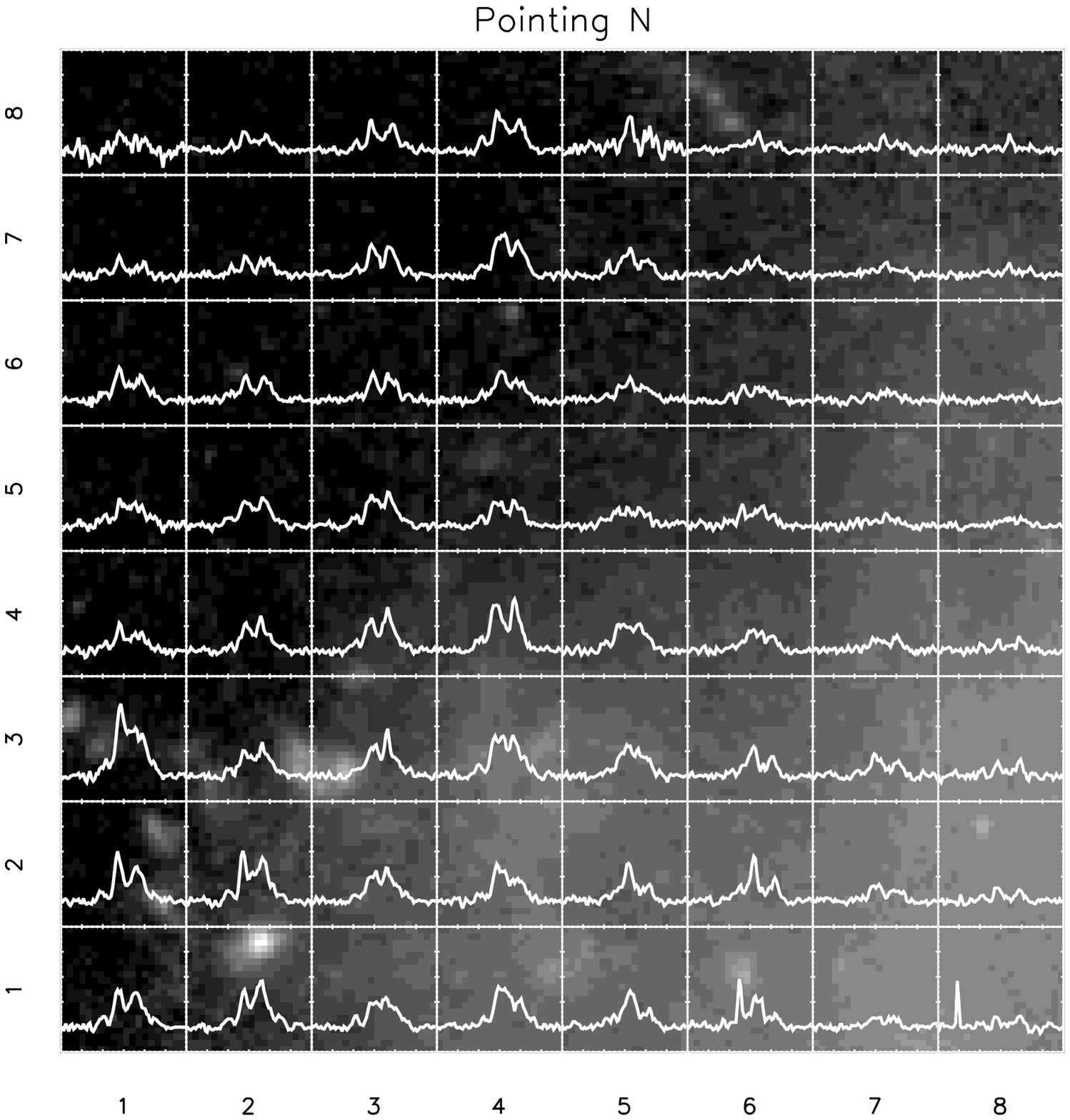}
      \caption{Same as Figure~\ref{hcg92s_only_todos} for pointing N.
The Y-axis scale is the same for all the spectra and ranges from $-1.53 \times 10^{-17}$ to $1.53 \times 10^{-16}$~erg~s$^{-1}$~cm$^{-2}$~\AA$^{-1}$.
}
         \label{hcg92n_only_todos}
   \end{figure}

\begin{figure}
   \centering
   \includegraphics[width=14cm]{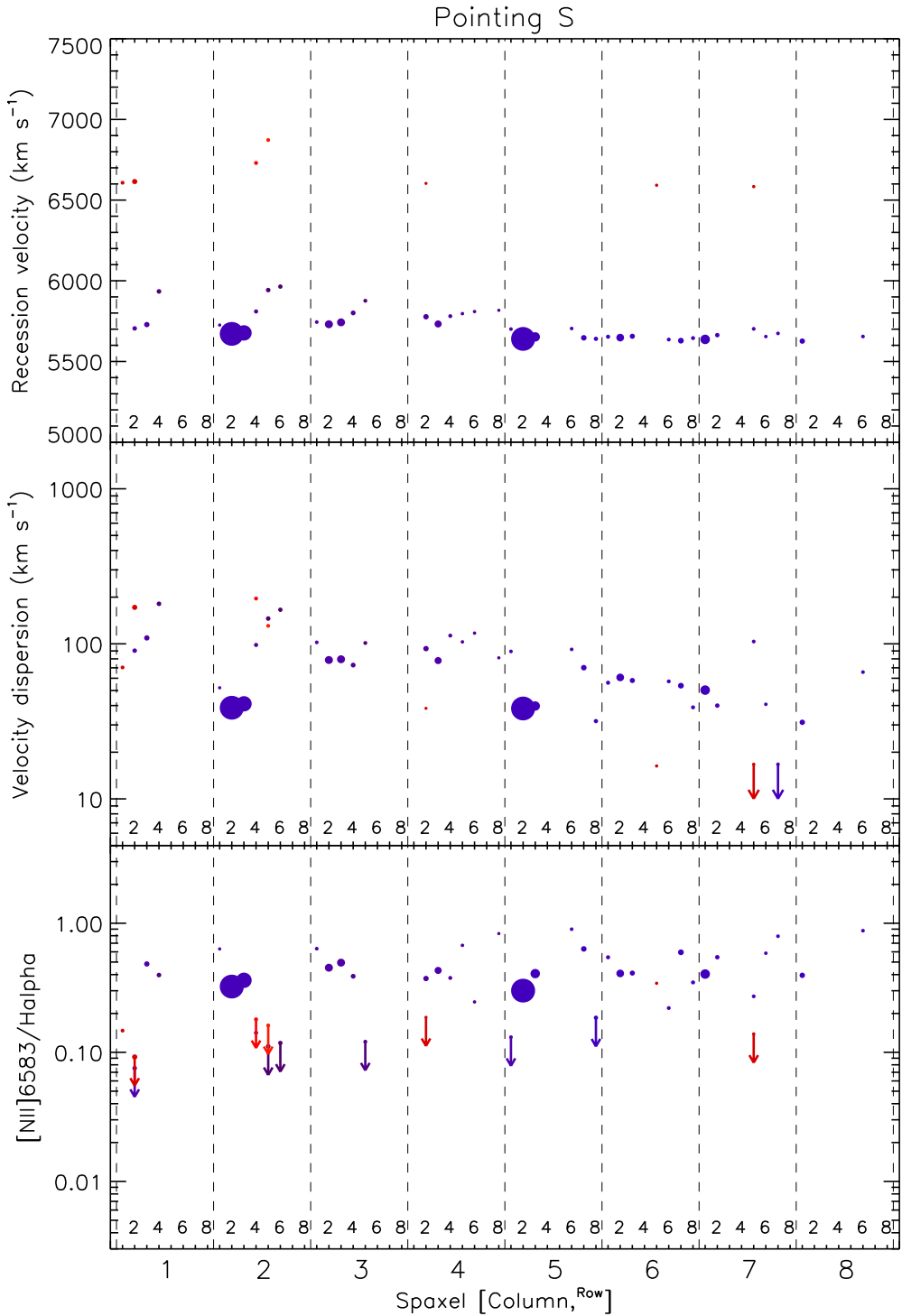}
      \caption{Recession velocity (top), velocity dispersion (middle) and [N{\sc ii}]$\lambda$6583\AA/H$\alpha$ flux ratio (bottom) for the spectra of pointing S.
Arrows correspond to upper limits in [N{\sc ii}]$\lambda$6583\AA/H$\alpha$ and $\sigma$.
Only components for which the intensity peak of the H$\alpha$ line is above $5\Sigma_{bkg}$ are plotted.
The major and minor ticks of the X-axis correspond to the column and the row of each spaxel respectively as it is illustrated in Figures~\ref{hcg92s_only_todos} 
to \ref{hcg92n_only_todos}.
For each spectrum, each component is represented for a filled dot.
In the three panels, the color of the dots is related to the recession velocity of the component, where bluer corresponds to lower velocity, and redder corresponds to higher velocity,
and the size of each dot is proportional to the flux of the H$\alpha$ line of the corresponding component.
}
         \label{hcg92s_compo}
   \end{figure}

\begin{figure}
   \centering
   \includegraphics[width=14cm]{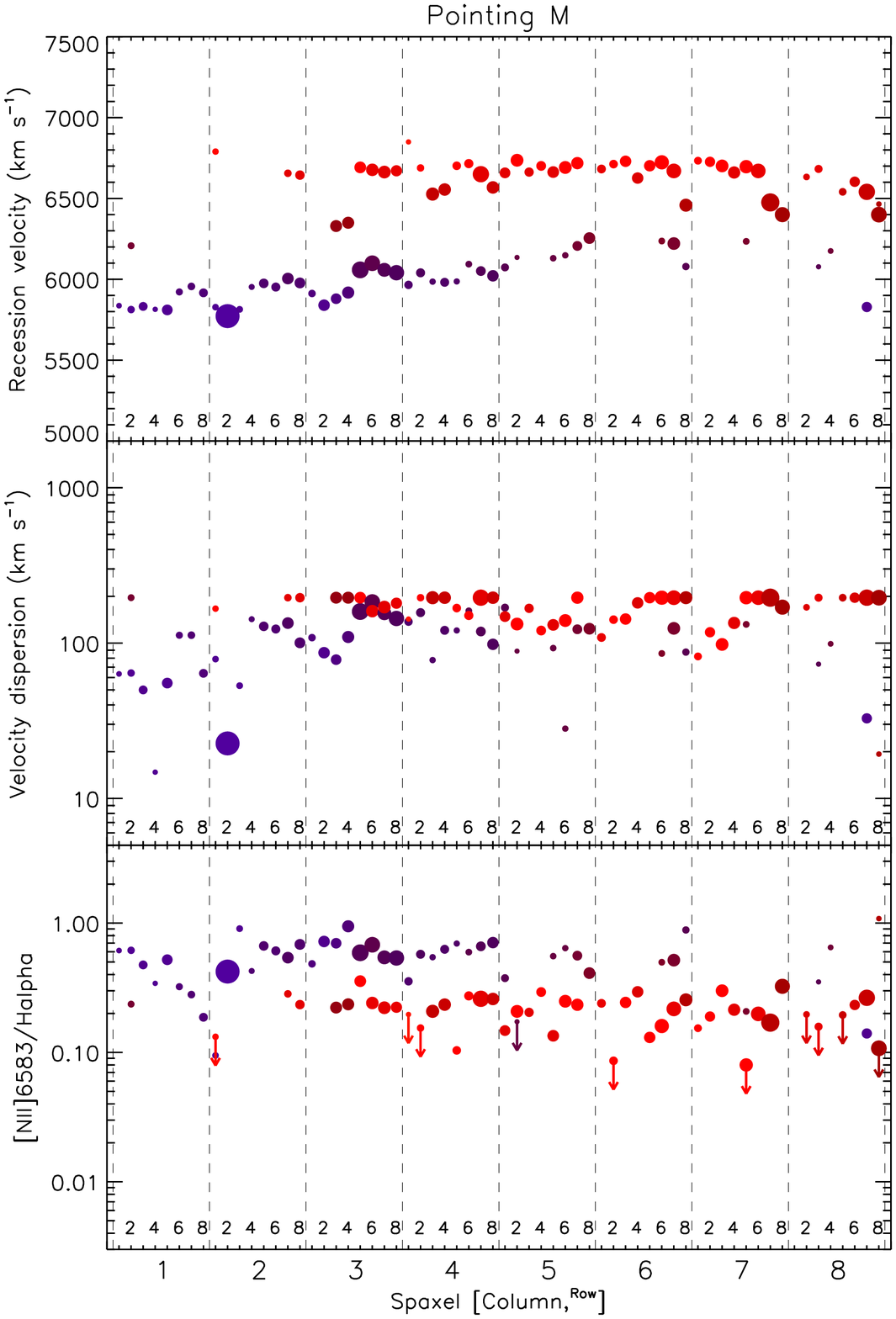}
      \caption{Same as Figure~\ref{hcg92s_compo} for pointing M.
}
         \label{hcg92m_compo}
   \end{figure}

\begin{figure}
   \centering
   \includegraphics[width=14cm]{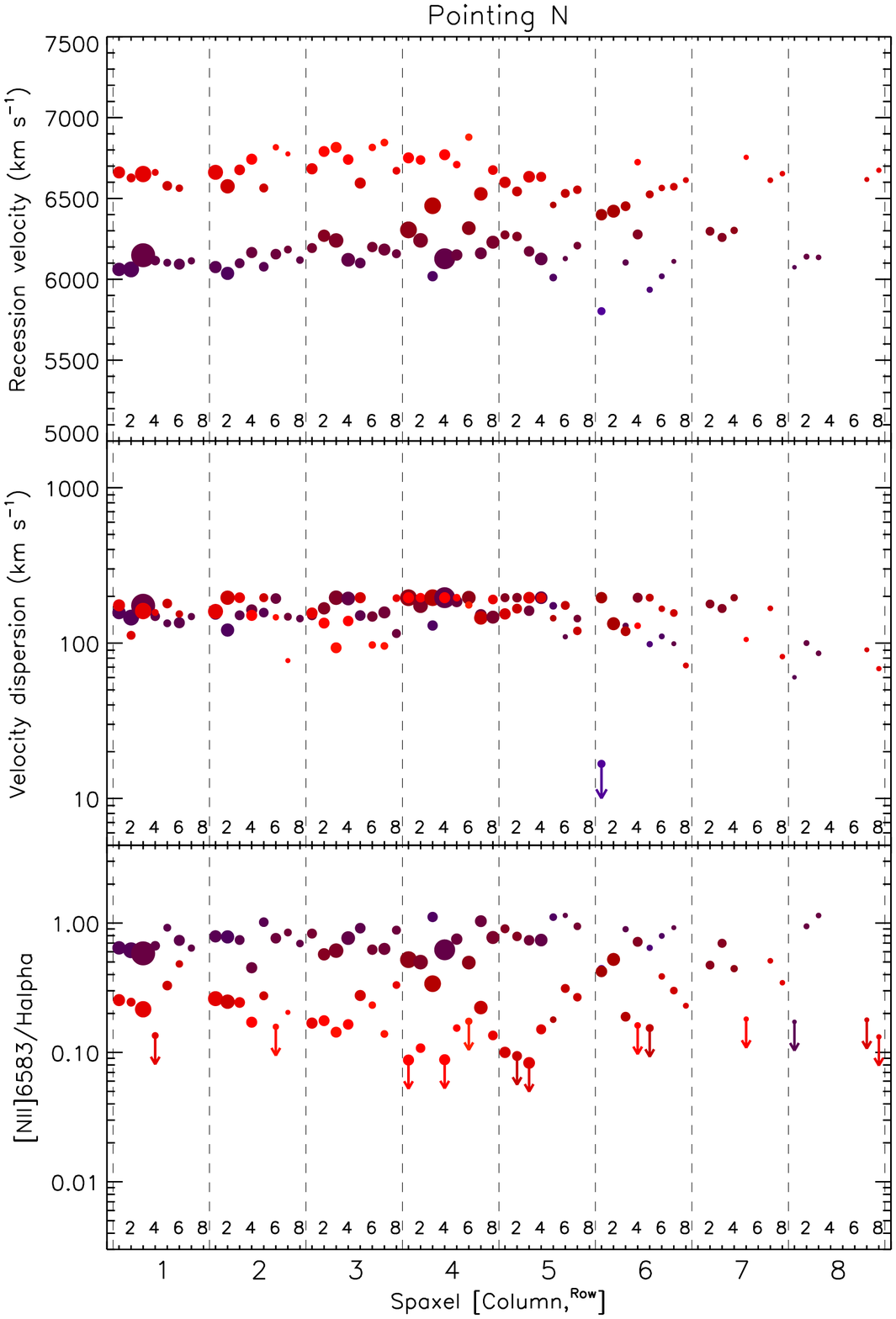}
      \caption{Same as Figure~\ref{hcg92s_compo} for pointing N.
}
         \label{hcg92n_compo}
   \end{figure}

\begin{figure}
   \centering
   \includegraphics[width=5cm]{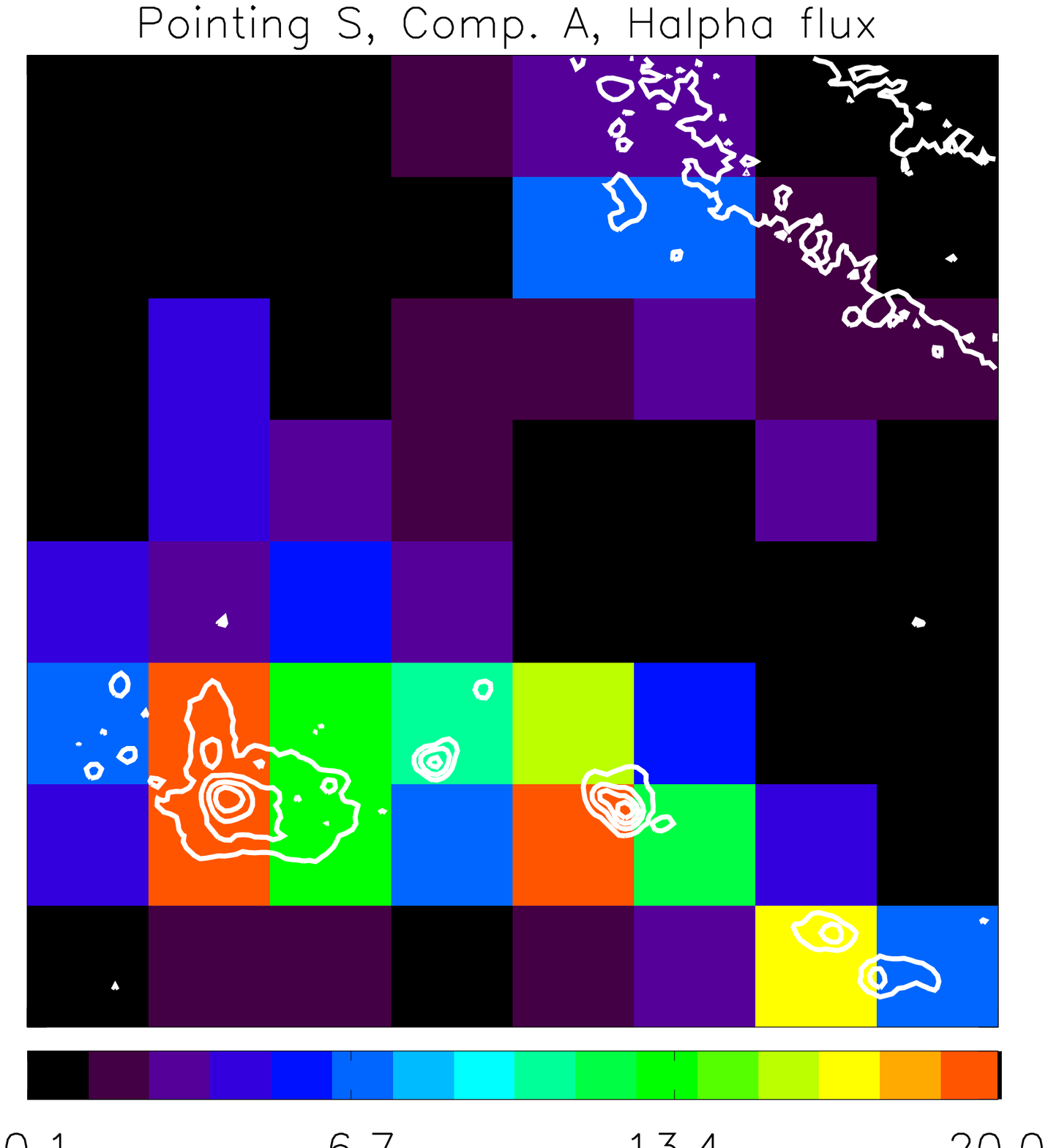}
   \includegraphics[width=5cm]{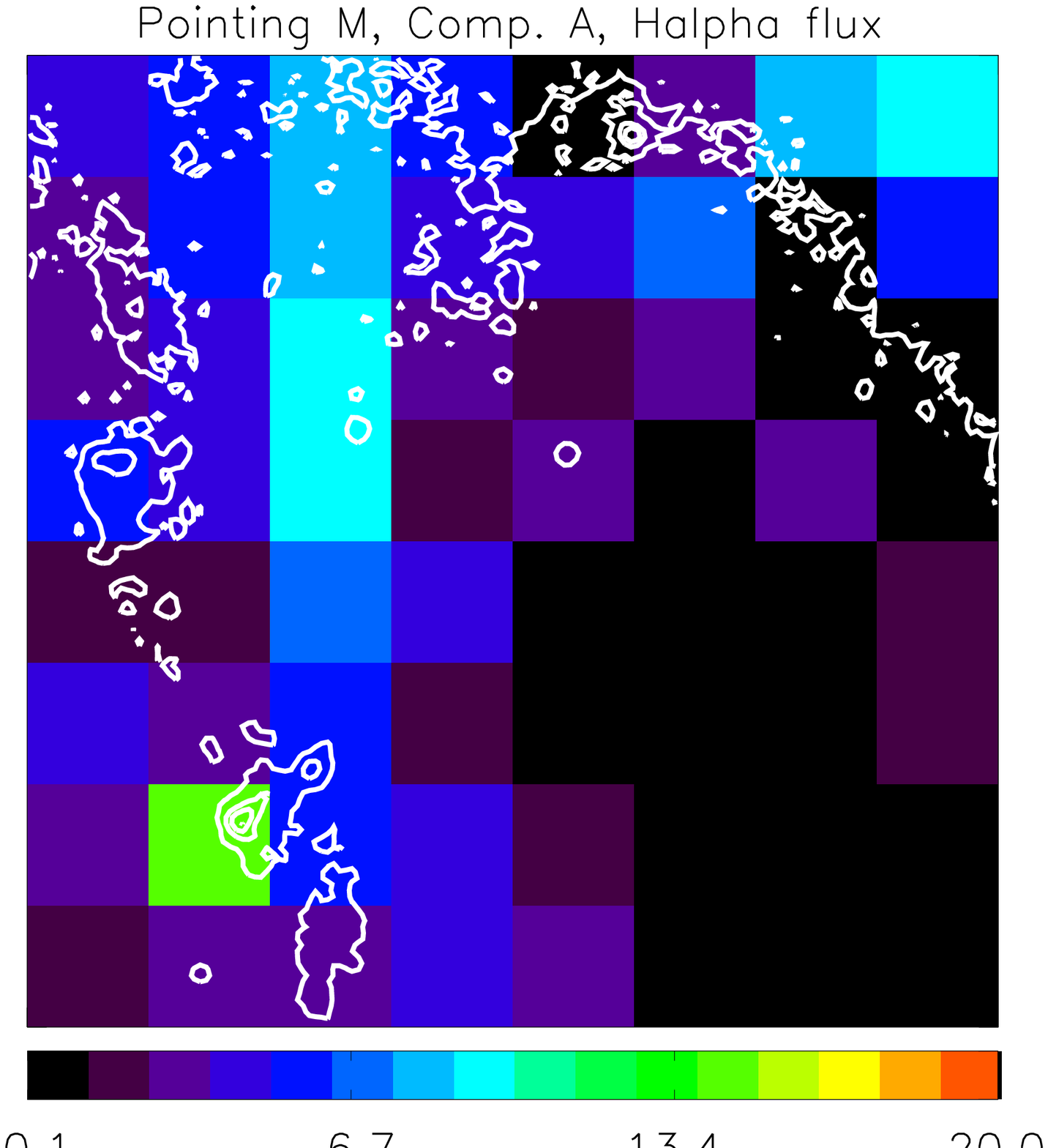}
   \includegraphics[width=5cm]{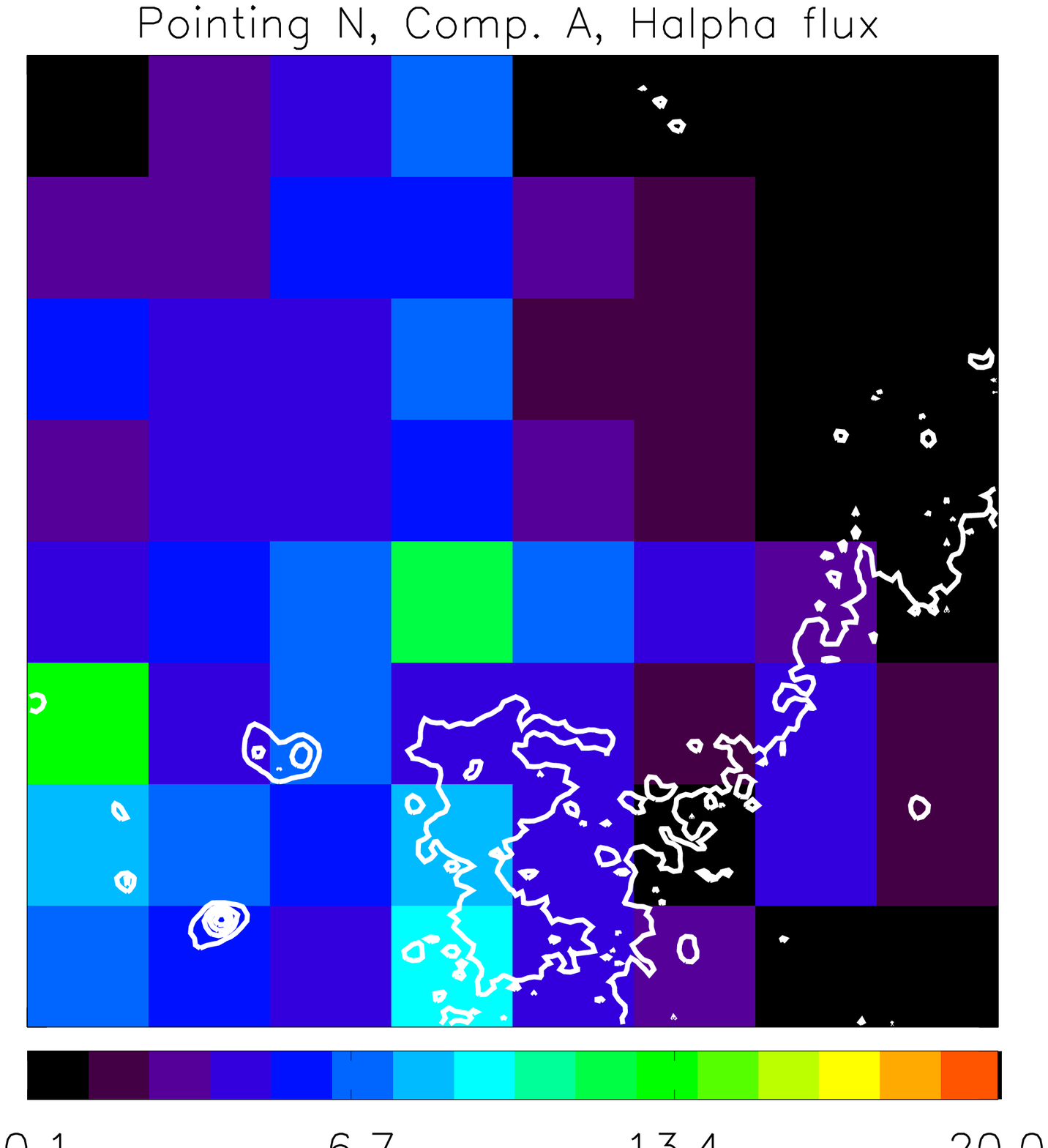}
   \includegraphics[width=5cm]{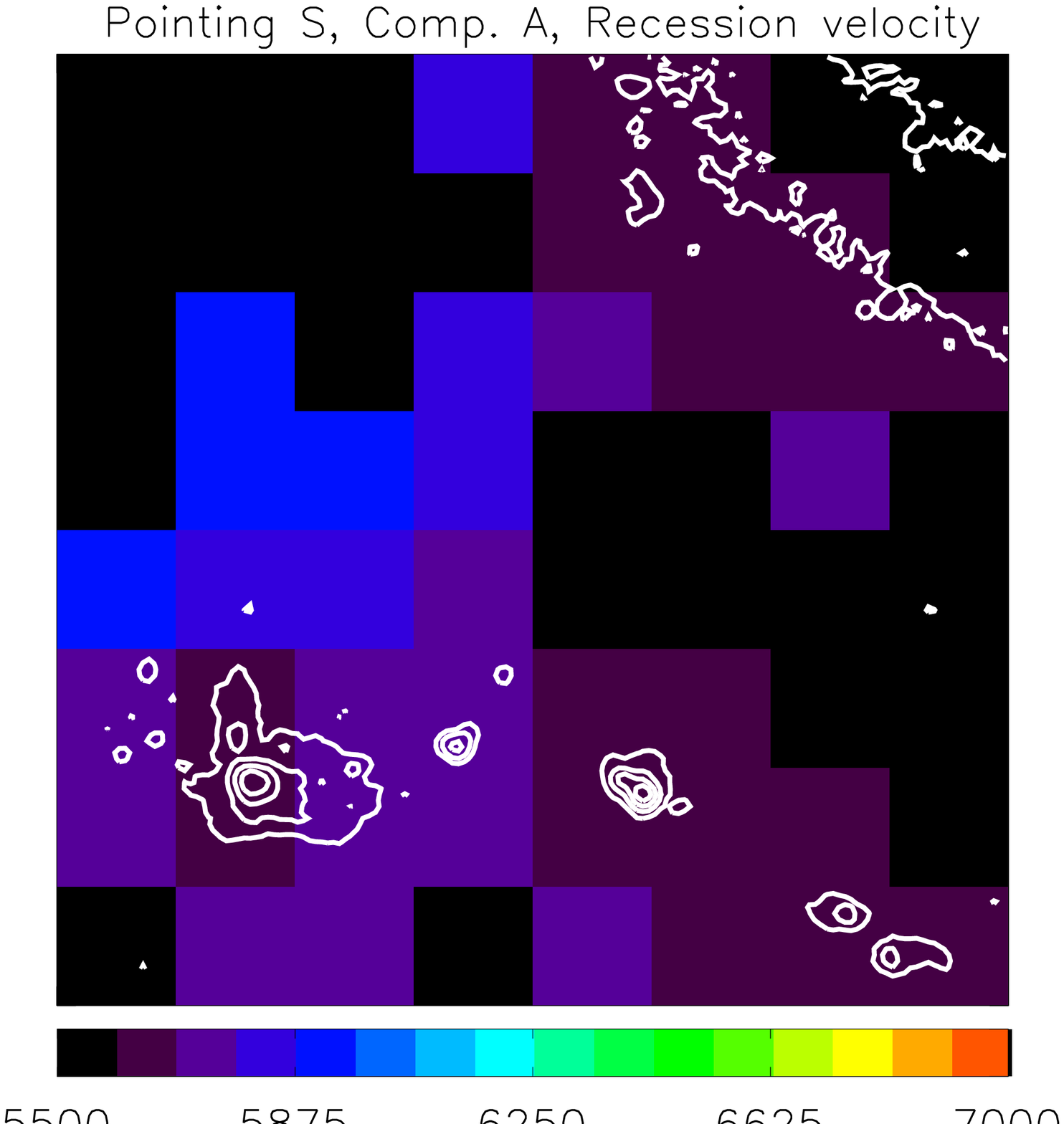}
   \includegraphics[width=5cm]{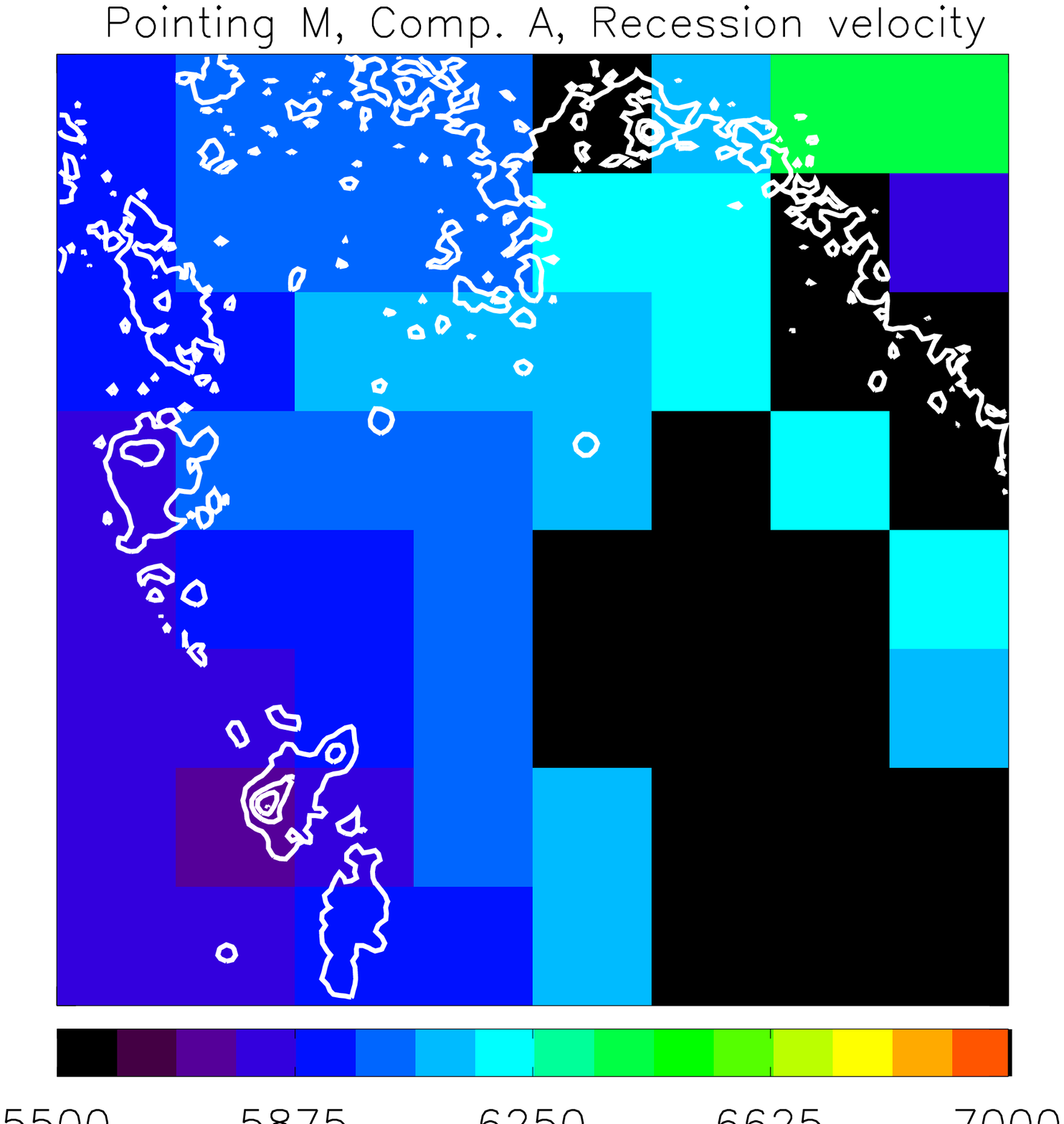}
   \includegraphics[width=5cm]{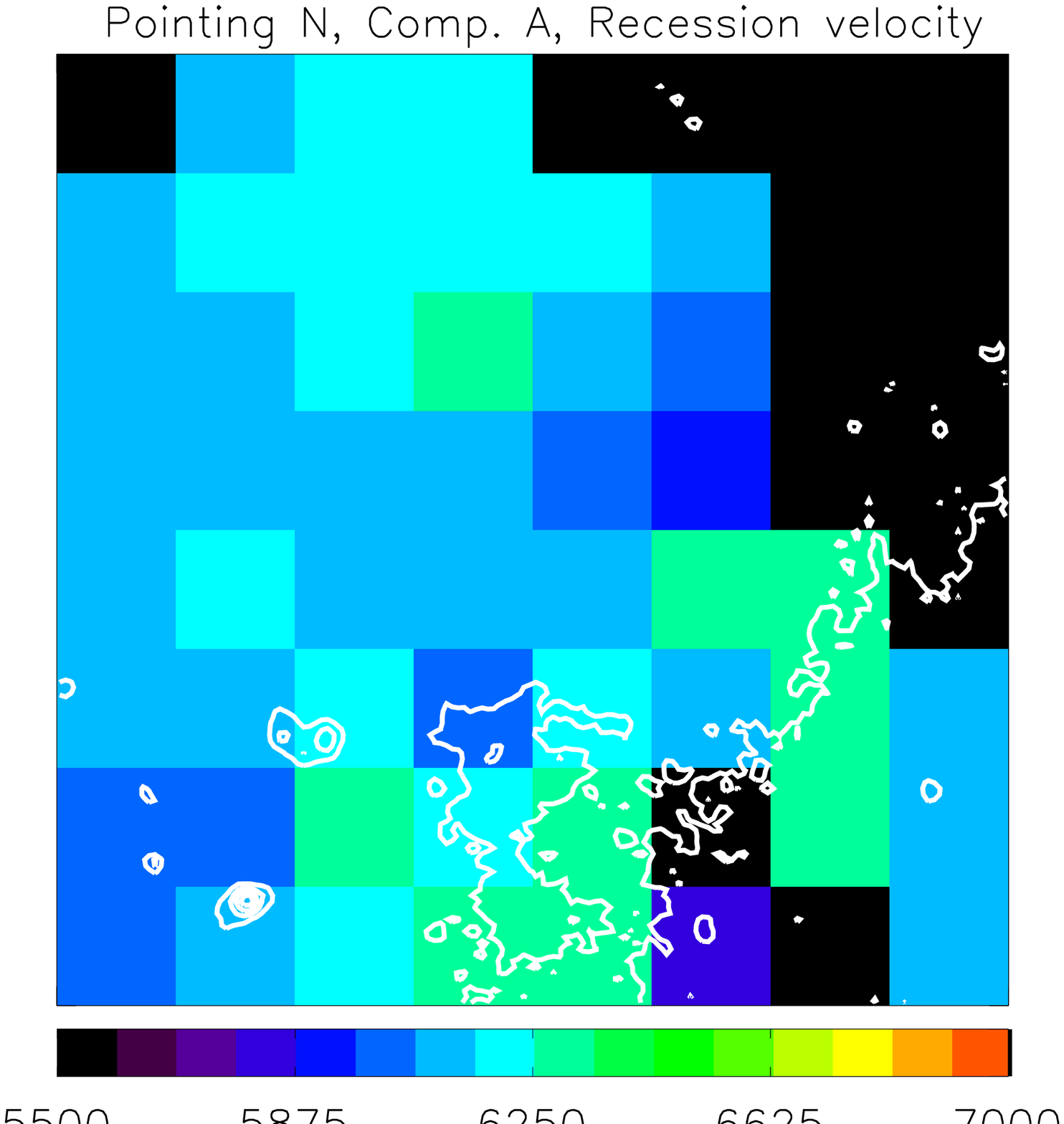}

   \includegraphics[width=5cm]{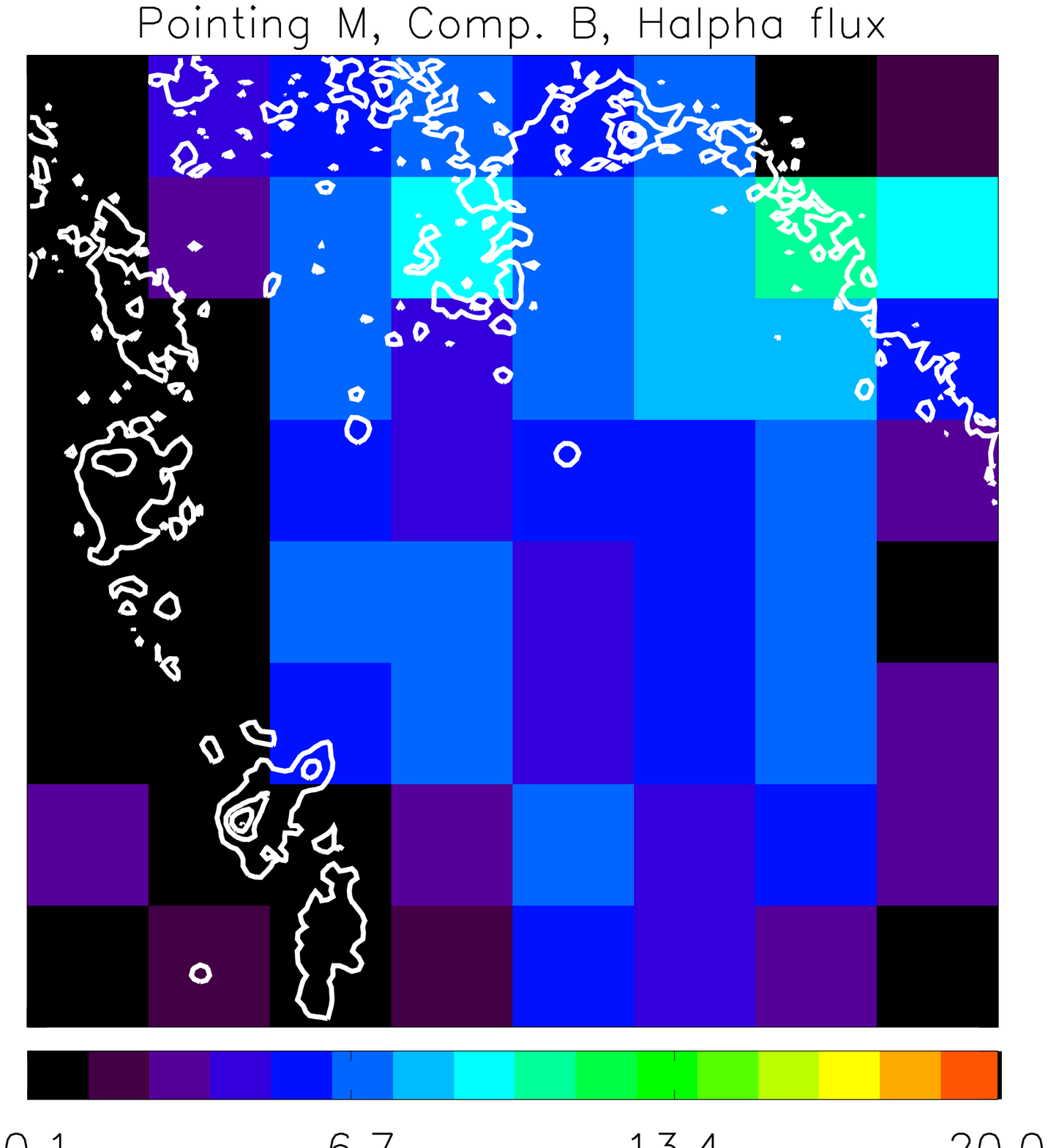}
   \includegraphics[width=5cm]{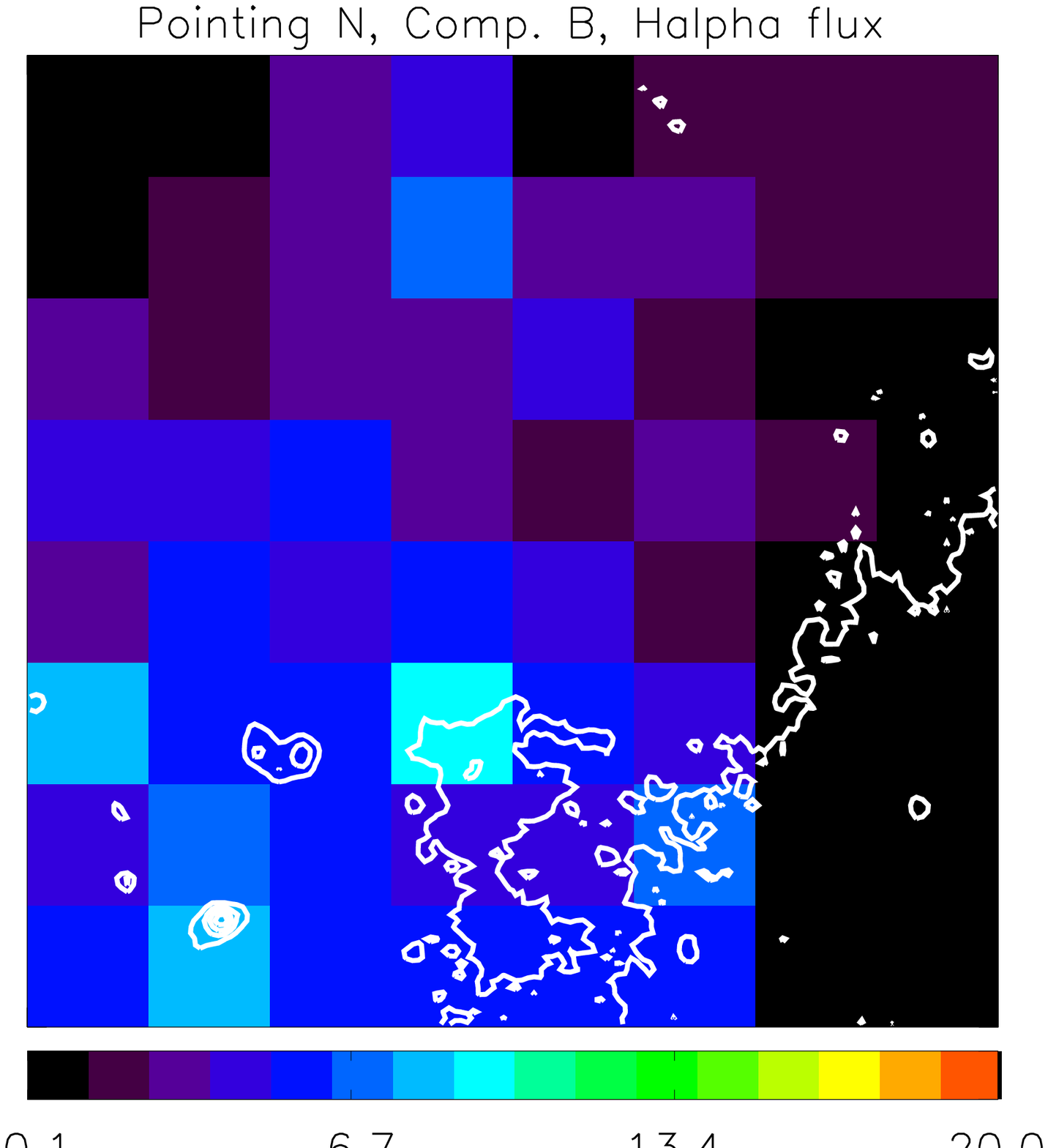}

   \includegraphics[width=5cm]{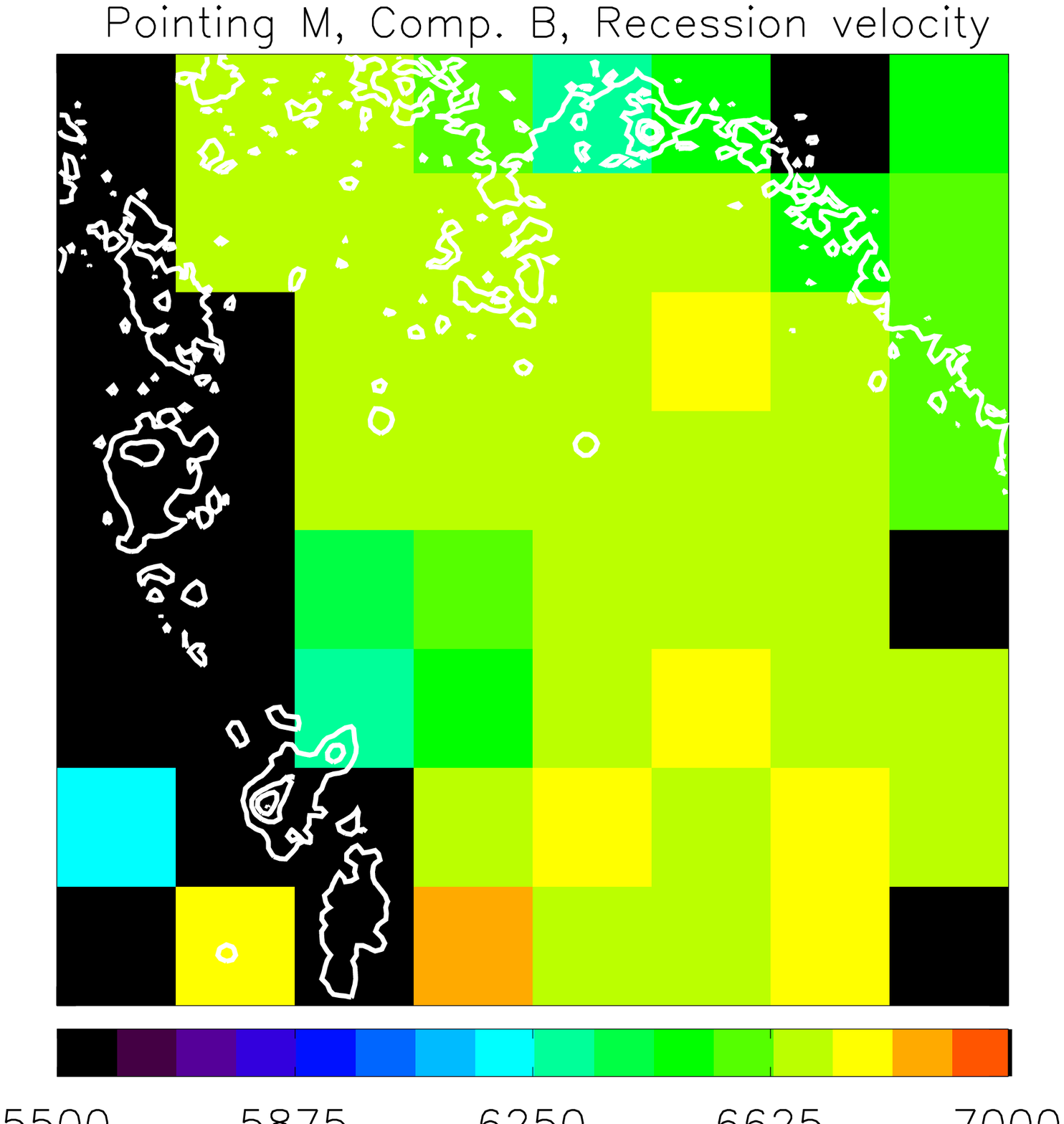}
   \includegraphics[width=5cm]{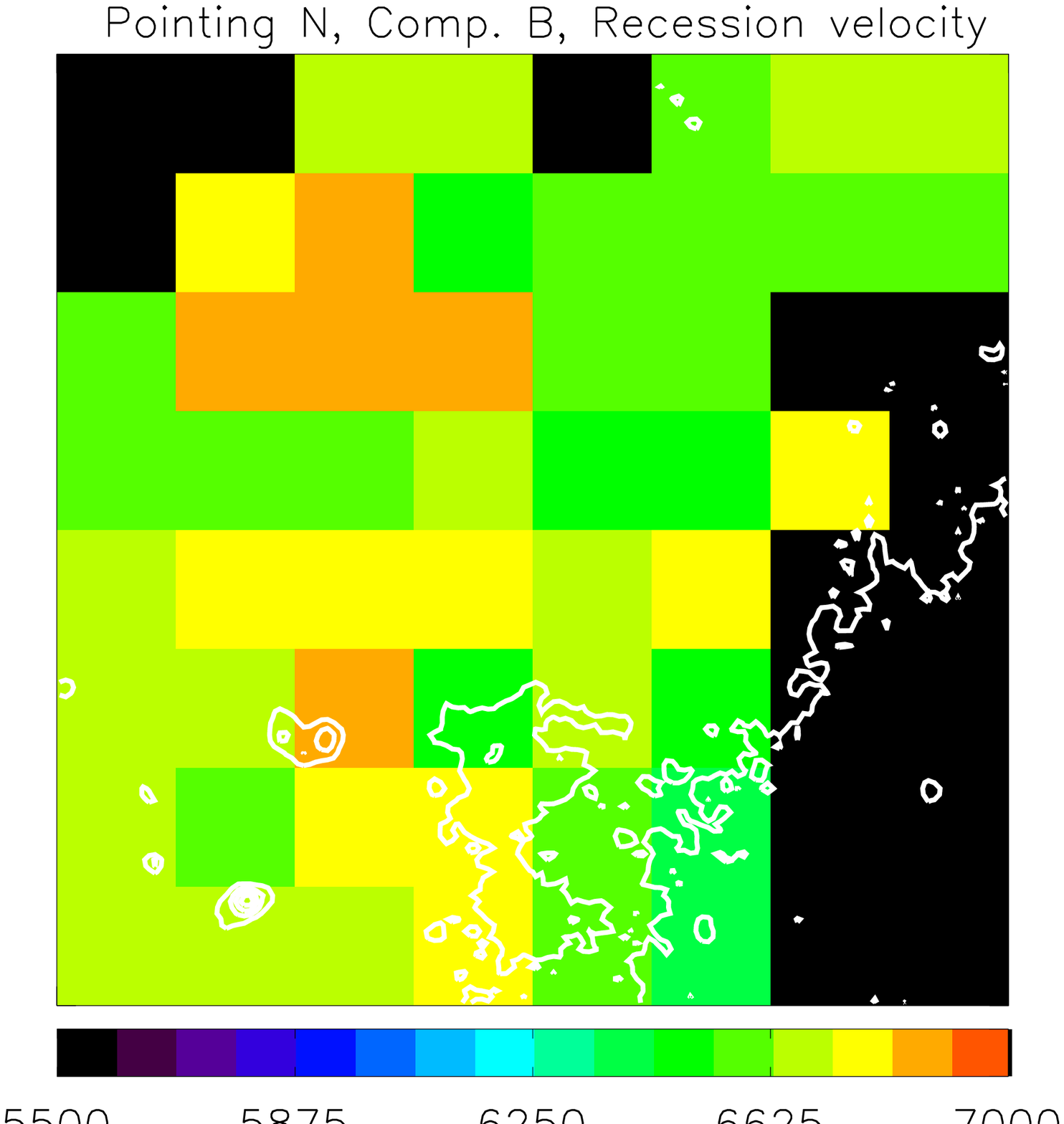}
      \caption{First row: H$\alpha$ flux maps of component A corresponding to pointings S (left), M (middle) and N (right), in units of $10^{-16}$~erg~s$^{-1}$~cm$^{-2}$.
Second row: Radial velocity maps of component A corresponding to pointings S (left), M (middle) and N (right), in units of km~s$^{-1}$.
Third row: H$\alpha$ flux maps of component B corresponding to pointings M (left) and N (right), in units of $10^{-16}$~erg~s$^{-1}$~cm$^{-2}$.
Fourth row: Radial velocity maps of component B corresponding to pointings M (left) and N (right), in units of km~s$^{-1}$.
White contours correspond to the $V$-band HST image: 18.99, 18.23, 17.48 and 16.73~mag~arcsec$^{-2}$. North is up, East is left.
}
         \label{hcg92s_maps}
   \end{figure}

\begin{figure}
   \centering
   \includegraphics[width=15cm]{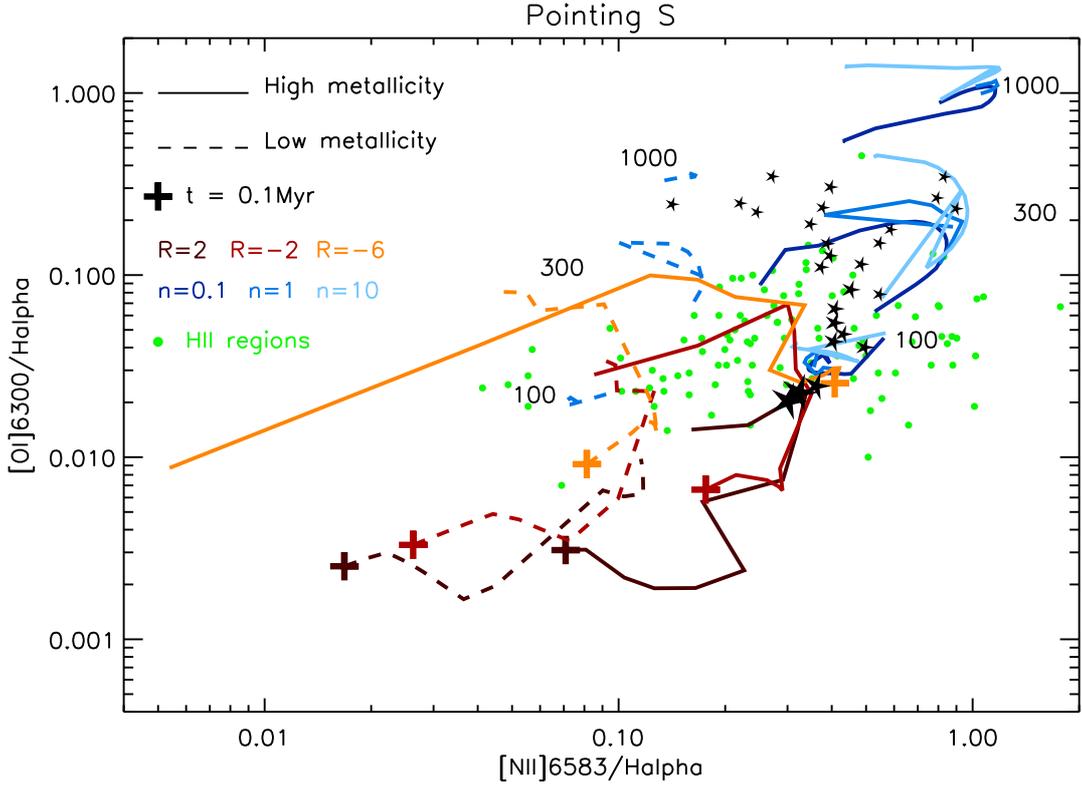}
      \caption{[N{\sc ii}]$\lambda$6583\AA/H$\alpha$ vs. [O{\sc i}]$\lambda$6300\AA/H$\alpha$ for the spaxels of pointing S.
Black stars and dots correspond to components A and B respectively.
The size of the dots is proportional to the flux of the H$\alpha$ line.
Only components for which the intensity peaks of the lines [O{\sc i}]$\lambda$6300\AA\ and [N{\sc ii}]$\lambda$6583\AA\ are above $\Sigma_{bkg}$ are plotted.
The brown lines correspond to the star formation models from Dopita et al. (2006) (solid lines for solar metallicity and dashed lines for 0.4 solar metallicity).
Each line corresponds to a temporal sequence where the beginning is indicated with a '$+$' and corresponds to an age of $t=0.1$~Myr,
and the opposite tip of the line corresponds to an age of $t=6$~Myr.
The values of the parameter $R$ are color codded as indicated in the legend.
The blue lines correspond to the shock$+$precursor models of Allen et al. (2008) (solid lines for solar metallicity and dashed lines for SMC metallicity).
The values of the pre-shock density are color coded and indicated in the legend in units of cm$^{-3}$.
For SMC metallicity only models with $n_{e} = 1$~cm$^{-3}$ are plotted.
The numbers close to the shock models indicate the shock velocity in km~s$^{-1}$.
Small green dots correspond to the samples of H{\sc ii} regions from van Zee et al. (1998) and van Zee \& Haynes (2006).
}
         \label{hcg92s_o1n2}
   \end{figure}

\begin{figure}
   \centering
   \includegraphics[width=15cm]{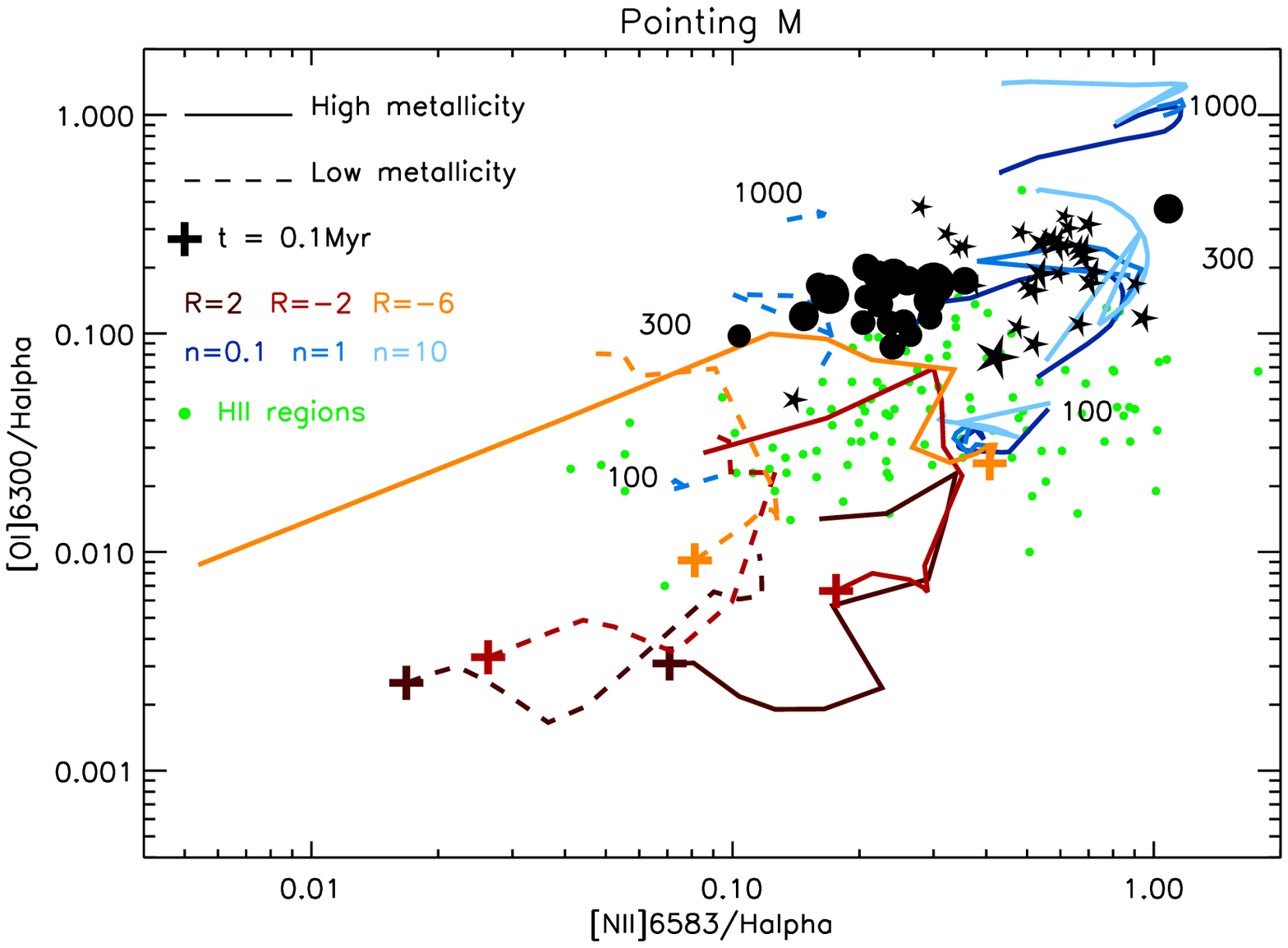}
      \caption{Same as Figure~\ref{hcg92s_o1n2} for pointing M.
}
         \label{hcg92m_o1n2}
   \end{figure}

\begin{figure}
   \centering
   \includegraphics[width=15cm]{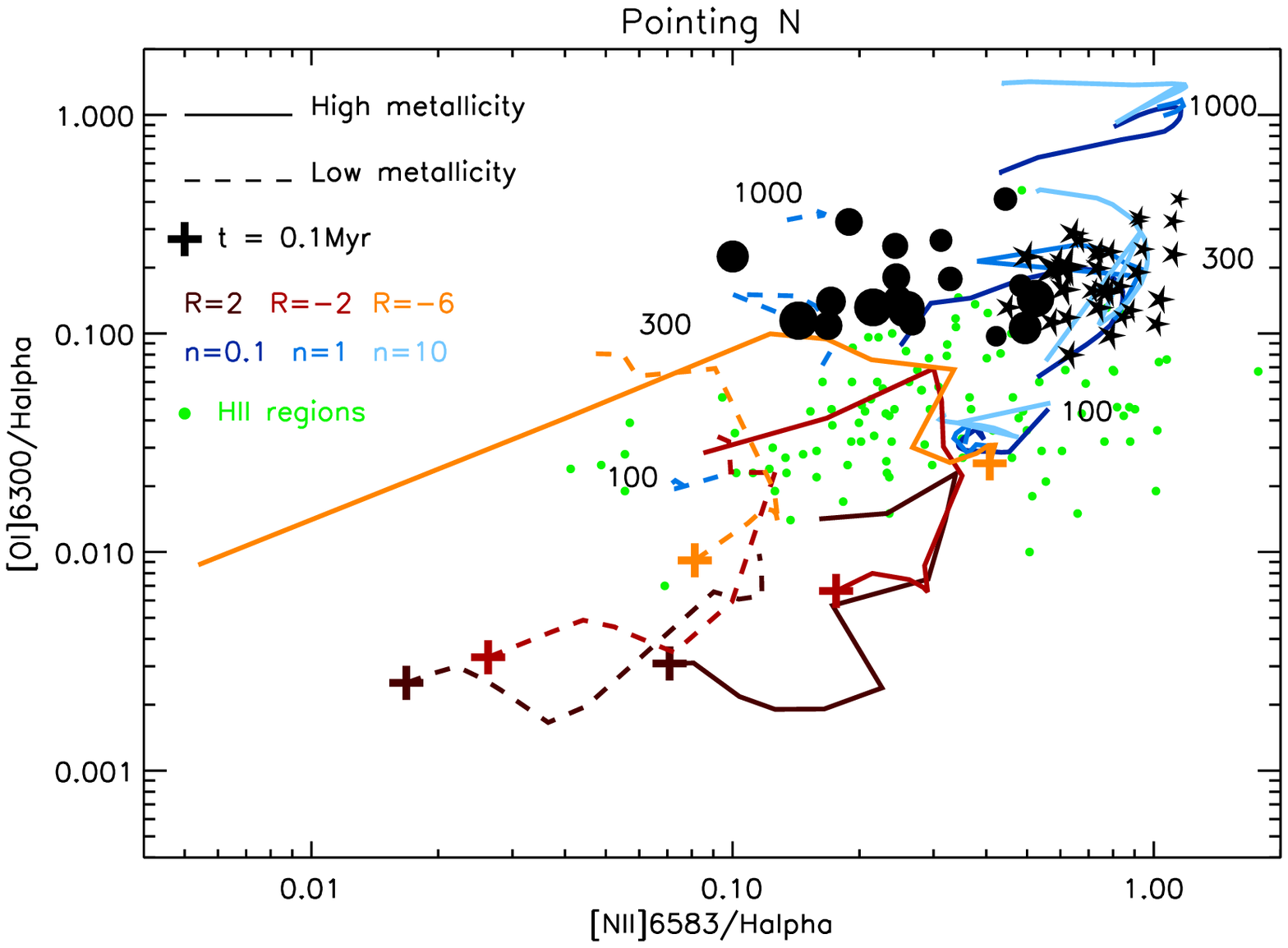}
      \caption{Same as Figure~\ref{hcg92s_o1n2} for pointing N.
}
         \label{hcg92n_o1n2}
   \end{figure}

\begin{figure}
   \centering
   \includegraphics[width=10cm]{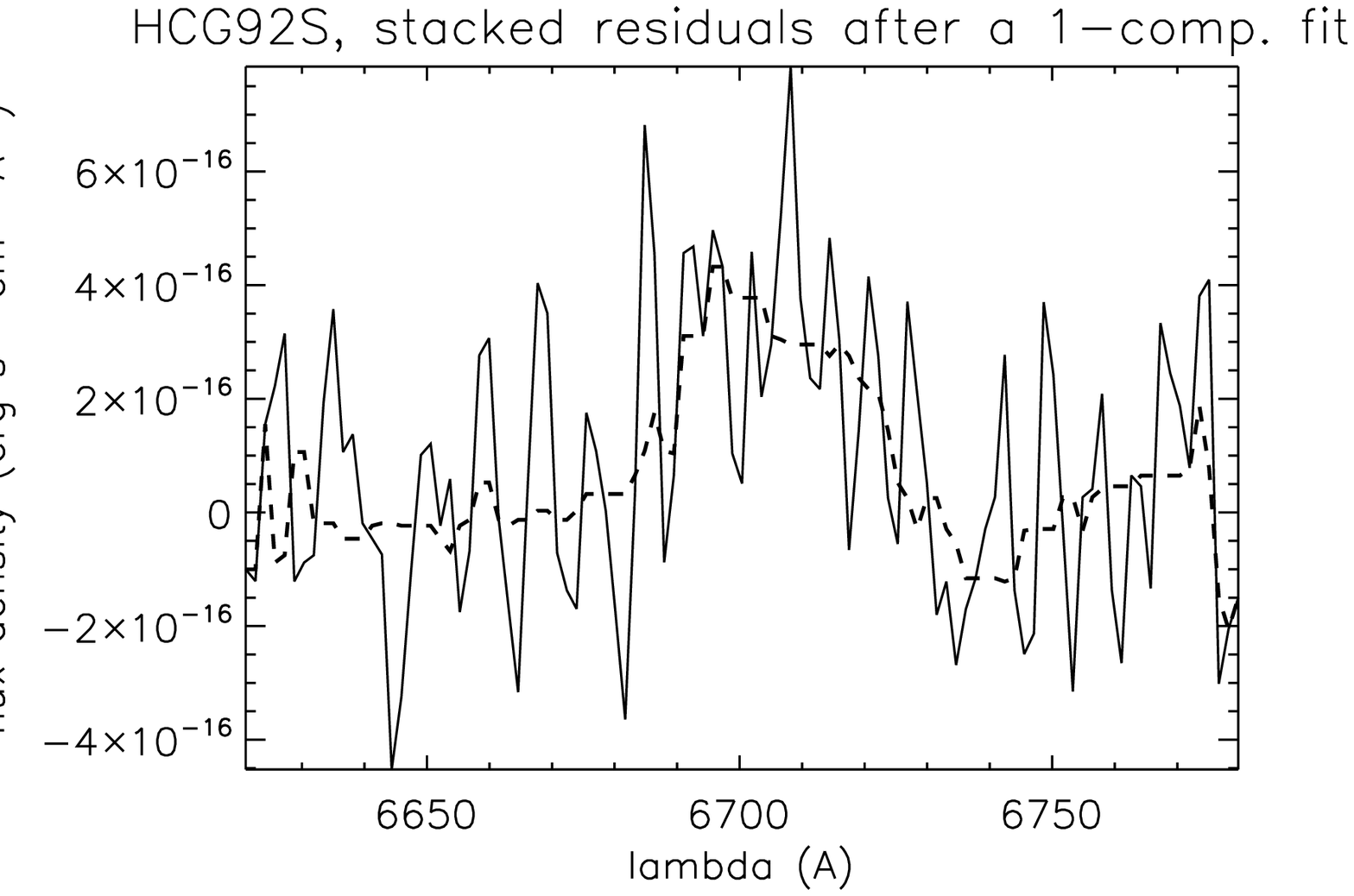}
   \includegraphics[width=10cm]{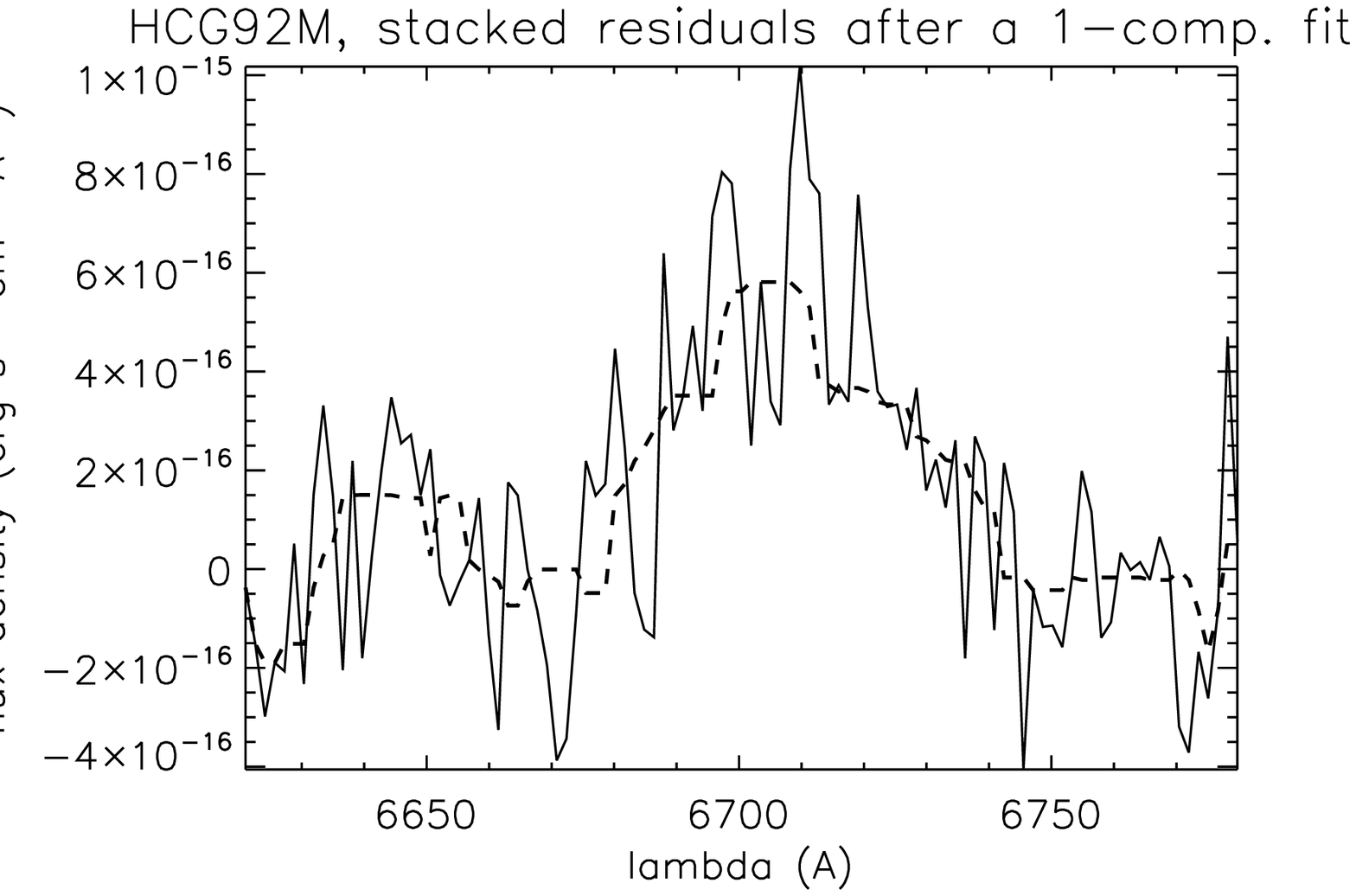}
   \includegraphics[width=10cm]{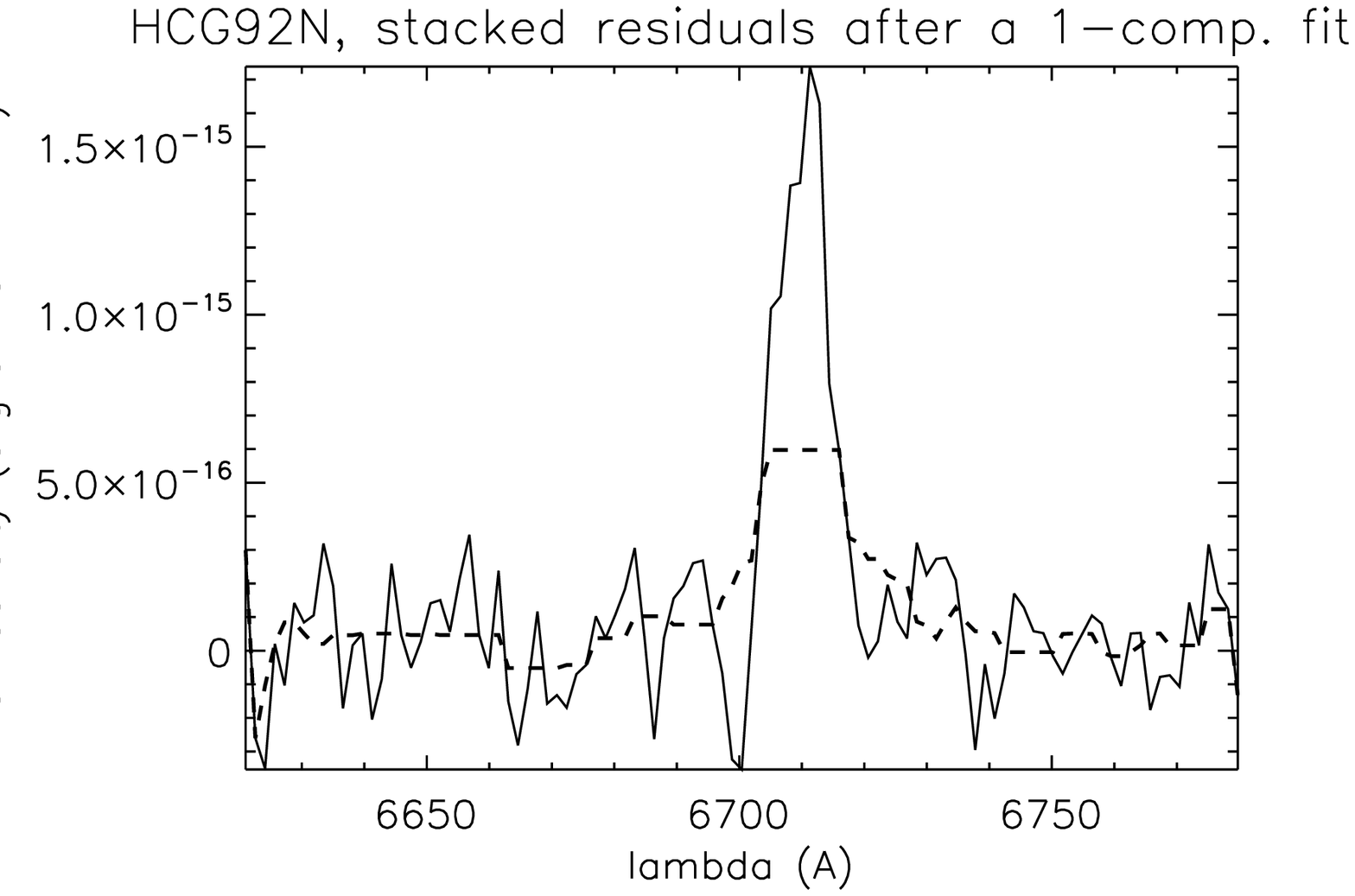}
      \caption{Stacked residuals of the 64 spaxels of the three pointings after a 1-component fit.
The dashed line corresponds to the median along 15 pixels in the X-axis.
}
         \label{resi_1com}
   \end{figure}

\begin{figure}
   \centering
   \includegraphics[width=10cm]{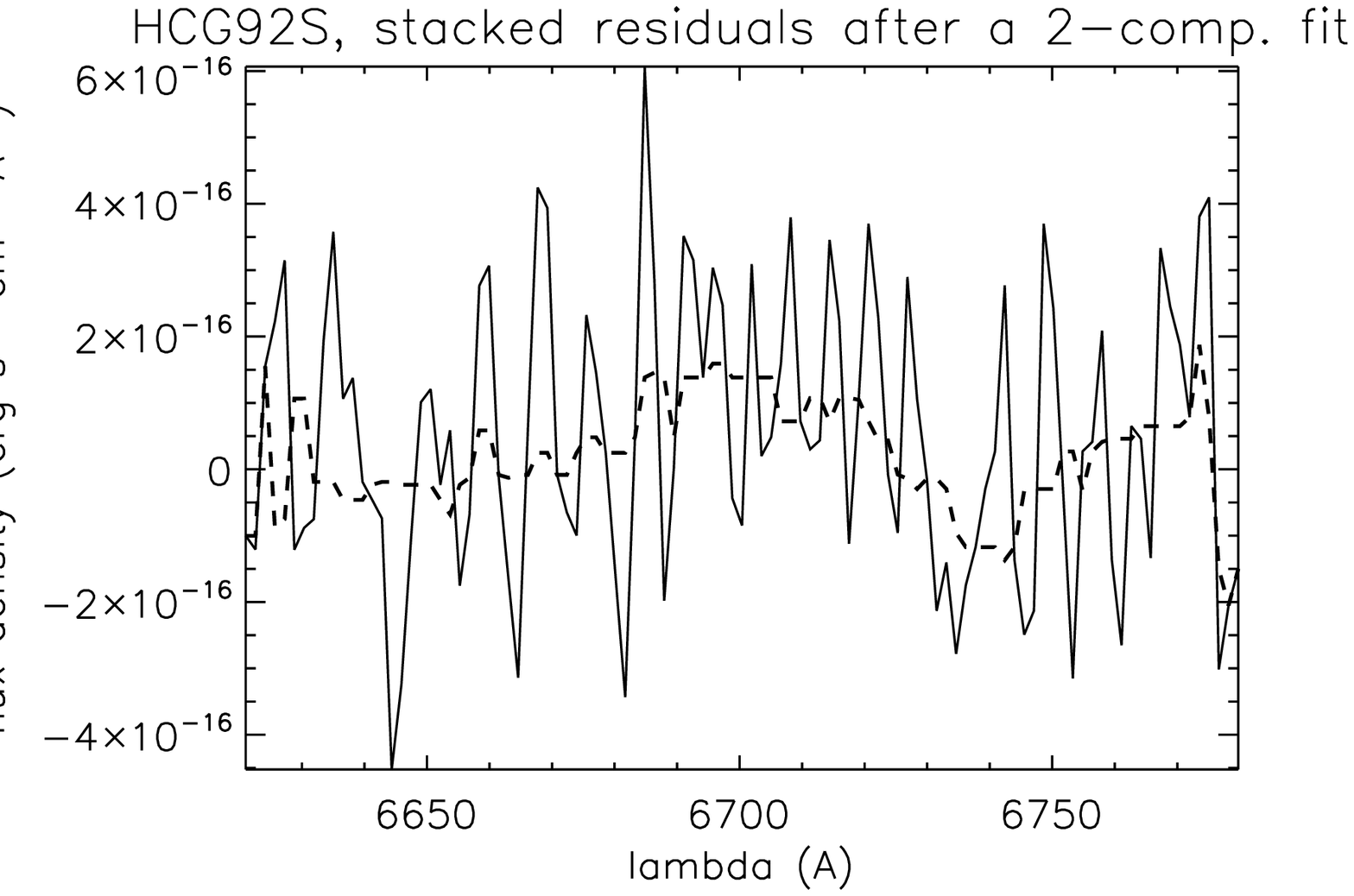}
   \includegraphics[width=10cm]{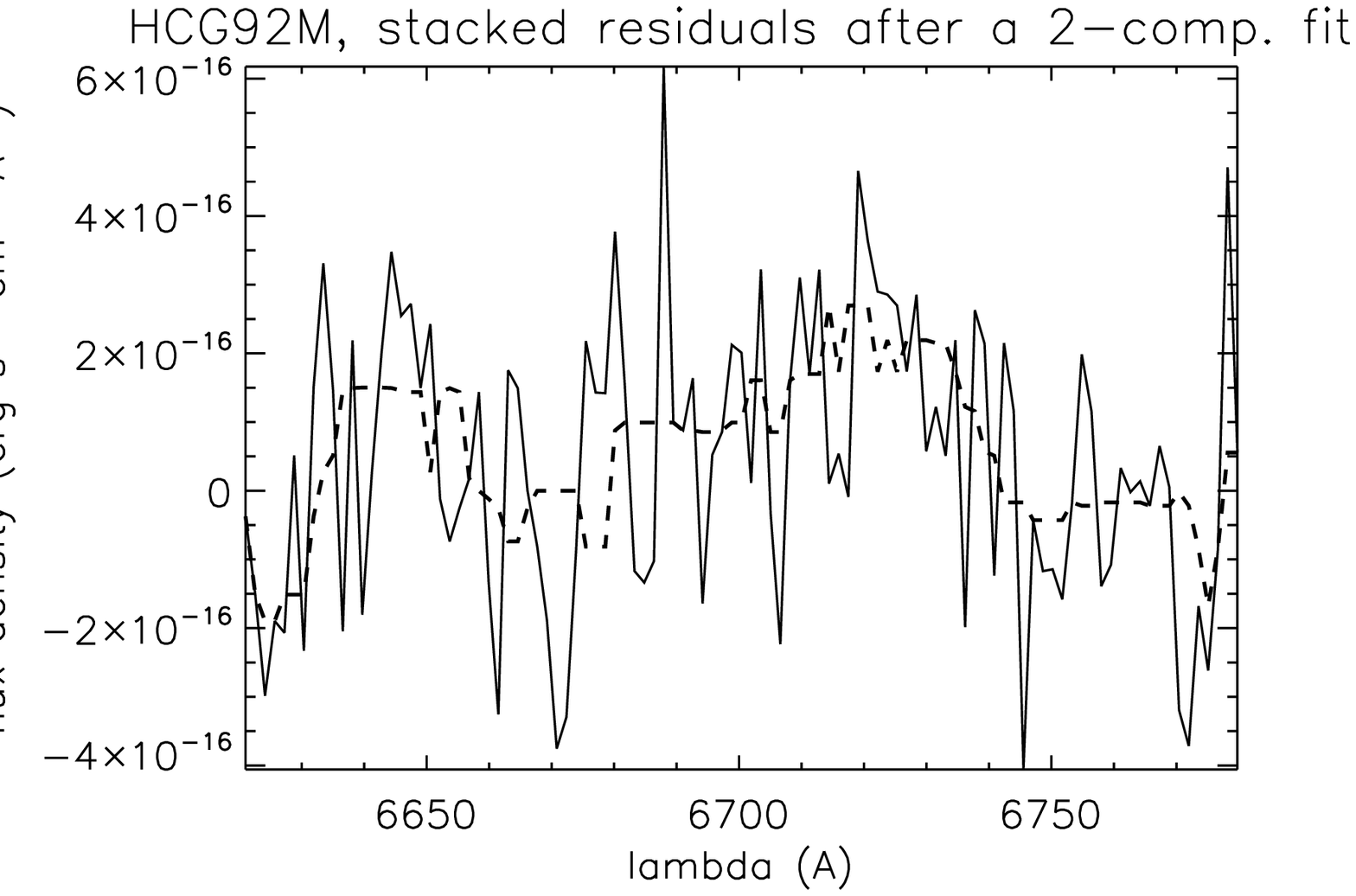}
   \includegraphics[width=10cm]{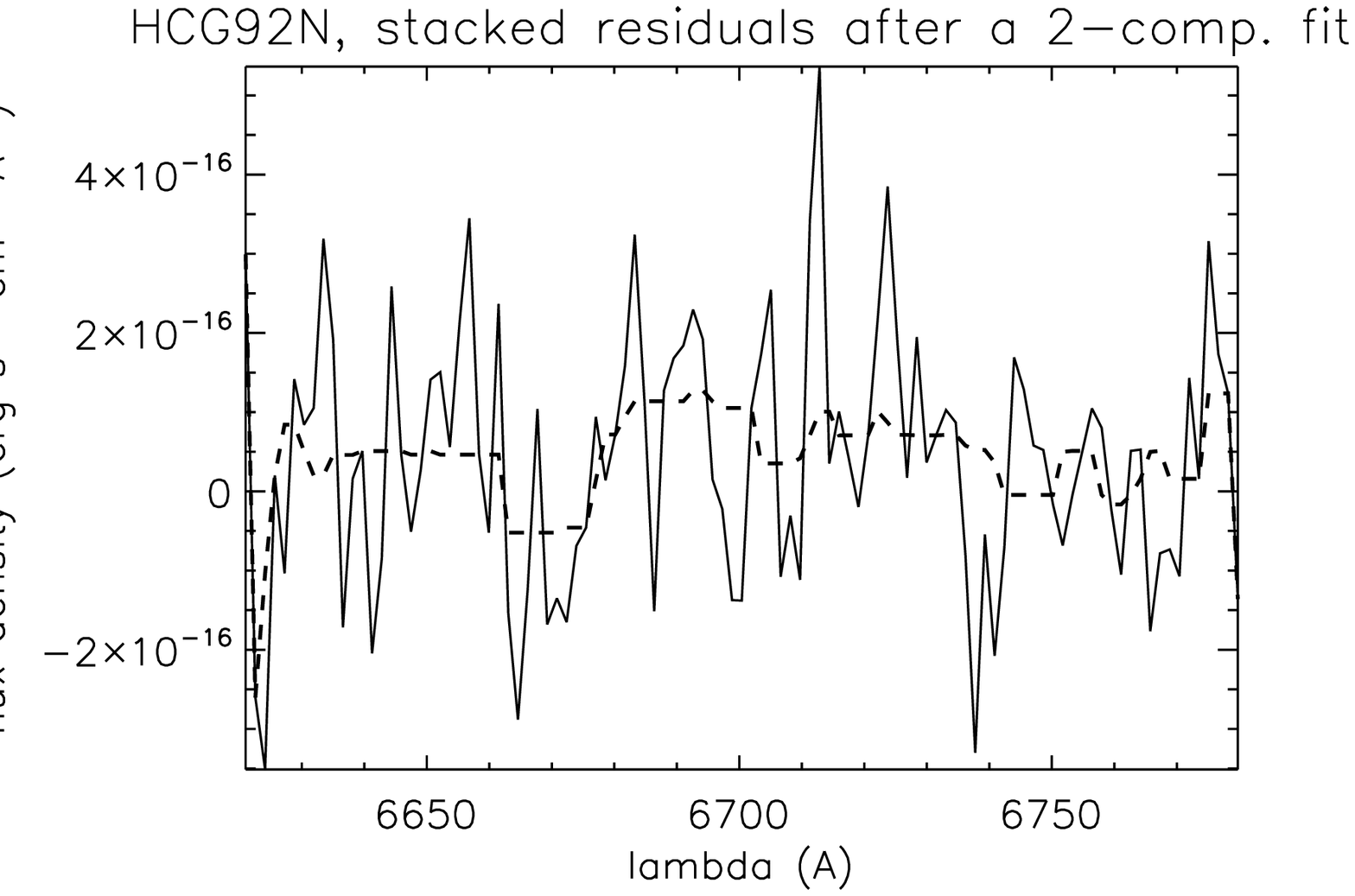}
      \caption{Stacked residuals of the 64 spaxels of the three pointings after a 2-component fit.
The dashed line corresponds to the median along 15 pixels in the X-axis.
}
         \label{resi_2com}
   \end{figure}

\begin{figure}
   \centering
   \includegraphics[width=20cm]{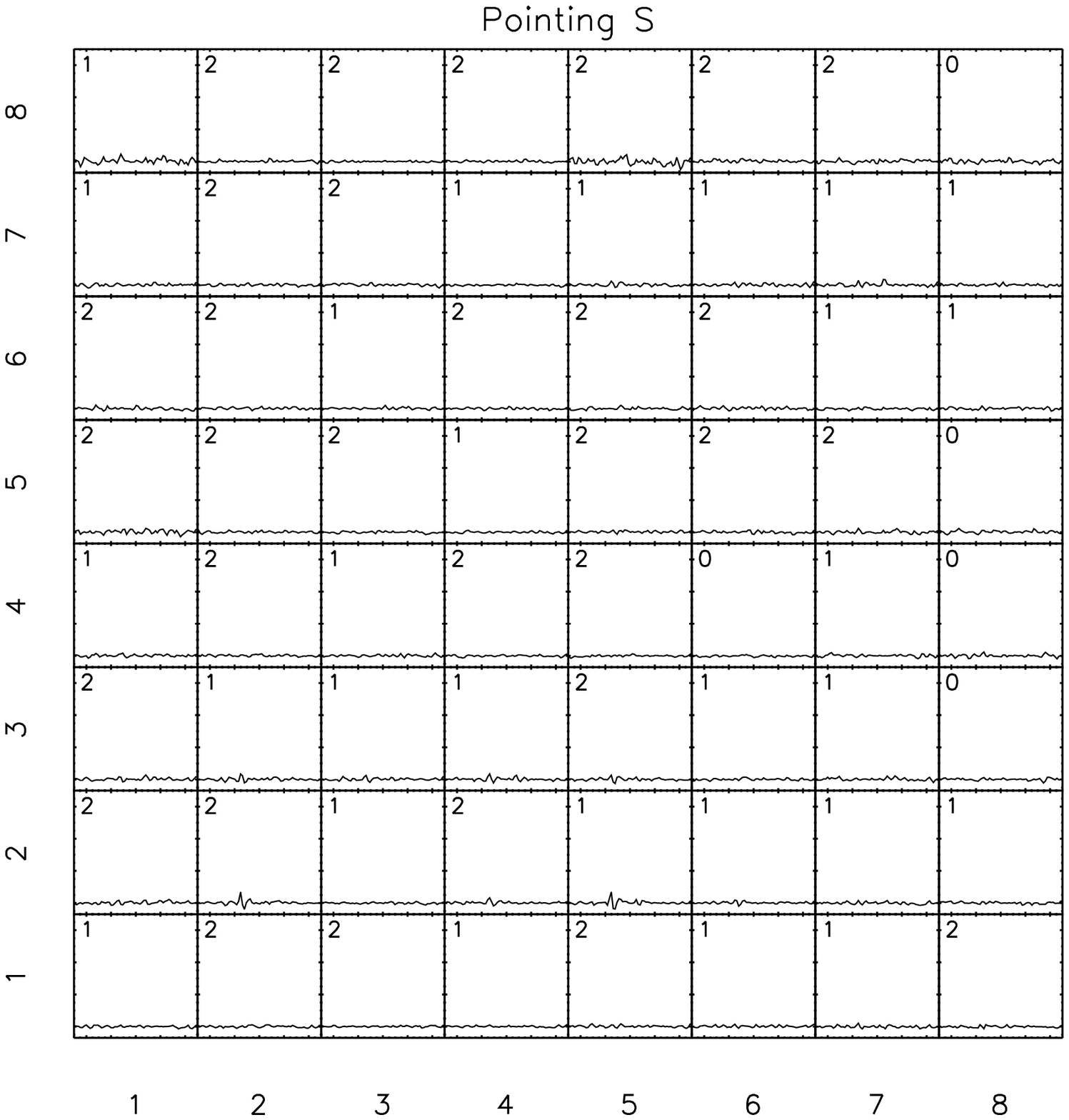}
      \caption{Spatial arrangement of the residua after the fitting procedure of pointing S.
The horizontal axis of all spectra ranges from 6650\AA\ to 6750\AA.
The vertical axis scale is the same as in Figure~\ref{hcg92s_only_todos}.
The number to the top left corner of each panel indicates the number of components resulting from the fitting procedure.
A spaxel labeled with '0' means that 0 components were assigned to this spaxel.
}
         \label{hcg92s_resi}
   \end{figure}

\begin{figure}
   \centering
   \includegraphics[width=20cm]{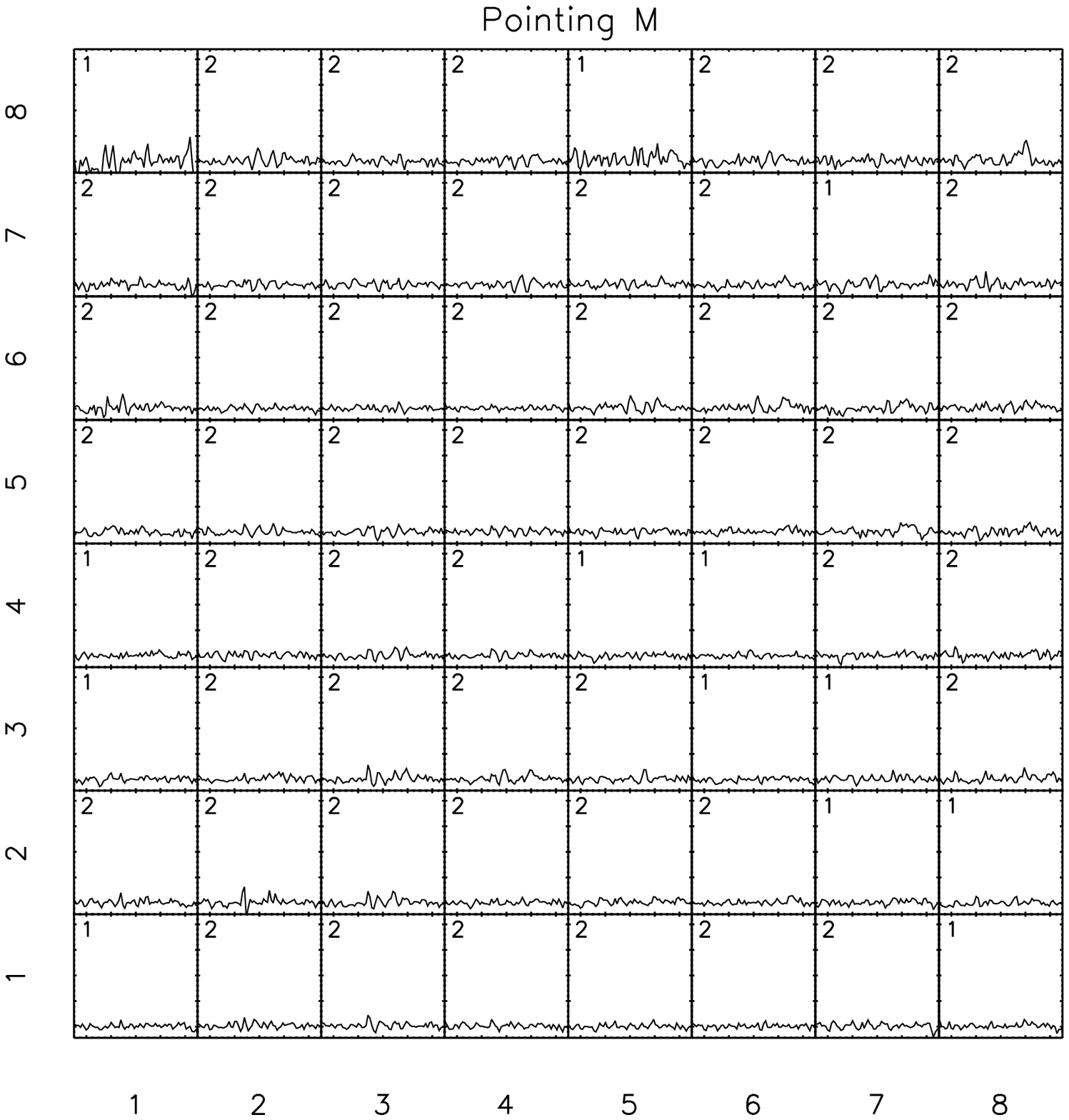}
      \caption{Spatial arrangement of the residua after the fitting procedure of pointing S.
The horizontal axis of all spectra ranges from 6650\AA\ to 6750\AA.
The vertical axis scale is the same as in Figure~\ref{hcg92m_only_todos}.
The number to the top left corner of each panel indicates the number of components resulting from the fitting procedure.
A spaxel labeled with '0' means that 0 components were assigned to this spaxel.
}
         \label{hcg92m_resi}
   \end{figure}

\begin{figure}
   \centering
   \includegraphics[width=20cm]{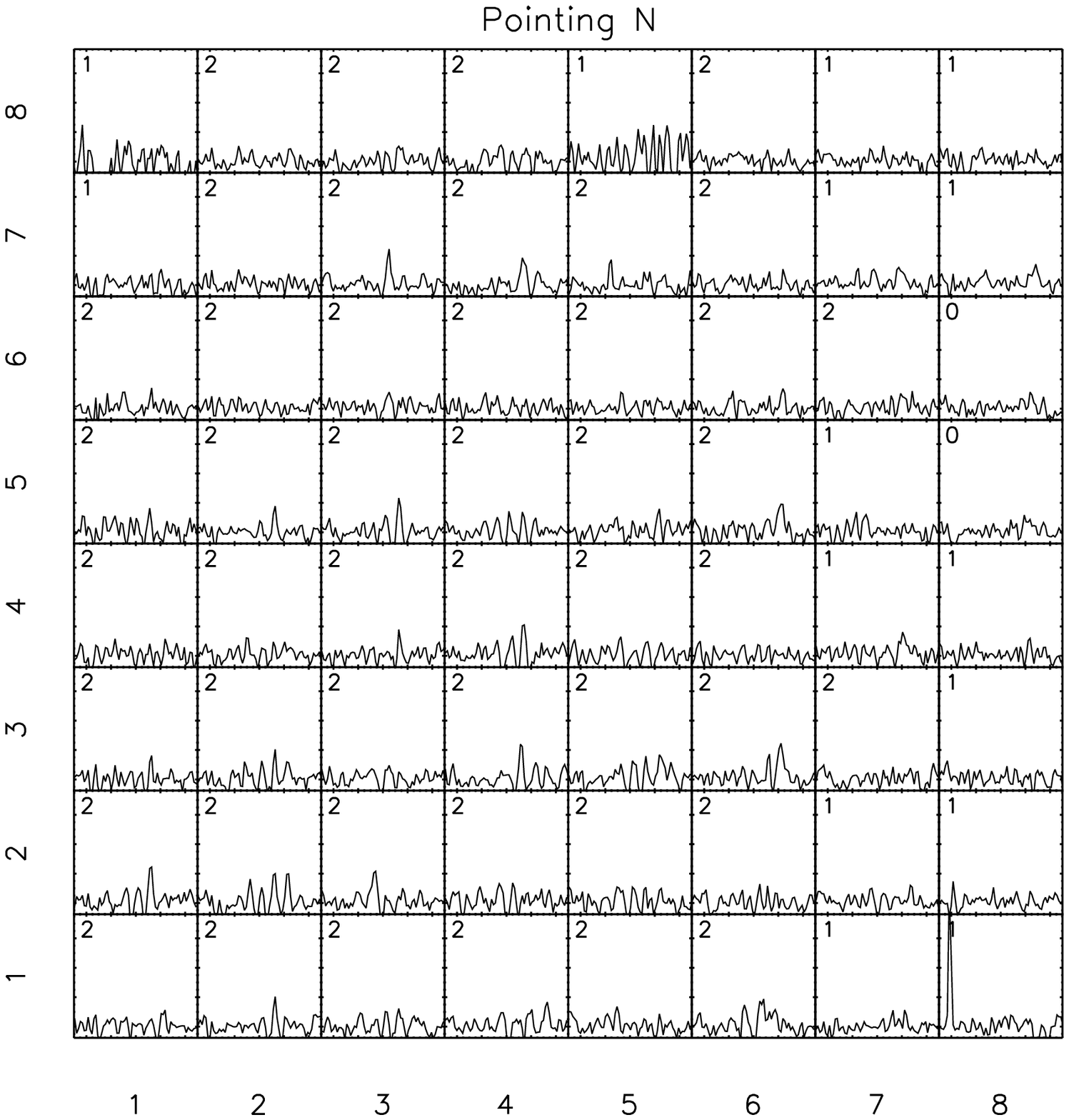}
      \caption{Spatial arrangement of the residua after the fitting procedure of pointing S.
The horizontal axis of all spectra ranges from 6650\AA\ to 6750\AA.
The vertical axis scale is the same as in Figure~\ref{hcg92n_only_todos}.
The number to the top left corner of each panel indicates the number of components resulting from the fitting procedure.
A spaxel labeled with '0' means that 0 components were assigned to this spaxel.
}
         \label{hcg92n_resi}
   \end{figure}

As mentioned in a previous section, figures~\ref{hcg92s_only_todos} to \ref{hcg92n_only_todos} 
suggest the possibility that in some spaxels the emission comes from more than a single kinematical component.
In order to decide the number of components we first try with a single component. 
Figure~\ref{resi_1com} shows the stacked spectra of the residuals after a 1-component fit for our three pointings.
In the three cases the stacked residuals show an emission feature around 6710\AA.
This feature is broad for pointings S and M, and narrow and well defined in pointing N.
The existence of these residual features could imply that we are missing a further component so
we try again the fitting procedure with two components.
Figure~\ref{resi_2com} shows stacked residuals after this 2-component fit and in this case we do not
see any emission feature, the result being reasonably flat.
For this reason, we decide to fix the number of kinematical components in two.
Figures~\ref{hcg92s_resi} to \ref{hcg92n_resi} show the residuals in the [6650,6750]\AA\ spectral interval, after
the subtraction of the 2-component H$\alpha$ $+$ [N{\sc ii}] fitting.
The residuals are quite flat, implying that our choice is reasonable.

Once the fit of these three lines is finished, we fit the [O{\sc i}]$\lambda$6300\AA\ line in the red spectra 
and the [O{\sc iii}]$\lambda$4959,5007\AA\ and H$\beta$ lines in the blue spectra to gaussians with the redshifts 
and velocity dispersions resulting from the fit to the corresponding red spectra.

\begin{figure}
   \centering
   \includegraphics[width=15cm]{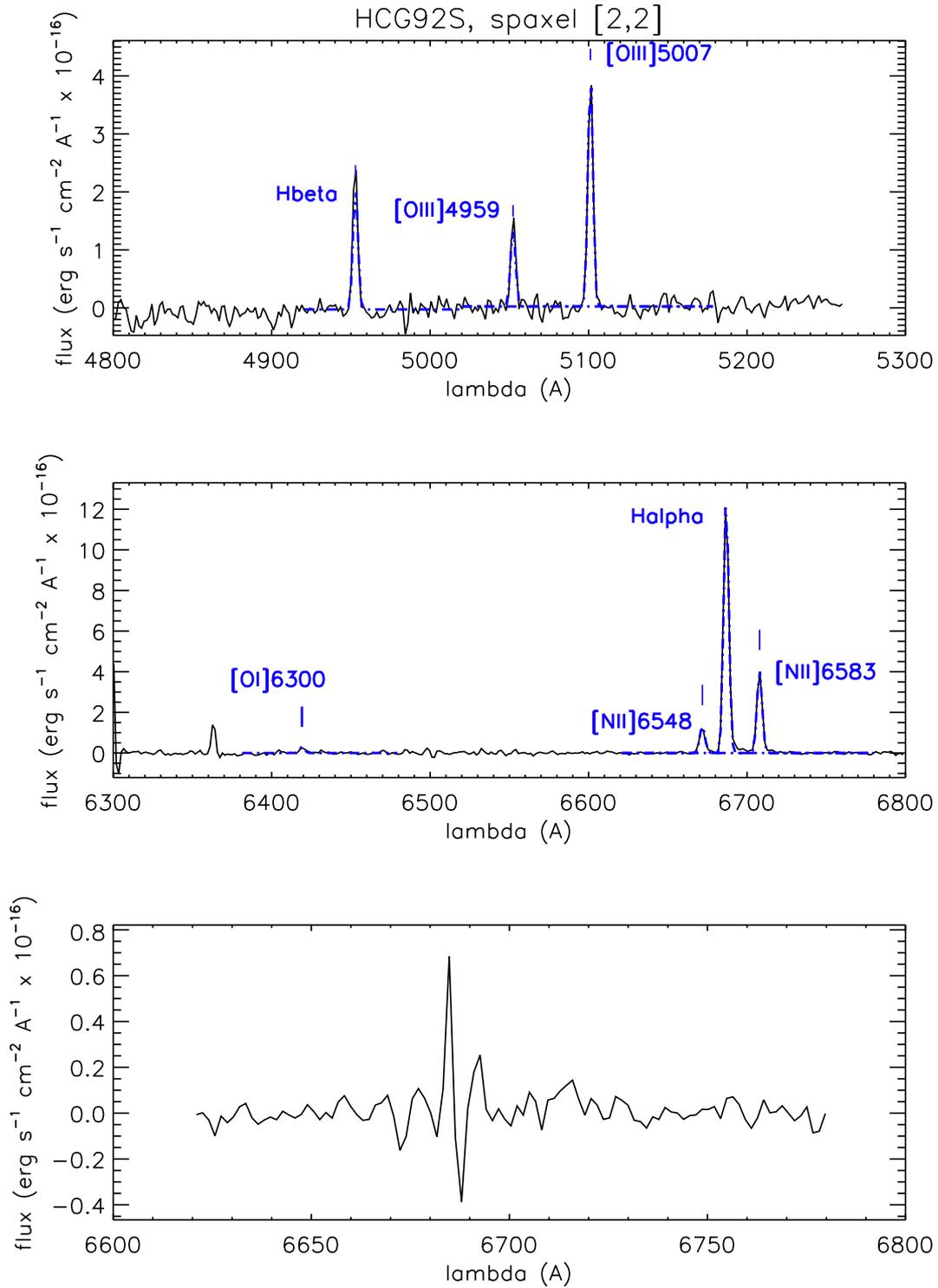}
      \caption{Top: Blue spectrum of spaxel S[2,2]; Middle: Red spectrum of spaxel S[2,2]. 
Bottom: Residua of the best fit to the spectrum of spaxel S[2,2] in the red wavelength range.
The best fit to the emission lines of components A and B for which the intensity peak is above $\Sigma_{bkg}$
are shown in blue and red respectively.
}
         \label{hcg92s_02_02}
   \end{figure}

\begin{figure}
   \centering
   \includegraphics[width=15cm]{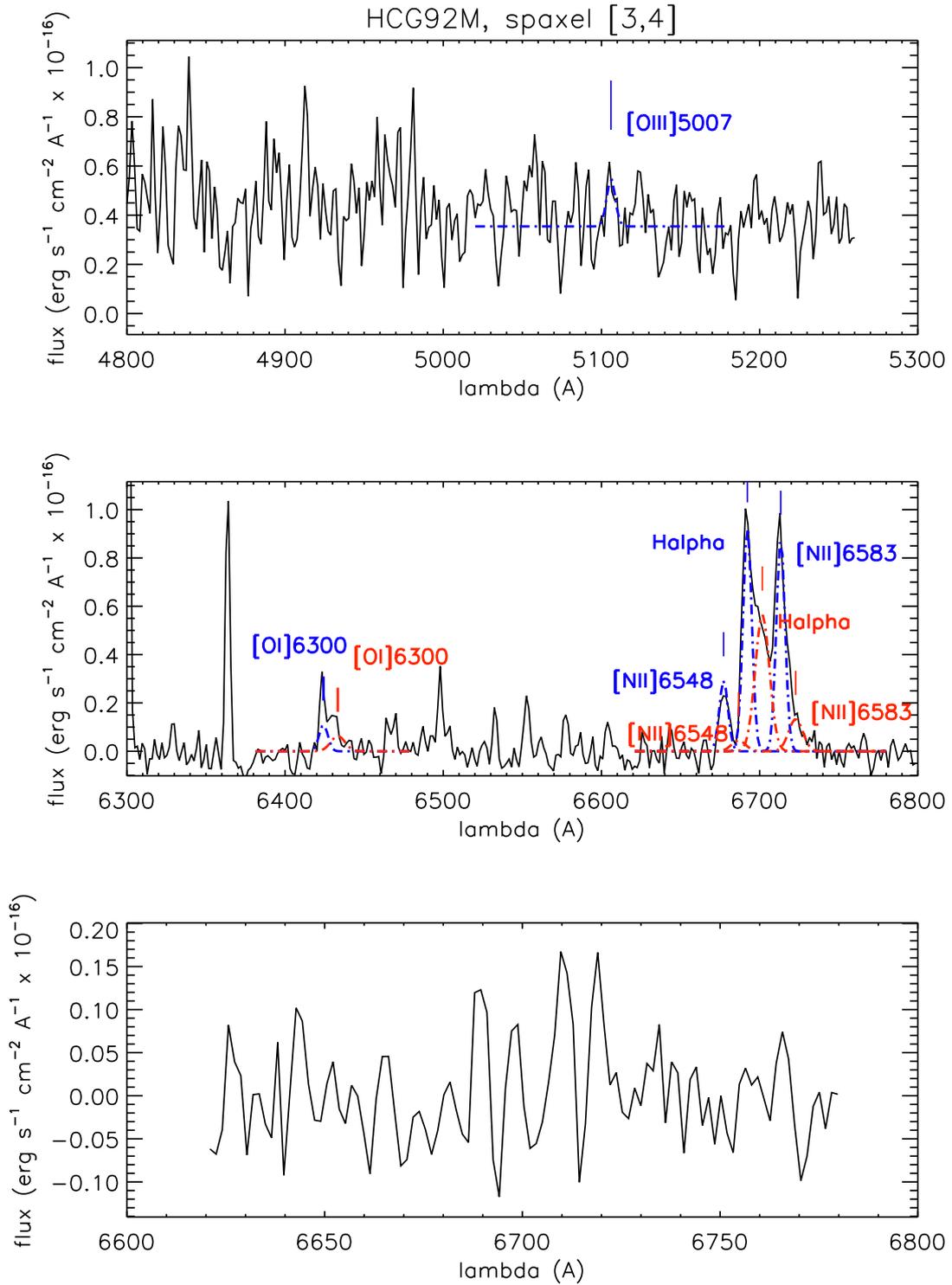}
      \caption{Same as figure~\ref{hcg92s_02_02} for spaxel M[3,4].
}
         \label{hcg92m_03_04}
   \end{figure}

\begin{figure}
   \centering
   \includegraphics[width=15cm]{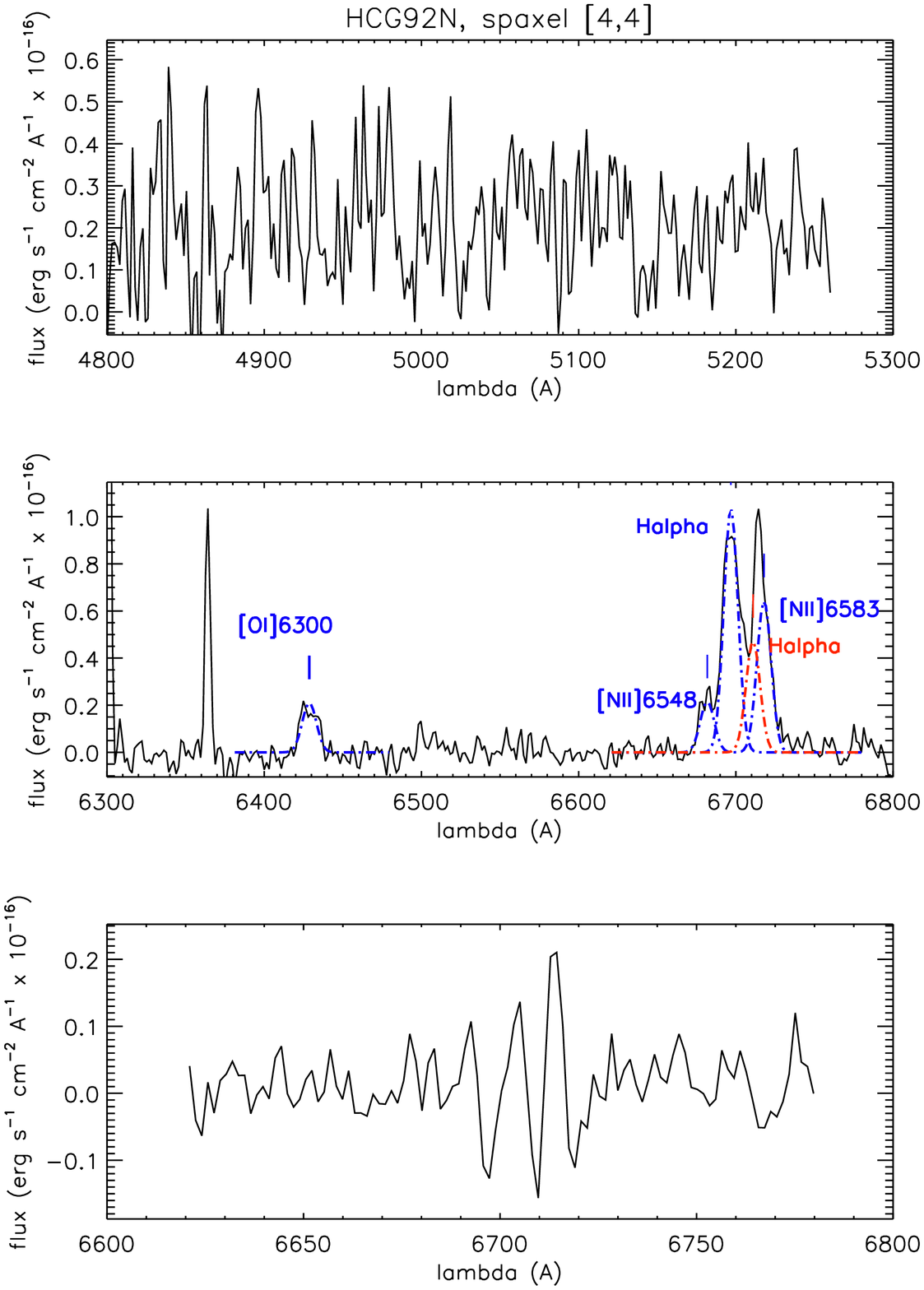}
      \caption{Same as figure~\ref{hcg92s_02_02} for spaxel N[4,4].
}
         \label{hcg92n_04_04}
   \end{figure}

To illustrate the quality of the fits, we show in figures~\ref{hcg92s_02_02} to \ref{hcg92n_04_04} three examples
of typical fits of each pointing.
Spaxel S[2,2] was fit to a single component and corresponds to one of the H{\sc ii} regions and shows
a typical H{\sc ii}-like spectrum.
Spaxels M[3,4] and N[4,4] were fit to two components and correspond to 
the shocked region.
The red spectra show broad lines unlike the blue spectra where few emission
lines were detected.
Figure~\ref{hcg92s_02_02} shows a conpicuous asymmetric residual at $\approx 6685$\AA, coincident with the strong H$\alpha$ line
detected in spaxel S[2,2].
This residual is unexplained and since we do not see it for other bright H$\alpha$ lines or arcs, it is unlikely be due to an instrumental effect.

In order to avoid spurious results due to low signal-to-noise, throughout this paper we will keep only 
those components for which the intensity peak of the H$\alpha$ line is above $5 \times \Sigma_{bkg}$. 
This means that the maximum number of components detected for individual spaxels is 2, but could be 1 or even zero 
depending on the signal-to-noise of the H$\alpha$ line.
For the rest of the lines, we fit a gaussian with $\sigma$ and redshift equal to those of the
corresponding H$\alpha$ line, and we measure the fluxes of all the lines for which the intensity peak
is above $1 \times \Sigma_{bkg}$.

\end{appendix}

\newpage

\onecolumn

\begin{table}
\begin{tabular}{lccccccc}
\hline
Pointing & R.A. (J2000.0) & Dec. (J2000.0) & Date & Exp. time (s) & Seeing (``) & Airmass & Conditions$^{\dagger}$ \\
\hline
S & 22 35 59.0 & 33 57 38.0 & Aug 24 & $3\times 1200$ & 1.5 & 1.11 &  NP \\
M & 22 35 59.5 & 33 57 55.0 & Aug 25 & $3\times 1200$ & 1.5 & 1.01 &  NP \\
N & 22 36 00.0 & 33 58 10.0 & Aug 22 & $3\times 1200$ & 1.2 & 1.12 &  P  \\
\hline
\end{tabular}
\caption{Diary of the observations. $^{\dagger}$ P (NP) stands for photometric (non photometric) atmospheric conditions.
\label{LOG}}
\end{table}

\newpage
\clearpage

\begin{table}
\begin{tabular}{lccccccccc}
\hline
Spaxel & $v_{rad}$ & $\sigma$ & $\Sigma_{bkg,blue}$ & $f$(H$\beta$) & $f$([O{\sc iii}]$\lambda$5007\AA\ & $\Sigma_{bkg,red}$ & $f$([O{\sc i}]$\lambda$6300\AA\ & $f$(H$\alpha$) & $f$([N{\sc ii}]$\lambda$6583\AA\ \\
       & & & & & /$f$(H$\beta$) & & /$f$(H$\alpha$) & & /$f$(H$\alpha$) \\
S[ 1, 1]a   & 6607.5 &  70.4 &  1.38e-17 & $<$ 2.44e-15 & --- &  6.02e-18 & $<$  0.13 &  9.54e-15 &    0.15 \\
S[ 1, 2]a   & 5704.8 &  90.4 &  1.70e-17 & $<$ 3.86e-15 & --- &  5.51e-18 &    0.11 &  1.65e-14 &    0.08 \\
S[ 1, 2]b   & 6615.0 & 171.9 &  1.70e-17 & $<$ 7.34e-15 & --- &  5.51e-18 & $<$  0.11 &  2.58e-14 &    0.09 \\
S[ 1, 3]a   & 5728.5 & 109.2 &  1.95e-17 & $<$ 5.33e-15 & --- &  5.68e-18 &    0.11 &  2.52e-14 &    0.48 \\
S[ 1, 4]a   & 5934.0 & 181.2 &  1.67e-17 & $<$ 7.59e-15 & --- &  6.03e-18 &    0.30 &  1.99e-14 &    0.40 \\
S[ 1, 5]   &  ---  &  ---  &  ---  &  ---  &  ---  &  ---  &  ---  &  ---  &  ---  \\
S[ 1, 6]   &  ---  &  ---  &  ---  &  ---  &  ---  &  ---  &  ---  &  ---  &  ---  \\
S[ 1, 7]   &  ---  &  ---  &  ---  &  ---  &  ---  &  ---  &  ---  &  ---  &  ---  \\
S[ 1, 8]   &  ---  &  ---  &  ---  &  ---  &  ---  &  ---  &  ---  &  ---  &  ---  \\
S[ 2, 1]a   & 5725.5 &  52.1 &  1.47e-17 & $<$ 1.92e-15 & --- &  5.45e-18 & $<$  0.17 &  4.26e-15 &    0.63 \\
S[ 2, 2]a   & 5670.9 &  38.7 &  1.28e-17 &   2.03e-14 &    1.94 &  4.46e-18 &    0.02 &  1.22e-13 &    0.32 \\
S[ 2, 3]a   & 5676.9 &  41.0 &  1.88e-17 &   1.20e-14 &    1.76 &  5.35e-18 &    0.02 &  7.64e-14 &    0.36 \\
S[ 2, 4]a   & 5809.8 &  98.3 &  1.50e-17 & $<$ 3.71e-15 & --- &  5.02e-18 &    0.24 &  1.38e-14 &    0.14 \\
S[ 2, 4]b   & 6730.2 & 196.0 &  1.50e-17 &   9.96e-15 & $<$  1.59 &  5.02e-18 & $<$  0.21 &  1.37e-14 & $<$  0.18 \\
S[ 2, 5]a   & 5942.4 & 145.6 &  1.37e-17 & $<$ 4.99e-15 & --- &  5.74e-18 &    0.24 &  1.88e-14 &    0.11 \\
S[ 2, 5]b   & 6872.4 & 130.7 &  1.37e-17 & $<$ 4.48e-15 & --- &  5.74e-18 & $<$  0.20 &  1.16e-14 &    0.16 \\
S[ 2, 6]a   & 5964.0 & 165.9 &  1.68e-17 & $<$ 6.99e-15 & --- &  5.03e-18 &    0.28 &  1.77e-14 & $<$  0.12 \\
S[ 2, 7]   &  ---  &  ---  &  ---  &  ---  &  ---  &  ---  &  ---  &  ---  &  ---  \\
S[ 2, 8]   &  ---  &  ---  &  ---  &  ---  &  ---  &  ---  &  ---  &  ---  &  ---  \\
S[ 3, 1]a   & 5744.1 & 102.2 &  1.37e-17 & $<$ 3.51e-15 & --- &  4.69e-18 & $<$  0.12 &  9.65e-15 &    0.64 \\
S[ 3, 2]a   & 5730.6 &  78.6 &  1.29e-17 &   8.82e-15 &    1.30 &  5.15e-18 &    0.08 &  4.55e-14 &    0.45 \\
S[ 3, 3]a   & 5742.3 &  79.5 &  1.41e-17 &   8.75e-15 &    1.45 &  5.33e-18 &    0.04 &  4.55e-14 &    0.50 \\
S[ 3, 4]a   & 5800.8 &  72.9 &  1.11e-17 & $<$ 2.02e-15 & --- &  4.95e-18 &    0.15 &  1.80e-14 &    0.39 \\
S[ 3, 5]a   & 5876.1 & 101.2 &  1.53e-17 & $<$ 3.89e-15 & --- &  5.63e-18 & $<$  0.17 &  1.18e-14 & $<$  0.12 \\
S[ 3, 6]   &  ---  &  ---  &  ---  &  ---  &  ---  &  ---  &  ---  &  ---  &  ---  \\
S[ 3, 7]   &  ---  &  ---  &  ---  &  ---  &  ---  &  ---  &  ---  &  ---  &  ---  \\
S[ 3, 8]   &  ---  &  ---  &  ---  &  ---  &  ---  &  ---  &  ---  &  ---  &  ---  \\
S[ 4, 1]   &  ---  &  ---  &  ---  &  ---  &  ---  &  ---  &  ---  &  ---  &  ---  \\
S[ 4, 2]a   & 5777.1 &  93.1 &  1.34e-17 & $<$ 3.13e-15 & --- &  4.67e-18 &    0.11 &  2.47e-14 &    0.37 \\
S[ 4, 2]b   & 6603.6 &  38.4 &  1.34e-17 & $<$ 1.29e-15 & --- &  4.67e-18 & $<$  0.18 &  2.41e-15 & $<$  0.19 \\
S[ 4, 3]a   & 5732.4 &  78.0 &  1.66e-17 &   4.50e-15 &    1.68 &  4.69e-18 &    0.05 &  3.77e-14 &    0.43 \\
S[ 4, 4]a   & 5780.7 & 113.0 &  1.37e-17 & $<$ 3.88e-15 & --- &  5.45e-18 &    0.23 &  1.26e-14 &    0.38 \\
S[ 4, 5]a   & 5796.3 & 102.9 &  1.39e-17 & $<$ 3.58e-15 & --- &  5.23e-18 & $<$  0.18 &  7.72e-15 &    0.67 \\
S[ 4, 6]a   & 5809.2 & 117.2 &  1.52e-17 & $<$ 4.47e-15 & --- &  4.63e-18 & $<$  0.23 &  7.05e-15 &    0.25 \\
S[ 4, 7]   &  ---  &  ---  &  ---  &  ---  &  ---  &  ---  &  ---  &  ---  &  ---  \\
S[ 4, 8]a   & 5817.0 &  81.2 &  1.46e-17 & $<$ 2.96e-15 & --- &  4.76e-18 &    0.35 &  4.96e-15 &    0.83 \\
\hline
\end{tabular}
\caption{Properties of the emission lines for the spaxels of pointing S: 
(1) Spectrum Id. and component; 
(2) Recession velocity (km~s$^{-1}$); 
(3) Velocity dispersion at $\lambda=6563$\AA\ (km~s$^{-1}$); 
(4) Standard deviation of the background in the blue spectrum (erg~s$^{-1}$~cm$^{-2}$~\AA$^{-1}$); 
(5) Integrated flux of the H$\beta$ line (erg~s$^{-1}$~cm$^{-2}$); 
(6) [O{\sc iii}]$\lambda$5007/H$\beta$ line flux ratio; 
(7) Standard deviation of the background in the red spectrum (erg~s$^{-1}$~cm$^{-2}$~\AA$^{-1}$); 
(8) [O{\sc i}]$\lambda$6300/H$\alpha$ line flux ratio; 
(9) Integrated flux of the H$\alpha$ line (erg~s$^{-1}$~cm$^{-2}$); 
(10) [N{\sc ii}]$\lambda$6583/H$\alpha$ line flux ratio. 
Colons before numbers mean that the line intensity peaks are lower than 3 times the standard deviation of the corresponding background.}
\label{tabla_fit_hcg92s}
\end{table}

\newpage
\clearpage

\addtocounter{table}{-1}

\begin{table}
\begin{tabular}{lccccccccc}
\hline
Spaxel & $v_{rad}$ & $\sigma$ & $\Sigma_{bkg,blue}$ & $f$(H$\beta$) & $f$([O{\sc iii}]$\lambda$5007\AA\ & $\Sigma_{bkg,red}$ & $f$([O{\sc i}]$\lambda$6300\AA\ & $f$(H$\alpha$) & $f$([N{\sc ii}]$\lambda$6583\AA\ \\
       & & & & & /$f$(H$\beta$) & & /$f$(H$\alpha$) & & /$f$(H$\alpha$) \\
\hline
S[ 5, 1]a   & 5700.6 &  89.3 &  1.31e-17 &   5.18e-15 & $<$  1.48 &  5.29e-18 & $<$  0.13 &  9.04e-15 & $<$  0.13 \\
S[ 5, 2]a   & 5639.7 &  38.3 &  1.40e-17 &   1.48e-14 &    2.05 &  5.55e-18 &    0.02 &  1.21e-13 &    0.30 \\
S[ 5, 3]a   & 5652.3 &  39.7 &  1.39e-17 &   8.90e-15 &    1.40 &  4.95e-18 &    0.05 &  3.94e-14 &    0.41 \\
S[ 5, 4]   &  ---  &  ---  &  ---  &  ---  &  ---  &  ---  &  ---  &  ---  &  ---  \\
S[ 5, 5]   &  ---  &  ---  &  ---  &  ---  &  ---  &  ---  &  ---  &  ---  &  ---  \\
S[ 5, 6]a   & 5704.2 &  92.1 &  1.78e-17 & $<$ 4.12e-15 & --- &  6.43e-18 & $<$  0.23 &  8.96e-15 &    0.90 \\
S[ 5, 7]a   & 5646.6 &  70.1 &  1.82e-17 & $<$ 3.19e-15 & --- &  6.17e-18 & $<$  0.07 &  2.35e-14 &    0.63 \\
S[ 5, 8]a   & 5640.3 &  31.7 &  3.77e-17 & $<$ 3.00e-15 & --- &  1.84e-17 & $<$  0.20 &  7.87e-15 & $<$  0.19 \\
S[ 6, 1]a   & 5653.2 &  56.1 &  1.43e-17 & $<$ 2.01e-15 & --- &  4.67e-18 &    0.15 &  1.06e-14 &    0.54 \\
S[ 6, 2]a   & 5647.8 &  60.7 &  1.30e-17 &   9.13e-15 &    2.16 &  5.10e-18 &    0.07 &  3.82e-14 &    0.41 \\
S[ 6, 3]a   & 5656.2 &  58.0 &  1.55e-17 &   3.42e-15 &    2.93 &  5.13e-18 & $<$  0.05 &  1.83e-14 &    0.41 \\
S[ 6, 4]   &  ---  &  ---  &  ---  &  ---  &  ---  &  ---  &  ---  &  ---  &  ---  \\
S[ 6, 5]a   & 6592.2 &  16.3 &  1.71e-17 & $<$ 6.99e-16 & --- &  6.09e-18 & $<$  0.16 &  1.59e-15 &    0.34 \\
S[ 6, 6]a   & 5635.8 &  57.2 &  1.77e-17 & $<$ 2.54e-15 & --- &  7.19e-18 &    0.25 &  8.79e-15 &    0.22 \\
S[ 6, 7]a   & 5628.9 &  53.7 &  1.62e-17 &   5.58e-15 & $<$  1.02 &  7.43e-18 & $<$  0.05 &  2.20e-14 &    0.60 \\
S[ 6, 8]a   & 5644.8 &  38.9 &  1.49e-17 & $<$ 1.45e-15 & --- &  5.93e-18 &    0.19 &  7.77e-15 &    0.35 \\
S[ 7, 1]a   & 5636.7 &  50.3 &  1.19e-17 &   5.65e-15 &    1.27 &  6.02e-18 &    0.04 &  4.69e-14 &    0.40 \\
S[ 7, 2]a   & 5663.1 &  39.9 &  1.87e-17 & $<$ 1.88e-15 & --- &  5.88e-18 &    0.08 &  1.05e-14 &    0.55 \\
S[ 7, 3]   &  ---  &  ---  &  ---  &  ---  &  ---  &  ---  &  ---  &  ---  &  ---  \\
S[ 7, 4]   &  ---  &  ---  &  ---  &  ---  &  ---  &  ---  &  ---  &  ---  &  ---  \\
S[ 7, 5]a   & 5702.1 & 103.5 &  1.80e-17 & $<$ 4.67e-15 & --- &  6.56e-18 &    0.35 &  1.07e-14 &    0.27 \\
S[ 7, 5]b   & 6583.8 & $<$ 16.7 &  1.80e-17 & $<$-7.53e-16 & --- &  6.56e-18 & $<$  0.23 & -1.98e-15 & $<$  0.14 \\
S[ 7, 6]a   & 5654.4 &  40.7 &  2.01e-17 & $<$ 2.05e-15 & --- &  7.57e-18 & $<$  0.22 &  5.20e-15 &    0.59 \\
S[ 7, 7]a   & 5673.9 & $<$ 16.7 &  2.14e-17 &  -1.98e-15 & $<$  0.95 &  8.09e-18 &    0.27 & -2.99e-15 &    0.79 \\
S[ 7, 8]   &  ---  &  ---  &  ---  &  ---  &  ---  &  ---  &  ---  &  ---  &  ---  \\
S[ 8, 1]a   & 5626.2 &  31.2 &  1.71e-17 & $<$ 1.34e-15 & --- &  5.70e-18 &    0.13 &  1.32e-14 &    0.39 \\
S[ 8, 2]   &  ---  &  ---  &  ---  &  ---  &  ---  &  ---  &  ---  &  ---  &  ---  \\
S[ 8, 3]   &  ---  &  ---  &  ---  &  ---  &  ---  &  ---  &  ---  &  ---  &  ---  \\
S[ 8, 4]   &  ---  &  ---  &  ---  &  ---  &  ---  &  ---  &  ---  &  ---  &  ---  \\
S[ 8, 5]   &  ---  &  ---  &  ---  &  ---  &  ---  &  ---  &  ---  &  ---  &  ---  \\
S[ 8, 6]a   & 5654.7 &  65.7 &  1.71e-17 & $<$ 2.81e-15 & --- &  7.06e-18 & $<$  0.14 &  8.22e-15 &    0.88 \\
S[ 8, 7]   &  ---  &  ---  &  ---  &  ---  &  ---  &  ---  &  ---  &  ---  &  ---  \\
S[ 8, 8]   &  ---  &  ---  &  ---  &  ---  &  ---  &  ---  &  ---  &  ---  &  ---  \\
\hline
\end{tabular}
\caption{Continued.}
\end{table}

\newpage
\clearpage

\begin{table}
\begin{tabular}{lccccccccc}
\hline
Spaxel & $v_{rad}$ & $\sigma$ & $\Sigma_{bkg,blue}$ & $f$(H$\beta$) & $f$([O{\sc iii}]$\lambda$5007\AA\ & $\Sigma_{bkg,red}$ & $f$([O{\sc i}]$\lambda$6300\AA\ & $f$(H$\alpha$) & $f$([N{\sc ii}]$\lambda$6583\AA\ \\
       & & & & & /$f$(H$\beta$) & & /$f$(H$\alpha$) & & /$f$(H$\alpha$) \\
\hline
M[ 1, 1]a   & 5836.8 &  63.3 &  1.66e-17 & $<$ 2.63e-15 & --- &  3.81e-18 &    0.34 &  6.53e-15 &    0.61 \\
M[ 1, 2]a   & 5812.8 &  64.2 &  1.59e-17 & $<$ 2.55e-15 & --- &  4.83e-18 &    0.25 &  1.06e-14 &    0.62 \\
M[ 1, 2]b   & 6207.6 & 196.0 &  1.59e-17 & $<$ 7.79e-15 & --- &  4.83e-18 & $<$  0.22 &  1.25e-14 &    0.24 \\
M[ 1, 3]a   & 5832.3 &  50.0 &  1.62e-17 & $<$ 2.03e-15 & --- &  4.54e-18 &    0.11 &  1.20e-14 &    0.48 \\
M[ 1, 4]a   & 5814.3 &  14.8 &  1.61e-17 & $<$ 5.94e-16 & --- &  5.14e-18 &    0.24 &  1.89e-15 &    0.34 \\
M[ 1, 5]a   & 5810.7 &  55.3 &  2.14e-17 & $<$ 2.97e-15 & --- &  7.11e-18 &    0.09 &  1.66e-14 &    0.52 \\
M[ 1, 6]a   & 5922.0 & 112.3 &  1.50e-17 & $<$ 4.23e-15 & --- &  4.84e-18 &    0.29 &  1.19e-14 &    0.32 \\
M[ 1, 7]a   & 5955.9 & 112.4 &  2.74e-17 & $<$ 7.72e-15 & --- &  5.02e-18 &    0.38 &  1.33e-14 &    0.28 \\
M[ 1, 8]a   & 5916.3 &  63.9 &  4.24e-17 & $<$ 6.78e-15 & --- &  1.54e-17 & $<$  0.20 &  1.36e-14 &    0.19 \\
M[ 2, 1]a   & 5827.5 &  78.9 &  1.21e-17 &   3.99e-15 & $<$  1.67 &  3.44e-18 & $<$  0.09 &  9.50e-15 &    0.09 \\
M[ 2, 1]b   & 6789.9 & 166.7 &  1.21e-17 & $<$ 5.05e-15 & --- &  3.44e-18 & $<$  0.17 &  1.09e-14 &    0.13 \\
M[ 2, 2]a   & 5772.3 &  22.6 &  1.42e-17 &   2.10e-15 &    1.01 &  3.38e-18 &    0.08 &  2.27e-14 &    0.42 \\
M[ 2, 3]a   & 5814.0 &  53.2 &  1.62e-17 &   3.19e-15 & $<$  1.48 &  4.89e-18 &    0.17 &  8.21e-15 &    0.91 \\
M[ 2, 4]a   & 5952.6 & 142.5 &  1.25e-17 & $<$ 4.45e-15 & --- &  4.13e-18 & $<$  0.20 &  9.39e-15 &    0.43 \\
M[ 2, 5]a   & 5974.8 & 128.2 &  1.25e-17 &   4.85e-15 & $<$  2.05 &  4.48e-18 &    0.11 &  1.92e-14 &    0.67 \\
M[ 2, 6]a   & 5952.0 & 123.2 &  1.56e-17 & $<$ 4.81e-15 & --- &  4.66e-18 & $<$  0.11 &  1.75e-14 &    0.61 \\
M[ 2, 7]a   & 6004.8 & 134.5 &  1.45e-17 & $<$ 4.89e-15 & --- &  4.66e-18 &    0.26 &  2.52e-14 &    0.54 \\
M[ 2, 7]b   & 6655.8 & 196.0 &  1.45e-17 & $<$ 7.13e-15 & --- &  4.66e-18 & $<$  0.19 &  1.43e-14 &    0.28 \\
M[ 2, 8]a   & 5977.5 & 100.4 &  1.71e-17 &   5.02e-15 & $<$  1.65 &  5.48e-18 &    0.22 &  2.15e-14 &    0.68 \\
M[ 2, 8]b   & 6644.4 & 196.0 &  1.71e-17 & $<$ 8.39e-15 & --- &  5.48e-18 & $<$  0.12 &  2.00e-14 &    0.23 \\
M[ 3, 1]a   & 5912.1 & 108.4 &  1.36e-17 & $<$ 3.70e-15 & --- &  3.99e-18 &    0.29 &  1.29e-14 &    0.48 \\
M[ 3, 2]a   & 5839.8 &  86.7 &  1.43e-17 & $<$ 3.11e-15 & --- &  3.74e-18 &    0.19 &  2.23e-14 &    0.72 \\
M[ 3, 3]a   & 5880.0 &  78.2 &  1.26e-17 & $<$ 2.47e-15 & --- &  4.91e-18 &    0.32 &  1.93e-14 &    0.70 \\
M[ 3, 3]b   & 6329.4 & 196.0 &  1.26e-17 & $<$ 6.19e-15 & --- &  4.91e-18 &    0.17 &  2.74e-14 &    0.22 \\
M[ 3, 4]a   & 5918.1 & 109.5 &  1.76e-17 & $<$ 4.84e-15 & --- &  4.78e-18 &    0.12 &  2.55e-14 &    0.94 \\
M[ 3, 4]b   & 6349.8 & 196.0 &  1.76e-17 & $<$ 8.67e-15 & --- &  4.78e-18 &    0.11 &  2.78e-14 &    0.24 \\
M[ 3, 5]a   & 6058.2 & 159.7 &  1.70e-17 & $<$ 6.82e-15 & --- &  4.21e-18 &    0.26 &  4.11e-14 &    0.59 \\
M[ 3, 5]b   & 6692.1 & 196.0 &  1.70e-17 & $<$ 8.37e-15 & --- &  4.21e-18 &    0.17 &  2.69e-14 &    0.36 \\
M[ 3, 6]a   & 6099.3 & 183.9 &  1.39e-17 & $<$ 6.39e-15 & --- &  3.98e-18 &    0.24 &  3.84e-14 &    0.68 \\
M[ 3, 6]b   & 6677.1 & 160.6 &  1.39e-17 & $<$ 5.58e-15 & --- &  3.98e-18 &    0.19 &  2.86e-14 &    0.24 \\
M[ 3, 7]a   & 6058.5 & 155.7 &  1.51e-17 & $<$ 5.88e-15 & --- &  5.05e-18 &    0.26 &  3.23e-14 &    0.54 \\
M[ 3, 7]b   & 6663.3 & 170.6 &  1.51e-17 & $<$ 6.44e-15 & --- &  5.05e-18 &    0.18 &  2.94e-14 &    0.22 \\
M[ 3, 8]a   & 6040.8 & 144.0 &  1.50e-17 & $<$ 5.43e-15 & --- &  5.21e-18 &    0.19 &  3.64e-14 &    0.54 \\
M[ 3, 8]b   & 6671.1 & 180.6 &  1.50e-17 & $<$ 6.80e-15 & --- &  5.21e-18 &    0.14 &  2.51e-14 &    0.22 \\
M[ 4, 1]a   & 5965.5 & 137.4 &  1.36e-17 & $<$ 4.68e-15 & --- &  4.02e-18 &    0.25 &  1.61e-14 &    0.35 \\
M[ 4, 1]b   & 6849.6 & 142.8 &  1.36e-17 & $<$ 4.86e-15 & --- &  4.02e-18 & $<$  0.21 &  7.32e-15 & $<$  0.20 \\
M[ 4, 2]a   & 6040.2 & 157.1 &  1.41e-17 & $<$ 5.57e-15 & --- &  4.37e-18 &    0.27 &  1.83e-14 &    0.57 \\
M[ 4, 2]b   & 6688.8 & 196.0 &  1.41e-17 & $<$ 6.94e-15 & --- &  4.37e-18 & $<$  0.17 &  1.39e-14 & $<$  0.15 \\
M[ 4, 3]a   & 5985.6 &  77.7 &  1.45e-17 & $<$ 2.82e-15 & --- &  3.97e-18 & $<$  0.13 &  8.78e-15 &    0.54 \\
M[ 4, 3]b   & 6527.1 & 196.0 &  1.45e-17 & $<$ 7.11e-15 & --- &  3.97e-18 &    0.15 &  3.11e-14 &    0.21 \\
M[ 4, 4]a   & 5981.4 & 120.8 &  1.18e-17 & $<$ 3.56e-15 & --- &  4.24e-18 &    0.30 &  1.73e-14 &    0.63 \\
M[ 4, 4]b   & 6555.3 & 196.0 &  1.18e-17 &   9.51e-15 & $<$  1.54 &  4.24e-18 &    0.16 &  2.91e-14 &    0.23 \\
M[ 4, 5]a   & 5986.8 & 120.6 &  1.41e-17 & $<$ 4.25e-15 & --- &  3.86e-18 & $<$  0.19 &  9.65e-15 &    0.69 \\
M[ 4, 5]b   & 6702.0 & 168.1 &  1.41e-17 & $<$ 5.92e-15 & --- &  3.86e-18 & $<$  0.15 &  1.71e-14 &    0.10 \\
M[ 4, 6]a   & 6093.6 & 161.4 &  1.61e-17 & $<$ 6.51e-15 & --- &  3.46e-18 &    0.19 &  1.16e-14 &    0.60 \\
M[ 4, 6]b   & 6715.5 & 150.6 &  1.61e-17 & $<$ 6.07e-15 & --- &  3.46e-18 & $<$  0.10 &  1.86e-14 &    0.27 \\
M[ 4, 7]a   & 6051.0 & 118.7 &  1.26e-17 & $<$ 3.76e-15 & --- &  4.70e-18 &    0.25 &  1.95e-14 &    0.66 \\
M[ 4, 7]b   & 6650.1 & 196.0 &  1.26e-17 & $<$ 6.20e-15 & --- &  4.70e-18 &    0.17 &  4.06e-14 &    0.26 \\
M[ 4, 8]a   & 6021.6 &  98.2 &  1.39e-17 & $<$ 3.43e-15 & --- &  4.09e-18 &    0.17 &  2.30e-14 &    0.71 \\
M[ 4, 8]b   & 6568.2 & 196.0 &  1.39e-17 & $<$ 6.85e-15 & --- &  4.09e-18 & $<$  0.10 &  2.92e-14 &    0.26 \\
\hline
\end{tabular}
\caption{Same as Table~\ref{tabla_fit_hcg92s} for the spaxels of pointing M.}
\label{tabla_fit_hcg92m}
\end{table}

\newpage
\clearpage

\addtocounter{table}{-1}

\begin{table}
\begin{tabular}{lccccccccc}
\hline
Spaxel & $v_{rad}$ & $\sigma$ & $\Sigma_{bkg,blue}$ & $f$(H$\beta$) & $f$([O{\sc iii}]$\lambda$5007\AA\ & $\Sigma_{bkg,red}$ & $f$([O{\sc i}]$\lambda$6300\AA\ & $f$(H$\alpha$) & $f$([N{\sc ii}]$\lambda$6583\AA\ \\
       & & & & & /$f$(H$\beta$) & & /$f$(H$\alpha$) & & /$f$(H$\alpha$) \\
\hline
M[ 5, 1]a   & 6073.5 & 169.1 &  1.37e-17 & $<$ 5.80e-15 & --- &  4.42e-18 &    0.17 &  1.53e-14 &    0.38 \\
M[ 5, 1]b   & 6658.2 & 148.1 &  1.37e-17 & $<$ 5.08e-15 & --- &  4.42e-18 &    0.12 &  2.32e-14 &    0.15 \\
M[ 5, 2]a   & 6135.3 &  88.8 &  1.35e-17 & $<$ 3.01e-15 & --- &  4.72e-18 &    0.33 &  6.09e-15 & $<$  0.17 \\
M[ 5, 2]b   & 6736.2 & 132.6 &  1.35e-17 & $<$ 4.49e-15 & --- &  4.72e-18 &    0.20 &  2.86e-14 &    0.21 \\
M[ 5, 3]a   & 6663.0 & 167.4 &  1.66e-17 & $<$ 6.98e-15 & --- &  4.56e-18 & $<$  0.14 &  1.92e-14 &    0.20 \\
M[ 5, 4]a   & 6701.4 & 120.5 &  1.56e-17 & $<$ 4.72e-15 & --- &  5.19e-18 &    0.14 &  1.96e-14 &    0.29 \\
M[ 5, 5]a   & 6130.2 &  92.9 &  2.04e-17 & $<$ 4.74e-15 & --- &  5.48e-18 & $<$  0.18 &  9.82e-15 &    0.55 \\
M[ 5, 5]b   & 6663.3 & 131.0 &  2.04e-17 & $<$ 6.69e-15 & --- &  5.48e-18 & $<$  0.10 &  2.53e-14 &    0.13 \\
M[ 5, 6]a   & 6148.2 &  28.1 &  1.60e-17 & $<$ 1.13e-15 & --- &  5.06e-18 & $<$  0.12 &  4.63e-15 &    0.64 \\
M[ 5, 6]b   & 6691.8 & 139.7 &  1.60e-17 & $<$ 5.62e-15 & --- &  5.06e-18 & $<$  0.10 &  2.88e-14 &    0.25 \\
M[ 5, 7]a   & 6206.4 & 122.7 &  1.69e-17 & $<$ 5.20e-15 & --- &  5.11e-18 & $<$  0.11 &  1.97e-14 &    0.56 \\
M[ 5, 7]b   & 6718.5 & 196.0 &  1.69e-17 &   1.35e-14 & $<$  1.89 &  5.11e-18 & $<$  0.13 &  2.86e-14 &    0.23 \\
M[ 5, 8]a   & 6255.0 & 123.7 &  4.79e-17 & $<$ 1.48e-14 & --- &  1.58e-17 & $<$  0.14 &  2.51e-14 &    0.41 \\
M[ 6, 1]a   & 6682.2 & 108.7 &  1.62e-17 & $<$ 4.41e-15 & --- &  4.17e-18 &    0.09 &  1.65e-14 &    0.24 \\
M[ 6, 2]a   & 6712.5 & 141.5 &  1.46e-17 & $<$ 5.18e-15 & --- &  4.08e-18 & $<$  0.10 &  1.68e-14 & $<$  0.09 \\
M[ 6, 3]a   & 6729.9 & 142.8 &  1.41e-17 & $<$ 5.06e-15 & --- &  4.40e-18 & $<$  0.09 &  2.49e-14 &    0.24 \\
M[ 6, 4]a   & 6626.1 & 181.5 &  1.57e-17 & $<$ 7.12e-15 & --- &  3.74e-18 & $<$  0.13 &  2.57e-14 &    0.29 \\
M[ 6, 5]a   & 6702.3 & 196.0 &  1.50e-17 & $<$ 7.36e-15 & --- &  5.05e-18 & $<$  0.14 &  2.59e-14 &    0.13 \\
M[ 6, 6]a   & 6236.7 &  85.7 &  1.64e-17 &   3.93e-15 & $<$  2.05 &  4.66e-18 & $<$  0.18 &  1.01e-14 &    0.50 \\
M[ 6, 6]b   & 6723.6 & 196.0 &  1.64e-17 & $<$ 8.03e-15 & --- &  4.66e-18 &    0.17 &  3.44e-14 &    0.16 \\
M[ 6, 7]a   & 6221.1 & 124.4 &  1.66e-17 & $<$ 5.19e-15 & --- &  4.54e-18 &    0.16 &  2.78e-14 &    0.52 \\
M[ 6, 7]b   & 6670.2 & 196.0 &  1.66e-17 & $<$ 8.17e-15 & --- &  4.54e-18 & $<$  0.08 &  3.57e-14 &    0.22 \\
M[ 6, 8]a   & 6078.9 &  87.5 &  1.55e-17 & $<$ 3.39e-15 & --- &  5.28e-18 & $<$  0.14 &  1.17e-14 &    0.88 \\
M[ 6, 8]b   & 6458.4 & 196.0 &  1.55e-17 & $<$ 7.60e-15 & --- &  5.28e-18 & $<$  0.12 &  3.07e-14 &    0.26 \\
M[ 7, 1]a   & 6733.5 &  82.0 &  1.56e-17 & $<$ 3.21e-15 & --- &  4.91e-18 & $<$  0.11 &  1.27e-14 &    0.15 \\
M[ 7, 2]a   & 6726.6 & 117.5 &  1.95e-17 &   7.58e-15 & $<$  1.46 &  5.47e-18 & $<$  0.10 &  2.09e-14 &    0.19 \\
M[ 7, 3]a   & 6701.1 &  98.0 &  2.10e-17 & $<$ 5.16e-15 & --- &  6.18e-18 &    0.17 &  2.64e-14 &    0.30 \\
M[ 7, 4]a   & 6660.6 & 134.8 &  2.26e-17 & $<$ 7.65e-15 & --- &  5.12e-18 & $<$  0.10 &  2.71e-14 &    0.21 \\
M[ 7, 5]a   & 6234.6 & 132.0 &  1.81e-17 & $<$ 6.00e-15 & --- &  5.17e-18 & $<$  0.26 &  1.14e-14 &    0.21 \\
M[ 7, 5]b   & 6696.6 & 196.0 &  1.81e-17 & $<$ 8.92e-15 & --- &  5.17e-18 & $<$  0.14 &  3.18e-14 & $<$  0.08 \\
M[ 7, 6]a   & 6670.2 & 196.0 &  2.12e-17 & $<$ 1.04e-14 & --- &  6.81e-18 & $<$  0.10 &  3.55e-14 &    0.20 \\
M[ 7, 7]a   & 6475.5 & 196.0 &  2.15e-17 & $<$ 1.06e-14 & --- &  7.09e-18 &    0.15 &  4.62e-14 &    0.17 \\
M[ 7, 8]a   & 6399.9 & 170.4 &  2.14e-17 & $<$ 9.13e-15 & --- &  7.97e-18 & $<$  0.11 &  3.64e-14 &    0.32 \\
M[ 8, 1]   &  ---  &  ---  &  ---  &  ---  &  ---  &  ---  &  ---  &  ---  &  ---  \\
M[ 8, 2]a   & 6633.0 & 170.0 &  2.05e-17 & $<$ 8.72e-15 & --- &  5.32e-18 & $<$  0.25 &  1.15e-14 & $<$  0.20 \\
M[ 8, 3]a   & 6077.7 &  73.2 &  1.95e-17 & $<$ 3.58e-15 & --- &  5.10e-18 & $<$  0.22 &  5.87e-15 &    0.35 \\
M[ 8, 3]b   & 6682.5 & 196.0 &  1.95e-17 & $<$ 9.58e-15 & --- &  5.10e-18 & $<$  0.22 &  1.58e-14 & $<$  0.16 \\
M[ 8, 4]a   & 6175.2 &  98.9 &  1.80e-17 & $<$ 4.46e-15 & --- &  6.43e-18 & $<$  0.26 &  8.02e-15 &    0.65 \\
M[ 8, 5]a   & 6540.9 & 196.0 &  1.93e-17 & $<$ 9.47e-15 & --- &  5.85e-18 & $<$  0.28 &  1.48e-14 &    0.19 \\
M[ 8, 6]a   & 6603.3 & 196.0 &  1.80e-17 & $<$ 8.85e-15 & --- &  4.73e-18 & $<$  0.18 &  2.27e-14 &    0.23 \\
M[ 8, 7]a   & 5828.7 &  32.8 &  1.69e-17 & $<$ 1.39e-15 & --- &  5.77e-18 & $<$  0.07 &  1.09e-14 &    0.14 \\
M[ 8, 7]b   & 6540.9 & 196.0 &  1.69e-17 & $<$ 8.31e-15 & --- &  5.77e-18 & $<$  0.12 &  4.07e-14 &    0.26 \\
M[ 8, 8]a   & 6399.9 & 196.0 &  2.24e-17 & $<$ 1.10e-14 & --- &  8.50e-18 &    0.11 &  3.88e-14 & $<$  0.11 \\
M[ 8, 8]b   & 6465.3 &  19.3 &  2.24e-17 &   2.70e-15 & $<$  0.79 &  8.50e-18 &    0.37 &  2.57e-15 &    1.08 \\
\hline
\end{tabular}
\caption{Continued.}
\end{table}

\newpage
\clearpage

\begin{table}
\begin{tabular}{lccccccccc}
\hline
Spaxel & $v_{rad}$ & $\sigma$ & $\Sigma_{bkg,blue}$ & $f$(H$\beta$) & $f$([O{\sc iii}]$\lambda$5007\AA\ & $\Sigma_{bkg,red}$ & $f$([O{\sc i}]$\lambda$6300\AA\ & $f$(H$\alpha$) & $f$([N{\sc ii}]$\lambda$6583\AA\ \\
       & & & & & /$f$(H$\beta$) & & /$f$(H$\alpha$) & & /$f$(H$\alpha$) \\
\hline
N[ 1, 1]a   & 6061.2 & 157.3 &  1.68e-17 & $<$ 6.62e-15 & --- &  3.90e-18 &    0.29 &  2.91e-14 &    0.64 \\
N[ 1, 1]b   & 6660.9 & 175.0 &  1.68e-17 & $<$ 7.36e-15 & --- &  3.90e-18 &    0.13 &  2.57e-14 &    0.25 \\
N[ 1, 2]a   & 6061.2 & 146.0 &  1.45e-17 & $<$ 5.31e-15 & --- &  3.98e-18 &    0.20 &  3.55e-14 &    0.62 \\
N[ 1, 2]b   & 6627.0 & 112.2 &  1.45e-17 & $<$ 4.08e-15 & --- &  3.98e-18 &    0.18 &  1.58e-14 &    0.24 \\
N[ 1, 3]a   & 6148.8 & 173.9 &  1.43e-17 & $<$ 6.22e-15 & --- &  5.33e-18 &    0.20 &  5.87e-14 &    0.58 \\
N[ 1, 3]b   & 6651.0 & 160.9 &  1.43e-17 &   9.78e-15 & $<$  1.25 &  5.33e-18 &    0.13 &  3.74e-14 &    0.22 \\
N[ 1, 4]a   & 6115.8 & 148.9 &  1.40e-17 & $<$ 5.23e-15 & --- &  3.87e-18 &    0.27 &  1.84e-14 &    0.67 \\
N[ 1, 4]b   & 6661.2 & 157.4 &  1.40e-17 & $<$ 5.53e-15 & --- &  3.87e-18 &    0.17 &  1.13e-14 & $<$  0.13 \\
N[ 1, 5]a   & 6102.6 & 134.1 &  1.93e-17 & $<$ 6.47e-15 & --- &  6.47e-18 &    0.34 &  1.36e-14 &    0.92 \\
N[ 1, 5]b   & 6578.1 & 179.8 &  1.93e-17 & $<$ 8.68e-15 & --- &  6.47e-18 &    0.18 &  1.91e-14 &    0.33 \\
N[ 1, 6]a   & 6093.6 & 135.2 &  1.55e-17 & $<$ 5.24e-15 & --- &  4.01e-18 &    0.20 &  2.20e-14 &    0.73 \\
N[ 1, 6]b   & 6563.4 & 153.9 &  1.55e-17 & $<$ 5.96e-15 & --- &  4.01e-18 &    0.17 &  1.28e-14 &    0.48 \\
N[ 1, 7]a   & 6114.0 & 148.2 &  1.99e-17 &   8.95e-15 & $<$  1.91 &  4.84e-18 & $<$  0.18 &  1.21e-14 &    0.64 \\
N[ 1, 8]   &  ---  &  ---  &  ---  &  ---  &  ---  &  ---  &  ---  &  ---  &  ---  \\
N[ 2, 1]a   & 6075.6 & 154.8 &  1.29e-17 & $<$ 5.02e-15 & --- &  4.18e-18 &    0.10 &  2.58e-14 &    0.79 \\
N[ 2, 1]b   & 6661.8 & 159.9 &  1.29e-17 & $<$ 5.19e-15 & --- &  4.18e-18 &    0.13 &  3.35e-14 &    0.26 \\
N[ 2, 2]a   & 6036.6 & 121.3 &  1.34e-17 & $<$ 4.07e-15 & --- &  4.27e-18 &    0.15 &  2.74e-14 &    0.78 \\
N[ 2, 2]b   & 6575.1 & 196.0 &  1.34e-17 & $<$ 6.58e-15 & --- &  4.27e-18 &    0.14 &  3.24e-14 &    0.25 \\
N[ 2, 3]a   & 6099.0 & 151.0 &  1.30e-17 & $<$ 4.91e-15 & --- &  4.41e-18 &    0.24 &  1.93e-14 &    0.74 \\
N[ 2, 3]b   & 6676.8 & 196.0 &  1.30e-17 & $<$ 6.37e-15 & --- &  4.41e-18 &    0.25 &  2.22e-14 &    0.24 \\
N[ 2, 4]a   & 6165.6 & 163.4 &  1.33e-17 & $<$ 5.44e-15 & --- &  3.89e-18 &    0.13 &  2.31e-14 &    0.45 \\
N[ 2, 4]b   & 6742.8 & 150.8 &  1.33e-17 & $<$ 5.02e-15 & --- &  3.89e-18 &    0.14 &  2.25e-14 &    0.17 \\
N[ 2, 5]a   & 6077.7 & 157.1 &  1.36e-17 & $<$ 5.35e-15 & --- &  4.63e-18 &    0.11 &  1.88e-14 &    1.02 \\
N[ 2, 5]b   & 6564.6 & 196.0 &  1.36e-17 & $<$ 6.68e-15 & --- &  4.63e-18 & $<$  0.14 &  1.74e-14 &    0.27 \\
N[ 2, 6]a   & 6155.4 & 193.5 &  1.34e-17 & $<$ 6.48e-15 & --- &  4.16e-18 & $<$  0.11 &  2.20e-14 &    0.76 \\
N[ 2, 6]b   & 6816.6 & 146.6 &  1.34e-17 & $<$ 4.91e-15 & --- &  4.16e-18 & $<$  0.18 &  9.69e-15 & $<$  0.16 \\
N[ 2, 7]a   & 6183.9 & 147.8 &  1.45e-17 & $<$ 5.38e-15 & --- &  3.81e-18 & $<$  0.15 &  1.38e-14 &    0.85 \\
N[ 2, 7]b   & 6775.5 &  77.1 &  1.45e-17 & $<$ 2.81e-15 & --- &  3.81e-18 & $<$  0.21 &  5.33e-15 &    0.20 \\
N[ 2, 8]a   & 6118.8 & 143.8 &  1.82e-17 & $<$ 6.56e-15 & --- &  3.87e-18 & $<$  0.10 &  1.30e-14 &    0.69 \\
N[ 3, 1]a   & 6193.2 & 151.1 &  1.38e-17 & $<$ 5.23e-15 & --- &  3.84e-18 &    0.16 &  1.96e-14 &    0.83 \\
N[ 3, 1]b   & 6683.4 & 156.0 &  1.38e-17 & $<$ 5.39e-15 & --- &  3.84e-18 &    0.11 &  2.31e-14 &    0.17 \\
N[ 3, 2]a   & 6269.4 & 167.4 &  1.17e-17 & $<$ 4.90e-15 & --- &  4.28e-18 &    0.11 &  2.64e-14 &    0.57 \\
N[ 3, 2]b   & 6790.5 & 134.5 &  1.17e-17 & $<$ 3.94e-15 & --- &  4.28e-18 & $<$  0.07 &  2.17e-14 &    0.18 \\
N[ 3, 3]a   & 6240.3 & 196.0 &  1.36e-17 & $<$ 6.68e-15 & --- &  4.11e-18 &    0.16 &  3.29e-14 &    0.61 \\
N[ 3, 3]b   & 6816.0 &  93.3 &  1.36e-17 & $<$ 3.18e-15 & --- &  4.11e-18 &    0.11 &  1.99e-14 &    0.14 \\
N[ 3, 4]a   & 6120.6 & 193.7 &  1.49e-17 & $<$ 7.24e-15 & --- &  4.04e-18 &    0.16 &  3.07e-14 &    0.77 \\
N[ 3, 4]b   & 6740.7 & 138.8 &  1.49e-17 & $<$ 5.19e-15 & --- &  4.04e-18 & $<$  0.07 &  2.08e-14 &    0.16 \\
N[ 3, 5]a   & 6100.5 & 150.5 &  1.15e-17 & $<$ 4.35e-15 & --- &  4.33e-18 &    0.19 &  2.14e-14 &    0.91 \\
N[ 3, 5]b   & 6594.9 & 196.0 &  1.15e-17 & $<$ 5.66e-15 & --- &  4.33e-18 & $<$  0.08 &  2.34e-14 &    0.28 \\
N[ 3, 6]a   & 6199.5 & 148.3 &  1.21e-17 & $<$ 4.51e-15 & --- &  4.04e-18 &    0.12 &  2.07e-14 &    0.62 \\
N[ 3, 6]b   & 6815.7 &  97.3 &  1.21e-17 & $<$ 2.96e-15 & --- &  4.04e-18 & $<$  0.08 &  1.16e-14 &    0.23 \\
N[ 3, 7]a   & 6183.9 & 157.6 &  1.40e-17 & $<$ 5.53e-15 & --- &  4.20e-18 &    0.08 &  2.54e-14 &    0.63 \\
N[ 3, 7]b   & 6845.7 &  95.9 &  1.40e-17 & $<$ 3.36e-15 & --- &  4.20e-18 & $<$  0.07 &  1.21e-14 &    0.14 \\
N[ 3, 8]a   & 6158.1 & 115.2 &  1.54e-17 & $<$ 4.44e-15 & --- &  3.29e-18 &    0.13 &  1.61e-14 &    0.88 \\
N[ 3, 8]b   & 6671.4 & 194.7 &  1.54e-17 & $<$ 7.50e-15 & --- &  3.29e-18 & $<$  0.15 &  1.40e-14 &    0.33 \\
N[ 4, 1]a   & 6305.7 & 196.0 &  1.41e-17 & $<$ 6.92e-15 & --- &  4.12e-18 & $<$  0.04 &  3.93e-14 &    0.52 \\
N[ 4, 1]b   & 6750.9 & 196.0 &  1.41e-17 & $<$ 6.92e-15 & --- &  4.12e-18 &    0.25 &  2.32e-14 & $<$  0.09 \\
N[ 4, 2]a   & 6240.6 & 174.1 &  1.42e-17 & $<$ 6.21e-15 & --- &  3.67e-18 &    0.22 &  3.29e-14 &    0.50 \\
N[ 4, 2]b   & 6738.0 & 196.0 &  1.42e-17 & $<$ 6.99e-15 & --- &  3.67e-18 & $<$  0.08 &  1.90e-14 &    0.11 \\
N[ 4, 3]a   & 6019.2 & 129.8 &  1.22e-17 & $<$ 3.96e-15 & --- &  3.88e-18 &    0.23 &  2.01e-14 &    1.11 \\
N[ 4, 3]b   & 6454.8 & 196.0 &  1.22e-17 & $<$ 5.99e-15 & --- &  3.88e-18 & $<$  0.05 &  3.90e-14 &    0.34 \\
N[ 4, 4]a   & 6126.9 & 196.0 &  1.28e-17 & $<$ 6.27e-15 & --- &  4.14e-18 &    0.20 &  5.08e-14 &    0.62 \\
N[ 4, 4]b   & 6770.4 & 196.0 &  1.28e-17 & $<$ 6.27e-15 & --- &  4.14e-18 & $<$  0.09 &  2.31e-14 & $<$  0.09 \\
N[ 4, 5]a   & 6150.3 & 185.1 &  1.49e-17 & $<$ 6.90e-15 & --- &  3.64e-18 & $<$  0.08 &  2.39e-14 &    0.75 \\
N[ 4, 5]b   & 6709.2 & 196.0 &  1.49e-17 & $<$ 7.30e-15 & --- &  3.64e-18 & $<$  0.15 &  1.34e-14 &    0.15 \\
N[ 4, 6]a   & 6316.5 & 196.0 &  1.43e-17 & $<$ 7.03e-15 & --- &  4.97e-18 &    0.11 &  3.07e-14 &    0.49 \\
N[ 4, 6]b   & 6879.0 & 175.9 &  1.43e-17 & $<$ 6.31e-15 & --- &  4.97e-18 & $<$  0.15 &  1.26e-14 & $<$  0.17 \\
N[ 4, 7]a   & 6160.8 & 151.7 &  1.31e-17 & $<$ 4.97e-15 & --- &  4.04e-18 &    0.14 &  2.47e-14 &    1.03 \\
N[ 4, 7]b   & 6528.3 & 145.3 &  1.31e-17 & $<$ 4.76e-15 & --- &  4.04e-18 & $<$  0.05 &  2.91e-14 &    0.22 \\
N[ 4, 8]a   & 6229.5 & 146.9 &  1.44e-17 & $<$ 5.30e-15 & --- &  3.62e-18 & $<$  0.05 &  2.78e-14 &    0.77 \\
N[ 4, 8]b   & 6675.6 & 190.9 &  1.44e-17 & $<$ 6.89e-15 & --- &  3.62e-18 & $<$  0.09 &  1.94e-14 &    0.14 \\
\hline
\end{tabular}
\caption{Same as Table~\ref{tabla_fit_hcg92s} for the spaxels of pointing N.}
\label{tabla_fit_hcg92n}
\end{table}

\newpage
\clearpage

\addtocounter{table}{-1}

\begin{table}
\begin{tabular}{lccccccccc}
\hline
Spaxel & $v_{rad}$ & $\sigma$ & $\Sigma_{bkg,blue}$ & $f$(H$\beta$) & $f$([O{\sc iii}]$\lambda$5007\AA\ & $\Sigma_{bkg,red}$ & $f$([O{\sc i}]$\lambda$6300\AA\ & $f$(H$\alpha$) & $f$([N{\sc ii}]$\lambda$6583\AA\ \\
       & & & & & /$f$(H$\beta$) & & /$f$(H$\alpha$) & & /$f$(H$\alpha$) \\
\hline
N[ 5, 1]a   & 6274.8 & 196.0 &  1.39e-17 & $<$ 6.81e-15 & --- &  4.59e-18 & $<$  0.12 &  1.74e-14 &    0.90 \\
N[ 5, 1]b   & 6599.7 & 154.1 &  1.39e-17 & $<$ 5.36e-15 & --- &  4.59e-18 &    0.22 &  2.34e-14 &    0.10 \\
N[ 5, 2]a   & 6264.6 & 196.0 &  1.49e-17 & $<$ 7.33e-15 & --- &  4.40e-18 &    0.24 &  1.90e-14 &    0.79 \\
N[ 5, 2]b   & 6543.3 & 166.6 &  1.49e-17 & $<$ 6.23e-15 & --- &  4.40e-18 & $<$  0.10 &  1.96e-14 & $<$  0.09 \\
N[ 5, 3]a   & 6172.8 & 161.6 &  1.40e-17 & $<$ 5.68e-15 & --- &  4.24e-18 &    0.23 &  2.12e-14 &    0.74 \\
N[ 5, 3]b   & 6634.2 & 196.0 &  1.40e-17 & $<$ 6.89e-15 & --- &  4.24e-18 & $<$  0.08 &  2.51e-14 & $<$  0.08 \\
N[ 5, 4]a   & 6125.7 & 196.0 &  1.56e-17 & $<$ 7.68e-15 & --- &  4.46e-18 &    0.13 &  2.78e-14 &    0.74 \\
N[ 5, 4]b   & 6633.9 & 195.6 &  1.56e-17 & $<$ 7.66e-15 & --- &  4.46e-18 & $<$  0.12 &  2.04e-14 &    0.15 \\
N[ 5, 5]a   & 6010.8 & 173.8 &  1.34e-17 & $<$ 5.85e-15 & --- &  4.57e-18 &    0.33 &  1.35e-14 &    1.11 \\
N[ 5, 5]b   & 6459.9 & 144.6 &  1.34e-17 & $<$ 4.86e-15 & --- &  4.57e-18 & $<$  0.16 &  1.03e-14 &    0.18 \\
N[ 5, 6]a   & 6128.1 & 109.8 &  1.63e-17 & $<$ 4.48e-15 & --- &  4.06e-18 &    0.41 &  6.43e-15 &    1.15 \\
N[ 5, 6]b   & 6531.3 & 174.9 &  1.63e-17 & $<$ 7.13e-15 & --- &  4.06e-18 &    0.27 &  1.70e-14 &    0.31 \\
N[ 5, 7]a   & 6208.2 & 143.8 &  1.55e-17 & $<$ 5.59e-15 & --- &  4.04e-18 &    0.24 &  1.32e-14 &    0.94 \\
N[ 5, 7]b   & 6553.8 & 119.8 &  1.55e-17 & $<$ 4.66e-15 & --- &  4.04e-18 &    0.11 &  1.50e-14 &    0.27 \\
N[ 5, 8]   &  ---  &  ---  &  ---  &  ---  &  ---  &  ---  &  ---  &  ---  &  ---  \\
N[ 6, 1]a   & 5802.9 & $<$ 16.7 &  1.47e-17 &  -8.92e-16 & $<$  1.50 &  4.10e-18 & $<$  0.04 & -4.20e-15 &    0.44 \\
N[ 6, 1]b   & 6399.3 & 196.0 &  1.47e-17 & $<$ 7.21e-15 & --- &  4.10e-18 &    0.10 &  2.43e-14 &    0.42 \\
N[ 6, 2]a   & 6421.2 & 133.2 &  1.38e-17 & $<$ 4.60e-15 & --- &  4.07e-18 &    0.14 &  2.72e-14 &    0.52 \\
N[ 6, 3]a   & 6103.8 & 129.1 &  1.46e-17 & $<$ 4.72e-15 & --- &  3.77e-18 & $<$  0.14 &  9.29e-15 &    0.90 \\
N[ 6, 3]b   & 6452.1 & 119.2 &  1.46e-17 & $<$ 4.36e-15 & --- &  3.77e-18 &    0.32 &  1.83e-14 &    0.19 \\
N[ 6, 4]a   & 6277.5 & 196.0 &  1.40e-17 & $<$ 6.87e-15 & --- &  5.03e-18 &    0.16 &  2.01e-14 &    0.72 \\
N[ 6, 4]b   & 6724.5 & 129.3 &  1.40e-17 & $<$ 4.53e-15 & --- &  5.03e-18 & $<$  0.10 &  1.01e-14 & $<$  0.16 \\
N[ 6, 5]a   & 5935.5 &  98.4 &  1.53e-17 & $<$ 3.79e-15 & --- &  4.56e-18 & $<$  0.12 &  8.67e-15 &    0.64 \\
N[ 6, 5]b   & 6525.3 & 196.0 &  1.53e-17 & $<$ 7.54e-15 & --- &  4.56e-18 & $<$  0.14 &  1.45e-14 & $<$  0.15 \\
N[ 6, 6]a   & 6018.6 & 110.5 &  1.46e-17 & $<$ 4.05e-15 & --- &  3.75e-18 & $<$  0.14 &  7.91e-15 &    0.80 \\
N[ 6, 6]b   & 6564.9 & 166.4 &  1.46e-17 & $<$ 6.10e-15 & --- &  3.75e-18 & $<$  0.17 &  1.04e-14 &    0.39 \\
N[ 6, 7]a   & 6110.7 &  99.1 &  1.56e-17 & $<$ 3.88e-15 & --- &  4.40e-18 &    0.33 &  5.47e-15 &    0.92 \\
N[ 6, 7]b   & 6572.4 & 156.2 &  1.56e-17 & $<$ 6.11e-15 & --- &  4.40e-18 & $<$  0.14 &  1.34e-14 &    0.30 \\
N[ 6, 8]a   & 6613.5 &  71.7 &  1.64e-17 & $<$ 2.96e-15 & --- &  4.32e-18 & $<$  0.09 &  6.99e-15 &    0.23 \\
N[ 7, 1]   &  ---  &  ---  &  ---  &  ---  &  ---  &  ---  &  ---  &  ---  &  ---  \\
N[ 7, 2]a   & 6297.0 & 178.5 &  1.89e-17 & $<$ 8.45e-15 & --- &  5.43e-18 & $<$  0.14 &  1.69e-14 &    0.47 \\
N[ 7, 3]a   & 6258.9 & 167.0 &  1.64e-17 & $<$ 6.88e-15 & --- &  4.23e-18 & $<$  0.11 &  1.74e-14 &    0.70 \\
N[ 7, 4]a   & 6302.4 & 196.0 &  1.45e-17 & $<$ 7.12e-15 & --- &  5.17e-18 &    0.41 &  1.28e-14 &    0.44 \\
N[ 7, 5]a   & 6754.5 & 105.5 &  1.61e-17 & $<$ 4.27e-15 & --- &  4.37e-18 & $<$  0.16 &  6.38e-15 & $<$  0.18 \\
N[ 7, 6]   &  ---  &  ---  &  ---  &  ---  &  ---  &  ---  &  ---  &  ---  &  ---  \\
N[ 7, 7]a   & 6612.6 & 167.3 &  1.51e-17 & $<$ 6.35e-15 & --- &  3.08e-18 & $<$  0.26 &  8.03e-15 &    0.51 \\
N[ 7, 8]a   & 6653.4 &  81.9 &  1.44e-17 & $<$ 2.95e-15 & --- &  4.46e-18 & $<$  0.15 &  6.44e-15 &    0.35 \\
N[ 8, 1]a   & 6074.4 &  60.2 &  1.86e-17 & $<$ 2.81e-15 & --- &  4.14e-18 & $<$  0.15 &  3.62e-15 & $<$  0.17 \\
N[ 8, 2]a   & 6140.1 & 100.0 &  1.94e-17 & $<$ 4.86e-15 & --- &  4.30e-18 & $<$  0.12 &  7.88e-15 &    0.94 \\
N[ 8, 3]a   & 6135.6 &  85.9 &  1.68e-17 & $<$ 3.63e-15 & --- &  5.36e-18 & $<$  0.12 &  6.74e-15 &    1.14 \\
N[ 8, 4]   &  ---  &  ---  &  ---  &  ---  &  ---  &  ---  &  ---  &  ---  &  ---  \\
N[ 8, 5]   &  ---  &  ---  &  ---  &  ---  &  ---  &  ---  &  ---  &  ---  &  ---  \\
N[ 8, 6]   &  ---  &  ---  &  ---  &  ---  &  ---  &  ---  &  ---  &  ---  &  ---  \\
N[ 8, 7]a   & 6617.4 &  90.5 &  1.59e-17 & $<$ 3.60e-15 & --- &  4.29e-18 & $<$  0.19 &  5.45e-15 & $<$  0.18 \\
N[ 8, 8]a   & 6674.4 &  68.4 &  1.85e-17 & $<$ 3.17e-15 & --- &  4.22e-18 & $<$  0.14 &  5.49e-15 & $<$  0.13 \\
\hline
\end{tabular}
\caption{Continued.}
\end{table}

\newpage
\clearpage

\begin{table}
\begin{tabular}{lcc}
\hline
Spaxel & $12 + \log$O/H (O3N2) \\
\hline
S[2,2] & 8.50 \\
S[2,3] & 8.50 \\
S[5,2] & 8.45 \\
M[2,2] & 8.61 \\
\hline
\end{tabular}
\caption{Oxygen abundances for the brightest spaxels showing H{\sc ii}-like spectra in pointings S and M,
estimated following the Pettini \& Pagel (2004) method, using the 
([O{\sc iii}]$\lambda$5007\AA/H$\beta$)/([N{\sc ii}]$\lambda$6583\AA/H$\alpha$) (O3N2) indicator.
\label{metalicidades}}
\end{table}

\newpage
\clearpage

\onecolumn

\onecolumn

\newpage
\clearpage

\newpage
\clearpage

\newpage
\clearpage

\newpage
\clearpage

\newpage
\clearpage

\newpage
\clearpage

\newpage
\clearpage

\newpage
\clearpage

\newpage
\clearpage

\newpage
\clearpage

\newpage
\clearpage

\newpage
\clearpage

\newpage
\clearpage

\newpage
\clearpage

\newpage
\clearpage

\newpage
\clearpage

\newpage
\clearpage

\newpage
\clearpage

\end{document}